\documentclass[
  journal=pasa,
  manuscript=research-paper, %% or "review"
  year=202X,
  volume=YY,
]{cup-journal}

\usepackage[breaklinks,colorlinks,urlcolor=blue,citecolor=blue,linkcolor=blue]{hyperref}
\usepackage{amsmath}
\usepackage{amssymb,microtype,siunitx,booktabs}
\usepackage{array}
\usepackage{subcaption}
\usepackage{caption}
\usepackage{longtable}
\usepackage{lscape}
\usepackage{verbatim}
\usepackage{orcidlink}
\usepackage{threeparttable}
\usepackage[normalem]{ulem}

\newcommand\hi{\mbox{H\,{\sc i}}\ }
\newcommand\hix{\mbox{H\,{\sc i}}}

\title{A FLASH \hi absorption search in compact sources in the pilot ASKAP interplanetary scintillation field}

\author{Paris E. Gordon-Hall \orcidlink{0009-0003-9392-4711}}
\affiliation{Sydney Institute for Astronomy, School of Physics A28, The University of Sydney, NSW 2006, Australia}
\alsoaffiliation{ATNF, CSIRO Space and Astronomy, PO Box 76, Epping, NSW 1710, Australia}

\author{Emily F. Kerrison\orcidlink{0000-0002-0011-6922}}
\affiliation{Sydney Institute for Astronomy, School of Physics A28, The University of Sydney, NSW 2006, Australia}
\alsoaffiliation{ATNF, CSIRO Space and Astronomy, PO Box 76, Epping, NSW 1710, Australia}

\author{Elizabeth K. Mahony\orcidlink{0000-0002-5053-2828}}
\affiliation{ATNF, CSIRO Space and Astronomy, PO Box 76, Epping, NSW 1710, Australia}

\author{Elaine M. Sadler \orcidlink{0000-0002-1136-2555}}
\affiliation{Sydney Institute for Astronomy, School of Physics A28, The University of Sydney, NSW 2006, Australia}
\alsoaffiliation{ATNF, CSIRO Space and Astronomy, PO Box 76, Epping, NSW 1710, Australia}

\author{Hyein Yoon}
\affiliation{Korea Astronomy and Space Science Institute, 776 Daedeok-daero, Yuseong-gu, Daejeon 34055, Korea}
\alsoaffiliation{Sydney Institute for Astronomy, School of Physics A28, The University of Sydney, NSW 2006, Australia}

\author{J.N.H.S. Aditya \orcidlink{0000-0002-0268-0375}}
\affiliation{Sydney Institute for Astronomy, School of Physics A28, The University of Sydney, NSW 2006, Australia}
\alsoaffiliation{Shanghai Astronomical Observatory, Chinese Academy of Sciences, Shanghai 200030, China}

\author{James R. Allison \orcidlink{0000-0003-0436-4680}}
\affiliation{First Light Fusion Ltd., Unit 9/10 Oxford Pioneer Park, Mead Road, Yarnton, Kidlington OX5 1QU, UK}

\author{Rajan Chhetri}
\affiliation{ATNF, CSIRO Space and Astronomy, PO Box 1130, Bentley, WA 6102, Australia}

\author{Ron Ekers}
\affiliation{ATNF, CSIRO Space and Astronomy, PO Box 76, Epping, NSW 1710, Australia}

\author{Marcin Glowacki}
\affiliation{Institute for Astronomy, University of Edinburgh, Royal Observatory, Edinburgh EH9 3HJ, UK}
\alsoaffiliation{Inter-University Institute for Data Intensive Astronomy, Department of Astronomy, University of Cape Town, Cape Town,
South Africa}
\alsoaffiliation{International Centre for Radio Astronomy Research (ICRAR), Curtin University, Bentley, WA, Australia}

\author{John Morgan}
\affiliation{ATNF, CSIRO Space and Astronomy, PO Box 1130, Bentley, WA 6102, Australia}

\author{Vanessa A. Moss}
\affiliation{ATNF, CSIRO Space and Astronomy, PO Box 76, Epping, NSW 1710, Australia}

\author{William Roster}
\affiliation{Max Planck Institute for Extraterrestrial Physics, 85748 Garching near Munich, Germany}

\author{Mara Salvato}
\affiliation{Max Planck Institute for Extraterrestrial Physics, 85748 Garching near Munich, Germany}

\author{Renzhi Su}
\affiliation{Shanghai Astronomical Observatory, Chinese Academy of Sciences, Shanghai 200030, China}

\author{Matthew Whiting}
\affiliation{ATNF, CSIRO Space and Astronomy, PO Box 76, Epping, NSW 1710, Australia}

\email[P.E. Gordon-Hall]{pgor0063@uni.sydney.edu.au}

\received {dd Mmm YYYY}
\revised  {dd Mmm YYYY}
\accepted {dd Mmm YYYY}
\published{22 September 202X}

\begin{document}

\begin{abstract}
In this pilot study, we use data from the First Large Absorption Survey in \hi (FLASH) to search for redshifted \hi 21\,cm absorption at $0.4<z<1$ towards 157 bright radio sources with Interplanetary Scintillation (IPS) measurements at 820\,MHz in a single field observed with the Australian Square Kilometre Array Pathfinder (ASKAP) radio telescope. 
The ASKAP IPS measurements, which use only 2.5 minutes of observing time in total, allow us to estimate the compactness of these radio sources on sub-arcsecond scales at a frequency within the 712--1000\,MHz FLASH band. In particular, the Normalised Scintillation Index (NSI) for each source reflects the fraction of the flux density arising from compact components less than 0.1\,arcsec in diameter. 
We find that $18\pm5$\% of ASKAP sources with flux densities above 150\,mJy are highly compact with NSI $\geq0.8$ -- implying that at least 80\% of their radio emission arises from a single region smaller than about 800\,pc in size. The compactness of these sources makes them ideal probes for an \hi absorption search, since the covering factor for any \hi gas clouds along the line of sight is likely to be high.  About half of the sources with NSI $\geq0.8$ also have peaked radio spectral energy distributions (SEDs), consistent with previous IPS studies at lower frequencies. 
These pilot results imply that IPS measurements with ASKAP can provide a simple and powerful tool for identifying uniform samples of compact radio sources at frequencies of a few hundred MHz across large areas of sky. 
With FLASH, we detect two new \hi absorption lines against compact sources in the $\sim30$\,deg$^2$ region of sky covered by the IPS data; an associated \hi line at redshift $z=0.9540$ (with a matching optical redshift) in MRC\,2125-237 (NSI = 0.98), and a likely intervening line at $z=0.4632$ towards MRC\,2131-241 (NSI = 0.84). The radio sources with these two detections are both bright and compact, with peaked radio SEDs.
\end{abstract}

\section{INTRODUCTION}
\label{sec:int}

Cold neutral atomic hydrogen (\hix) is a tracer for the reservoir of fuel for star formation, and studies of \hi are therefore important in understanding the formation and evolution of galaxies. Absorption lines in the spectra of background quasars can be used to study the cosmic evolution of the baryon content of the Universe \citep{Peroux_2020}, and  \hi 21\,cm absorption lines can also probe the column density of cold \hi between the observer and a background source. Unlike emission, the detection limit for an \hi 21\,cm absorption line is independent of redshift, and is limited only by the spin temperature of the \hi gas, the availability of suitably-bright background radio sources that can be used as probes, and the receiver bandwidth (which determines the frequencies that \hi 21\,cm absorption can be observed). 

The First Large Absorption Survey in \hi (FLASH) is a wide-field radio survey for \hi 21\,cm line absorption in the intermediate redshift range $0.4<z<1.0$ \citep[look-back times of 4-8 Gyr;][]{Allison_2022, Yoon_2025}. FLASH covers 24,000 deg$^2$ of sky, and is one of nine key Survey Science Projects being carried out with the Australian Square Kilometre Array Pathfinder \citep[ASKAP;][]{Hotan_2021}.

\hi 21\,cm line absorption is classified as either associated or intervening. Associated lines are where the neutral gas causing the absorption is intrinsic to the radio source or quasar. Intervening lines are where the neutral gas is lying along the line of sight to the quasar in a separate galaxy. For associated \hi absorption lines the detection rate using FLASH pilot survey data was measured by \cite{Su_2022} and \cite{Aditya_2024} and was found to be 2.9\% and 1.8\% respectively \citep{Yoon_2025}. Intervening lines are usually narrower than associated lines as shown by \cite{Curran_2021}, and they used a literature compilation of known associated and intervening \hi absorption lines to show that the median linewidth for intervening absorption lines was significantly lower than the median linewidth for associated lines. In FLASH data, narrower lines are harder to detect and are possibly misidentified as spectral artefacts \citep[Section 9.3]{Yoon_2025}.

Several targeted searches in \hi found that radio continuum sources with \hi absorption were usually compact, including both compact steep-spectrum (CSS) and peaked-spectrum (PS) sources \citep[e.g.][]{Veron-Cetty_2000, Vermeulen_2003, Philstrom_2003, Gupta_2006, Chandola_2011, Gereb_2015, Maccagni_2017, Morganti_2018, Aditya_2024, Yoon_2025}. 

CSS and PS sources are compact and strong radio sources which are believed to be young radio AGN \citep{odea1998, odea2021}. The radio spectra of the intrinsically small PS sources can appear peaked due to either synchrotron self-absorption or free-free absorption \citep{odea2021}. 
The frequency of the peak in a synchrotron self-absorbed source scales inversely with the angular size of the source, so more compact radio sources are expected to peak at higher frequencies \citep{odea2021}. 
Some earlier publications have discussed separate classes of GHz-Peaked-Spectrum (GPS) sources, High Frequency Peakers (HFPs), and Megahertz Peaked-Spectrum (MPS) sources. In this work, we refer to all these objects as PS sources. 

Large and unbiased catalogues of PS radio sources are now becoming available through the combination of large-area radio continuum surveys across a wide frequency range \citep[e.g.][]{Callingham2017,Slob2022,Kerrison_2025}.  

Interplanetary scintillation (IPS) can produce rapid fluctuations in the intensity of radio waves as they travel through density variations in the solar wind \citep{Clarke_1964, Hewish_1964}. These fluctuations reduce rapidly for sources larger than some characteristic diameter ($\sim 0.1$ arcseconds at 800\,MHz), so measurements of IPS allow us to identify sub-arcsecond scale compact components in low-frequency radio sources. 

\cite{Morgan_2018} showed that the Murchison Widefield Array \citep[MWA;][]{Tingay_2013} could be used to carry out wide-field IPS measurements across large areas of sky at $160$\,MHz. This makes it possible to assemble large samples of compact low-frequency radio sources without the complexity associated with low-frequency long-baseline interferometry. 

\cite{Chhetri_2018} presented the first astrophysical application of the technique of wide-field IPS with the MWA. They used five minutes of data observed at 162 MHz with 0.5\,s time resolution to detect 2550 bright continuum sources across a single 900\,deg$^2$ field. They also introduced the concept of a {\it normalised scintillation index}\ (NSI) to remove the effects of solar elongation on the observed level of scintillation. 

They found that around 12\,\% of sources in this field demonstrated rapid fluctuations caused by IPS, indicating compact structures. They find 9\,\% of bright sources have an NSI $\geq 0.9$ which they define as strongly scintillating, implying that $\geq90$\,\% of their low-frequency emission arises from a sub-arcsecond compact component.

Interestingly, they also found that PS radio sources were the dominant population within strongly scintillating, low-frequency sources in their sample. 
In a follow-up paper, \cite{sadler2019} showed that the strongly scintillating MWA sources were a relatively distant population, with a median redshift of $z\sim1.5$, and that at least 30\,\% of them were likely to lie at $z>2$. Compact (scintillating) sources at these redshifts are ideal targets for \hi absorption studies.

\cite{Aditya_2024} found that two out of the three objects with detected \hi absorption in their sample show IPS at 162 MHz. Both of these objects were also compact with high NSIs. They suggest IPS as an efficient alternative to VLBI in finding compact sources which have an expected higher \hi detection rate \citep[e.g.][]{Curran_2013}.

\cite{Chhetri_2023} used measurements of a single bright source to show that the wide-field IPS technique could be extended to higher frequencies using ASKAP at 862\,MHz with the CRAFT coherent (CRACO) fast transient system \citep{wang2025}. 
Following this, \cite{Chhetri_2026} used all 36 phased array feed (PAF) beams of ASKAP to make the first wide-field IPS observations with ASKAP. 
%
%They found a roughly bimodal distribution of NSI values in the ASKAP sample, in contrast to what was seen in the MWA IPS data, and suggested that this might correspond to two distinct populations - one with compact hot spots embedded in extended lobes, and the other made up of purely compact sources associated with a galactic nucleus.

%\subsection{This paper} 
In this paper, we combine the \cite{Chhetri_2026} IPS catalogue for a single ASKAP field with FLASH \citep{Allison_2022} spectral-line  observations of the same field, to search for \hi absorption against the ASKAP IPS sources. 
%The layout of this paper is as follows. 
In Section \ref{sec:sample}, we discuss how the ASKAP IPS compact source sample was selected, and how it was crossmatched with FLASH to produce the final catalogue of sources focused on in this work. In Section \ref{sec:SED}, we explain the use of \texttt{RadioSED} to fit the radio spectral energy distributions with models that show if the source is likely PS or not. In Section \ref{sec:optical}, we show the Legacy Survey \citep{Legacy_paper} images for the key sources of this paper, and any optical redshifts and counterparts. In Section \ref{sec:HI}, we explain the process of searching for \hi absorption lines, and show the two detections found from the sample of sources searched. In Section \ref{sec:hi_properties}, we discuss the properties of the two \hi detections. In Section \ref{sec:discussion}, we discuss the nature of the sources with \hi detections, and the \hi detection rate. We also discuss the NSI and spectral index relationship, and in particular a source with unusual measurements, the connection between NSI and optical classification, as well as possible future work. In Section \ref{sec:conclusion}, we summarise the key findings in this paper. Throughout this paper, we assume a flat, $\Lambda$, cold dark matter ($\Lambda$CDM) cosmology with values from \cite{Planck_paper}; $H_0=67.4\,$km\,s$^{-1}$\,Mpc$^{-1}$, $\Omega_M=0.315$, and
$\Omega_\Lambda=1-0.315$.

\section{THE ASKAP IPS COMPACT SOURCE SAMPLE}
\label{sec:sample}
\subsection{ASKAP IPS data}
We use the ASKAP IPS data from \cite{Chhetri_2026} to estimate the compactness of the radio sources that have been searched for \hi absorption by FLASH.  
The ASKAP IPS observations used the CRACO system to measure flux densities at 110 ms intervals. These observations were made on 2023 February 2, with 16 observations around the Sun covering solar elongations between 7.5 and 12.5 degrees. The IPS observations were carried out as part of technical tests to determine the feasibility for expanding on IPS work in future, making use of the autonomous scheduling and science operations system implemented for ASKAP (Moss et al. (in prep.)). 

The ASKAP IPS data pipeline produces images made at 110 ms intervals (``snapshot images''), as well as single images for the full duration of the observation (called a ``standard image''), for each ASKAP beam separately. 
The ``variability image'' for each beam is then produced by first applying a filter to remove time variations from the ionosphere to emphasise timescales associated with IPS.
%filtering in the time domain to emphasise timescales associated with IPS, 
Then the standard deviation for each pixel in the snapshot images is taken along the time axis \citep[described further in][]{Morgan_2018, Chhetri_2018}. 
Sources with variance due to IPS have high standard deviation, meaning that only sub-arcsecond scale compact sources remain in the variability image. This makes the identification of compact sources much easier. 

The Aegean source finder \citep{Hancock2018} 
was used to identify the scintillating compact sources in the variability image. 
For each source, a scintillation index was derived by comparing the scintillation flux density from the variability image with the mean flux density from the standard image. A normalised scintillation index (NSI) was then calculated by dividing the observed scintillation index by the predicted scintillation index (calculated using a simple empirical relationship \citep{Hewish_1969, Rickett_1973, Chhetri_2018}), making it possible to compare the scintillation properties of all the sources in the field by removing the effect of solar elongation \citep{Chhetri_2018}. \cite{Chhetri_2026} also estimated upper limits on the NSI for sources that were detected in the standard image, but not in the variability image. We note that some sources have NSIs above 1, and high NSI uncertainty which is potentially due to measurement issues. Finally, these authors measured a `compact' flux density for each source - representing the flux density contained within a region $\sim0.1$\,arcsec in size. This was derived by multiplying the peak flux density of each source by its measured NSI value. 

\begin{table*}[]
    \centering
    \begin{tabular}{lrrrrrrr}
\hline
Field & SBID & RA\_J2000 & Dec\_J2000 & Integration & Date & Sky coverage & Central \\ 
& & (hms) & (dms) & time (hrs) & observed & (deg$^2$) & frequency (MHz) \\
\hline
FLASH 350 & 55464 & 21:44:09.20 & -25:07:52.0 & 2 & 17-12-2023 & 6.4 x 6.4 & 855.5 \\
FLASH 405 & 62794 & 21:49:05.59 & -18:51:49.0 & 2 & 11-06-2024 & 6.4 x 6.4 & 855.5 \\ 
\hline         
    \end{tabular}
    \caption{Pointing centres, integration times, observation dates, sky coverage and central frequencies of the two ASKAP FLASH fields used in this study. The SBID is the observation number listed in the CASDA archive.}
    \label{tab:flash1}
\end{table*}

\subsection{FLASH data in the ASKAP IPS field}
\begin{figure}
    \centering
    \includegraphics[angle=0,width=1.0\hsize]{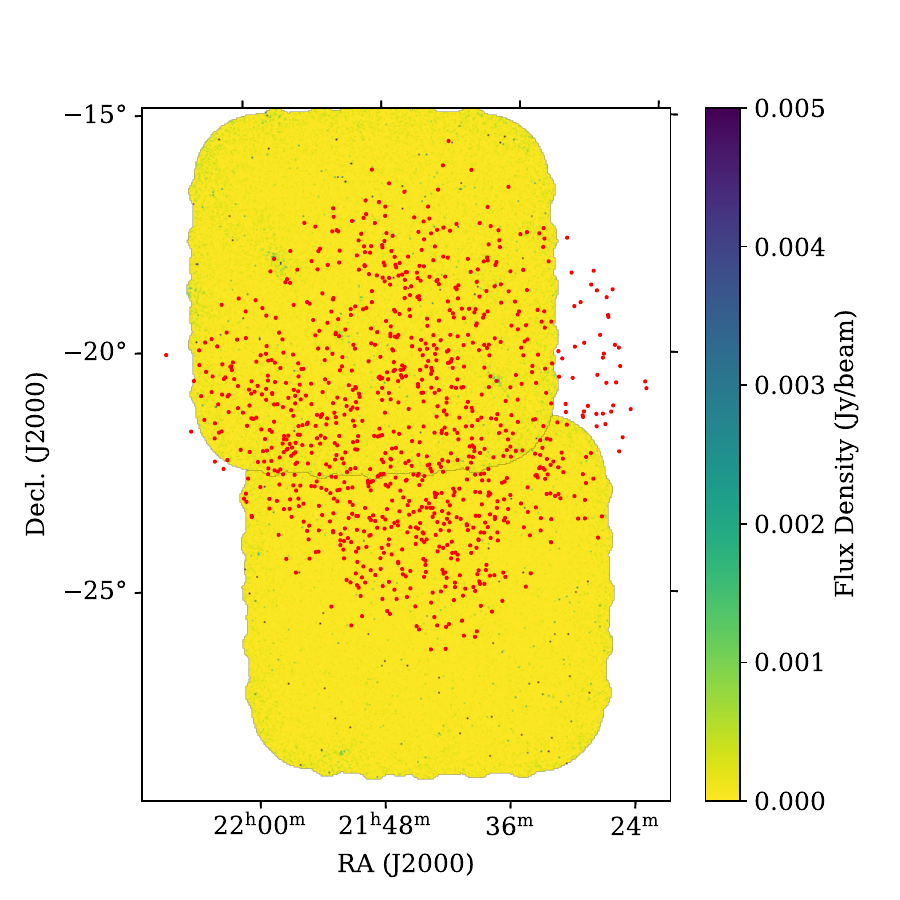}
    \caption{The radio continuum images of the two FLASH fields, SBID55464 and SBID62794, with the ASKAP IPS source locations overlaid on top shown by the red points. There is some overlap between the two FLASH fields.}
    \label{fig:sky_coverage}
\end{figure}

Two FLASH survey fields, listed in Table \ref{tab:flash1}, overlap the single ASKAP IPS field analysed by \cite{Chhetri_2026}. Figure \ref{fig:sky_coverage} shows the radio continuum sources in the two FLASH fields as well as the positions of all the ASKAP IPS sources (including those not covered by FLASH) shown in the Appendix in Tables \ref{tab:main_bright} and Table \ref{tab:main_add}. 

Processed FLASH data products for both fields are available from the public CSIRO ASKAP Science Data Archive (\href{https://data.csiro.au/domain/casdaObservation}{CASDA}), and include continuum source catalogues (at a central frequency of 856\,MHz and with 12-15\,arcsec resolution) produced by the Selavy source finder for each field \citep{Whiting_2012}. 

The quality of each processed FLASH observation was checked by members of the science team, who provided validation notes to accompany the CASDA data release. 
As discussed by \cite{Yoon_2025}, the FLASH data are largely free from terrestrial radio frequency interference (RFI), though small regions of data can be affected by spectral-line artefacts which was a factor in these datasets. 

The continuum island source catalogues (see \cite{Yoon_2025} for further description) for FLASH fields 350 (SBID 55464) and 405 (SBID 62794) were each crossmatched separately with a combined ASKAP-MWA IPS catalogue (prepared by \cite{Chhetri_2026}) as outlined in Table \ref{tab:crossmatch}. We selected a large search radius of 30 arcseconds to encompass all possible matches, and then only included the best match for each ASKAP IPS source. Since these FLASH fields overlap, some sources had a match in both FLASH fields. Both of these matches were included in the final catalogue.
The number of IPS continuum sources is smaller than the number of FLASH continuum detections because of the very short IPS observing time (2.5 minutes \citep{Chhetri_2026}); compared to the much deeper FLASH catalogue with an integration time of two hours per field \citep{Allison_2022, Yoon_2025}. There are also significantly fewer sources once only scintillating sources are selected in the variability image.
From the crossmatched catalogue, we selected only sources with a FLASH integrated flux density $S_{856 \text{MHz}}\geq30$\,mJy, which is the limit for spectrum extraction for most FLASH fields. This limit is imposed to balance processing time, data volume, and overall efficiency since the detection of \hi in the FLASH spectral-line data is not likely for fainter sources \citep{Yoon_2025}.  
\cite{Chhetri_2026} found that the ASKAP IPS  observations are complete for compact components above 150\,mJy/beam down to NSI~0.1 - i.e. they  detect all compact components with flux density $S_{820\text{MHz}}>150$\,mJy. Sources with high NSI values can be detected down to total flux densities well below 150\,mJy. 

Finally, we selected two separate FLASH-IPS samples: 
\begin{enumerate}[label=(\alph*)]
    \item 
    {\bf The bright-source sample}.  This sample contains the 88 sources that lie within the FLASH footprint and have both peak flux density $\geq150$\,mJy and continuum S/N $\geq20$ in the ASKAP IPS catalogue. The S/N limit ensures that all the included sources are in regions of the ASKAP IPS field where the upper limits on NSI are restrictive, so that this sample can be used for statistical analyses. The bright-source sample contains 59 objects with detected NSI, and 29 with upper limits in NSI.  
    \item 
    {\bf The additional sample}. This sample contains a further 69 sources that lie within the FLASH area and have flux density $\geq30$\,mJy, but do not meet the criteria for inclusion in the bright-source sample. Only sources with detected scintillation, and a measured NSI value are included in the additional sample, since the upper limits on NSI are not generally restrictive in this flux density range.
\end{enumerate}

\begin{table}
    \centering
    \begin{tabular}{l r}
    \hline
Sample & Number \\
    \hline
       ASKAP-MWA IPS catalogue  & 1026 \\
       FLASH Field 350 (SBID 55464)  & 15901 \\
       FLASH Field 405 (SBID 62794) & 11192 \\
       Crossmatch of FLASH and IPS (includes duplicate matches) & 1137 \\
       Unique sources $\geq$30\,mJy with NSI detection & 128\\
       \hspace*{0.4cm} (after removing duplicates) & \\
\hline
       Bright-source sample ($\geq150$\,mJy, includes NSI upper limits) & 88 \\
       Additional sample (other FLASH sources, NSI det. only) & 69 \\
       \hline
    \end{tabular}
    \caption{A summary of the number of sources in each crossmatching step, after filtering for bright sources with an ASKAP NSI, and in the final two samples of sources searched for \hi absorption. }
    \label{tab:crossmatch}
\end{table}

\section{RADIO SED FITTING}
\label{sec:SED}

\subsection{Fitting method} 
To determine which of our sources had a peaked radio spectrum, we fitted the radio Spectral Energy Distributions (SEDs) for each of the 157 sources in the bright-source and additional samples. 

The SED fitting was done using radio flux density measurements from the 888\,MHz 
Rapid ASKAP Continuum Survey (RACS) \citep{McConnell_2020} as well as several other large-area, publicly available continuum surveys. These included the Galactic and Extragalactic All-Sky MWA Extended Survey (GLEAM-X) (76-227 MHz) \citep{Hurley-Walker_2022}, the NRAO VLA Sky Survey (NVSS) (1.4 GHz) \citep{condon1998}, the Very Large Array Sky Survey (VLASS) (3\,GHz) \citep{Lacy_2020}, the Australia Telescope 20 GHz Survey (AT20G) (5, 8, 20 GHz) \citep{Murphy_2010}. 

Modelling of the SEDs was done with \texttt{RadioSED}, a package designed to be used with broadband radio continuum data \citep{Kerrison2024}. This code performs Bayesian inference using nested sampling \citep{Skilling_2004}. Four models are implemented including a simple power law, a log-space parabola (described in \cite{Dallacasa_2000} and \cite{Orienti_2007}), a functional form for a PS source, which reduces to a synchrotron self-absorbed source under the condition k=2.5 \citep{Snellen_1998}, and a `retriggered' model comprising the functional form from \cite{Snellen_1998} combined with an additional power law component. \texttt{RadioSED} infers which of these analytical models is the best fit for each SED. 

Tables \ref{tab:main_bright} and \ref{tab:main_add} list the full data table, displaying the source name, alternate NED name, object type, coordinates, photometric redshift, spectroscopic redshift, ASKAP NSI, and SED result. 

\subsection{SED properties of the bright-source sample}

Table \ref{tab:bright_sed} shows the distribution of SED fits for the 88 sources in the bright-source sample with $\rm{S}\geq150$\,mJy, which represents the general galaxy population at this flux density level. Since we include sources that only have an NSI upper limit in the bright-source sample, which are assumed to be extended sources, and peaked sources are only compact sources, we find as expected that most of the sources in this sample are best-fitted by power law radio spectra. 

\begin{table}[]
    \centering
    \begin{tabular}{lrrrrrr}
\hline
  SED class     & Number &  Fraction \\
       \hline
Power law & 72 & 82\% \\
Peaked & 12 &  14\% \\
Flat/inverted & 4 & 4\%  \\
Total & 88 & \\
       \hline
    \end{tabular}
    \caption{ Radio SED distribution for the 88 sources in the bright-source ($\rm{S}\geq150$\,mJy) sample.  }
    \label{tab:bright_sed}
\end{table}

Peaked radio SEDs are found in 14\% %14$\pm4$\%
of sources in Table \ref{tab:bright_sed}. This is roughly consistent with the $\sim10\%$ fraction found in bright-source samples at GHz frequencies \citep{odea1998} and the value of 9\% % $9\pm3\%$ 
for PS sources in the MWA IPS sample \citep{Chhetri_2018,sadler2019}. 
We see no evidence for a variation in the PS fraction with flux density in the current (relatively small) sample of 88 sources. 
%In comparison, \cite{Planck_collab} found 10\% of the sources in their sample of high-frequency radio sources were PS and compact. 

Before moving to a more general discussion of the relationship between source size and radio SED, we first summarise the distribution of NSI values within this flux density-limited sample.  The relationship is discussed in detail for a larger ASKAP IPS sample by \cite{Chhetri_2026}, but Table \ref{tab:bright_nsi} shows the breakdown for the two IPS samples used in the current paper to search for \hi absorption lines. 

\begin{table}[]
    \centering
    \begin{tabular}{lrccrrrrr}
\hline
  Source type & NSI  & Bright-source &  Additional  \\
              &      & \multicolumn{1}{c}{sample} & \multicolumn{1}{c}{sample} \\
\hline
Compact   & $\geq0.8$ & 16 (18\%) &  45 (65\%) \\  
Partly resolved   & 0.4--0.8  & 16 (18\%) & 20 (29\%)    \\
Extended  & $\leq0.4$ & 48 (55\%) & 4 (6\%)\\
\multicolumn{2}{l}{Unclear (NSI upper limit $>0.4$)} & 8 (9\%) & \\
Total &  & 88 & 69 \\
\multicolumn{2}{l}{Median NSI} & $0.26\pm0.05$ & \\%$0.88\pm0.16$ \\
       \hline
    \end{tabular}
    \caption{ASKAP NSI distributions for the 88 sources in the bright-source sample and the 69 sources in the additional sample.}
    \label{tab:bright_nsi}
\end{table}

The median ASKAP NSI value for the bright-source sample is $0.26\pm0.05$, which we can consider to be representative of the overall source population at this flux density level. For the fainter additional sample, the median NSI is $0.88\pm0.16$. We know that this will be biased towards sources with high NSI values (compact sources) since NSI upper limits are not included in the additional sample. For this reason, the additional sample's median NSI is not included in Table \ref{tab:bright_nsi}.

\subsection{SED properties of compact and resolved sources}

Table \ref{tab:sed_nsi} shows the distribution of radio SED classes, fitted as either peaked, power law, or flat (which included sources with an inverted radio spectrum),
for the bright-source sample and additional sample, split by NSI value into compact (NSI $\geq0.8$), slightly resolved ($0.4<\text{NSI}<0.8$), or extended (NSI $\leq0.4)$ sources. As explained in Section \ref{sec:int}, the NSI represents the fraction of the low-frequency emission from the source arising from a sub-arcsecond compact component. We include the additional sample here as well as the bright-source sample so that our statistics were not biased towards bright sources.

As seen in Table \ref{tab:sed_nsi}, almost half of the compact sources have peaked radio spectra, as was also found by \cite{Chhetri_2018} at the lower MWA frequency of 162\,MHz, while all of the extended sources are fitted by a power law. Only 9\% of the compact sources have flat or inverted radio spectra. Note that IPS cannot yet be used to differentiate between a slightly resolved compact structure or a point source embedded in a completely resolved extended structure. A source with a slightly resolved NSI could be a double with unresolved components, or a single compact source with some extended emission resolved out \citep[see][Figure 5 for source morphologies consistent with a particular NSI]{Morgan_2019}.

\begin{table}[]
    \centering
    \begin{tabular}{lrrrrrr}
\hline
& & \multicolumn{3}{c}{Radio SED fit} \\
Source type  & NSI  & Peaked & Power law & Flat/Inv & Total\\
    \hline
    Compact & $\geq0.8$ & 27 & 26 & 5 & 58\\
            &     & (47\%) & (45\%) & (9\%) & (37\%)\\  
    Partly & 0.4--0.8 & 5 & 31 & 3 & 39\\
    resolved  &  & (13\%) & (79\%) & (8\%) & (25\%)\\
Extended & $\leq0.4$ & 0 & 60 & 0 & 60\\
          &    & (0\%) & (100\%) & (0\%) & (38\%)\\
    \hline
    \end{tabular}
    \caption{Radio SED distribution for compact, resolved, and extended sources with NSI detections or upper limits. This table includes all 157 sources from the bright-source sample and additional sample.}
    \label{tab:sed_nsi}
\end{table}

\section{OPTICAL IDENTIFICATIONS AND REDSHIFTS}
\label{sec:optical}

\subsection{Optical counterparts of the FLASH-IPS sources}
We used the FLASH optical ID catalogue made by \cite{Roster_2026} to identify optical counterparts for the FLASH-IPS sources listed in Tables \ref{tab:main_bright} and \ref{tab:main_add}.  
These authors used machine learning techniques to identify the counterparts of $\sim80,000$ unresolved (angular size $<10-15$\,arcsec) radio sources from the FLASH survey by crossmatching the radio positions with optical images from the DESI Legacy Imaging Surveys Data Release 10 \citep[LS10;][]{dey2019}. LS10 provides four-band $griz$ images and catalogues for much of the southern sky, as well as classification of each catalogued object as either an optical point source (PSF) or one of four classes of galaxies: round exponential galaxies with a variable radius ("REX"), deVaucouleurs ("DEV") profiles (elliptical galaxies), exponential ("EXP") profiles (spiral galaxies), and Sersic ("SER") profiles. The catalogue made by \cite{Roster_2026} also includes photometric redshift estimates derived using the PICZL code \citep{roster2024}. The optical identifications and photometric redshifts from the \cite{Roster_2026} catalogue were accepted if the optical counterpart probability $p_{\rm{any}}$ had a value above 0.29. This cutoff provides a good compromise with roughly 80\% purity and 80\% completeness, corresponding to an
estimated chance-association rate of $\sim20$\% \citep{Roster_2026}.  

Unfortunately, the LS10 imaging in the ASKAP IPS field is incomplete, with images for much of the region only available in one or two of the four bands. For this reason, we grouped the four optical galaxy classes together, and classified each object as either a quasar (LS10 PSF classification), or a galaxy. The photometric redshifts listed in Tables \ref{tab:main_bright} and \ref{tab:main_add} should also be regarded as indicative in many cases, especially for those with $z>2$. 

We also searched the NASA/IPAC Extragalactic Database (NED) for optical spectroscopic redshifts of sources in the FLASH-IPS sample. This yielded 16 redshift measurements (9 galaxies and 7 quasars). Redshifts for a further 11 quasars are available from Gaia spectra as listed in the Quaia catalogue \citep{quaia2024}. 

Optical counterparts and photometric redshift estimates are available for 54 of the 88 sources in the bright-source sample (Table \ref{tab:main_bright}), and 48 of the 69 sources in the additional sample (Table \ref{tab:main_add}). 

We can be confident that all the radio sources in Tables \ref{tab:main_bright} and \ref{tab:main_add} are genuine; they are many times brighter than the $\sim0.5$\,mJy detection limit of the FLASH images and also appear in catalogues from the earlier NVSS \citep{condon1998} and RACS \citep{hale2021} radio surveys. However, about 30\% of radio sources in Tables \ref{tab:main_bright} and \ref{tab:main_add} do not have a listed optical counterpart, and there are several reasons for this:
\begin{itemize}
\item 
{\it Extended radio sources: }\ Around 10\% of FLASH sources are extended on scales larger than 15-30\,arcsec, and resolved into two or more components in ASKAP images. Since identifying optical counterparts for these sources is more challenging \citep[see e.g.][]{best2005,ching2017}, they were not included in the \cite{Roster_2026} catalogue.
\item 
{\it Ambiguous optical counterpart:}\ In a few cases, two similar objects are visible on the LS10 images, and the PICZL code is unable to determine which is the more likely counterpart. It is also possible that both objects could contribute to the observed radio emission. 
\item 
{\it Low-probability optical counterpart:}\ Here, a  potential counterpart is visible on the LS10 images, but is unlikely to be associated with the radio source. The true counterpart may be a more distant galaxy that is not visible, or the radio-optical offset may be large enough that we cannot be confident that the association is genuine.
\item 
{\it Blank field:}\ The counterpart is fainter than 
the detection limit of the LS10 optical images, typically $>21.5$\,mag in $g$ and $r$ in the ASKAP IPS region \citep{saxena2024}.
\item
{\it Foreground stars: }\ In a small number of cases (e.g. J214032.5-211559 and J214011.6-182529 in Tables \ref{tab:main_bright} and \ref{tab:main_add}), the presence of a bright Galactic foreground star means that reliable photometry is not available for nearby objects. 
\end{itemize}

\subsection{Redshift distribution}
\begin{figure}
    \centering
    \includegraphics[angle=0,width=0.9\hsize]{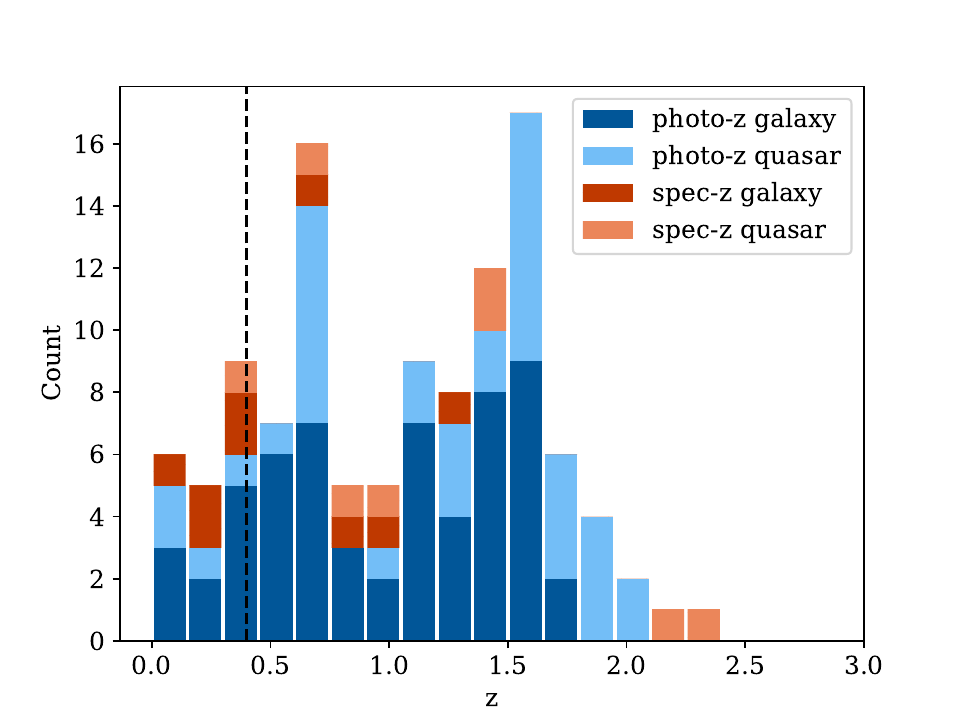}
    \caption{Histogram of the photometric and spectroscopic redshifts for sources divided into whether they were classified as either quasars or galaxies. Photometric is shown in blue, and spectroscopic is shown in striped orange with the paler colours showing quasars and darker colours showing galaxies. The two photometric redshift outliers are not shown in the diagram. The dashed vertical line is at $z=0.4$.}
    \label{fig:redshift_hist}
\end{figure}

Figure \ref{fig:redshift_hist} shows a stacked histogram of the photometric redshifts and spectroscopic redshifts separately and split into either quasars or galaxies, as was also done by \cite{Kerrison_2025}. A small number of galaxies with PICZL photometric redshifts of $z>2$ are not included in this diagram. Unlike quasars, most galaxies with $z>1.5$ are too faint to be visible in the LS10 images \citep{Roster_2026}. The high photometric redshifts estimated for these few objects are likely to be unreliable due to the low SNR and missing bands in the LS10 photometry. 

\section{\hi ABSORPTION-LINE SEARCH}
\label{sec:HI}

We use the FLASH spectral-line data to search for \hi 21\,cm absorption against the sources listed in Tables \ref{tab:main_bright} and \ref{tab:main_add}. This is an untargeted search in the sense that we search across the full FLASH redshift range ($0.4<z<1.0$) for all objects, and do not apply any form of optical preselection. 

\subsection{\hi absorption search}
Spectral-line profiles and linefinder results were collected from the FLASH database after validation by the FLASH team. FLASHfinder is a Bayesian line-finding tool \citep{Allison_2012} designed to make an automated search for \hi absorption lines in large ASKAP spectral-line datasets \citep[e.g.][]{Glowacki_2019, Allison_2020, Sadler_2020, Allison_2021, Mahony_2022, Renzhi_2022, Aditya_2024, Yoon_2025}. FLASHfinder is run on the extracted spectra from each SBID, and a single Gaussian component is fit to each candidate line. The output data file includes measurements and uncertainties of: the redshift of the peak of the candidate line, the peak optical depth of the line, the integrated optical depth of the line, the FWHM linewidth with a single Gaussian fit of the line, and the Bayesian evidence value ln\,(B). The ln\,(B) parameter quantifies how likely it is that the line detection exists compared to the null hypothesis, which is where no line is present. The median rms noise level of the spectra is around $5.5 \text{ mJy beam}^{-1}$ \citep{Yoon_2025}.

The output file contains all candidate detections which could be astronomical signals or artefacts, and only candidates with a $\ln\,(B)$ value above 30 were considered as robust detections in this paper. As explained by \cite{Yoon_2025}, a conservative approach is to consider lines for which $\ln\,(B) \geq 30$ because when the original spectra are inverted and the linefinder is re-run, only spectral artefacts and noise should be seen. As shown in their Figure 11, $\ln\,(B) \sim 30$ is an upper envelope to most of the linefinder points from the inverted spectrum. 
However, they acknowledge that choosing this value of ln\,(B) has likely resulted in some true absorption lines being missed. Since the ASKAP IPS catalogue only covers two FLASH SBIDs, ln\,(B) values down to 10 were checked by visual inspection to identify potential detections. An expectation is that only one or two spectra will contain \hi absorption in an SBID, while spectral artefacts will typically occur in the same frequency position in multiple spectra.

\subsection{\hi detections} 
Two strong \hi absorption lines were detected, in J212805.5-232946 (with ln\,B = 127) and J213437.6-235537 (with ln\,B = 198).  

\begin{figure*}[t!]
    \centering
    \begin{subfigure}[t]{0.5\linewidth}
        \centering
        \includegraphics[height=7cm]{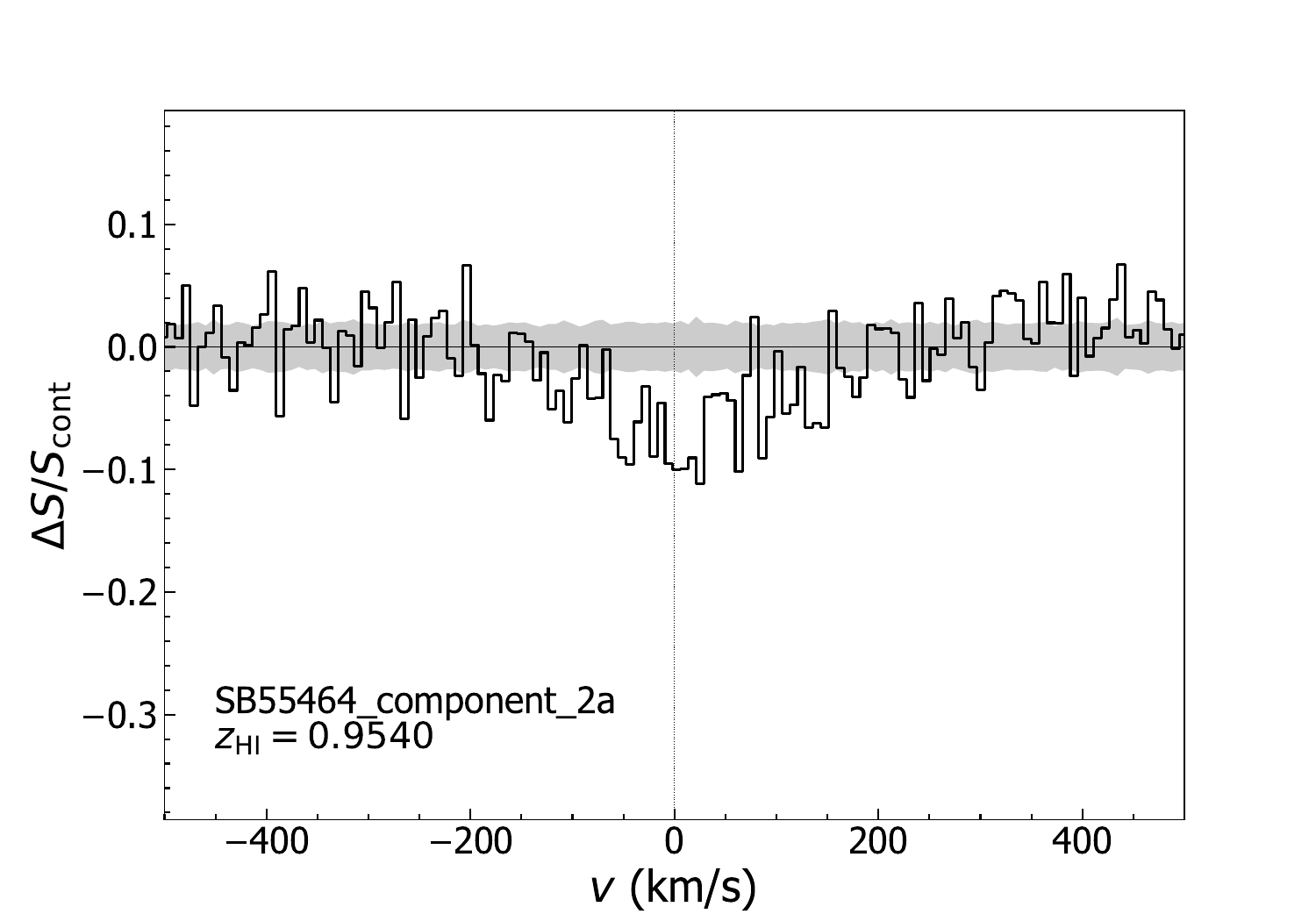}
    \end{subfigure}%
    ~ 
    \begin{subfigure}[t]{0.5\linewidth}
        \centering
        \includegraphics[height=6cm]{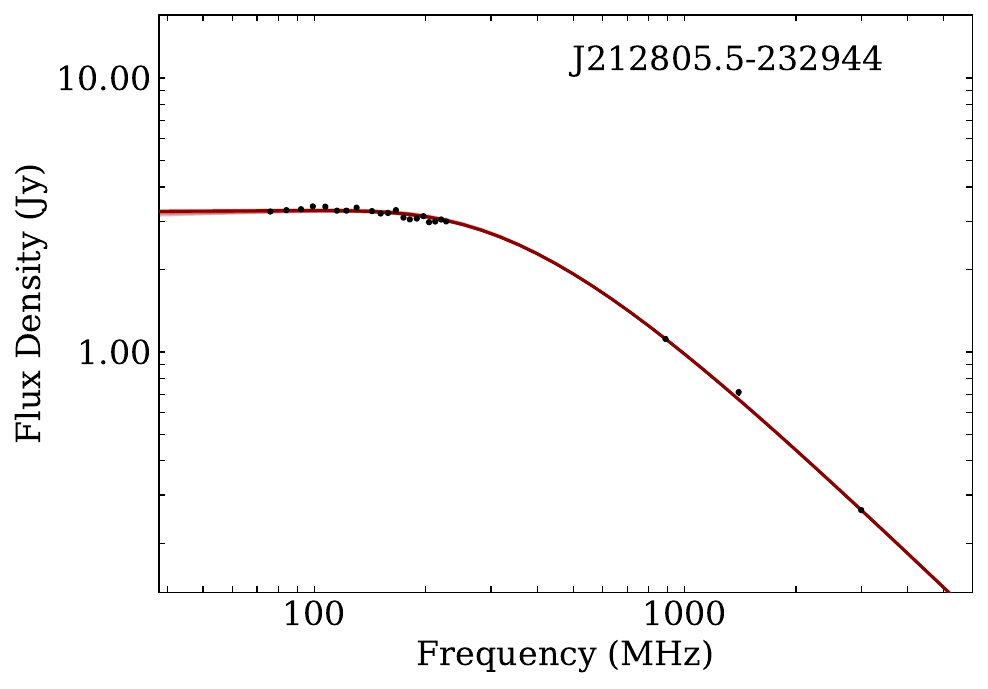}
    \end{subfigure}
    \caption{The \hi absorption spectral-line profile on the left, and radio SED fit for J212805.5-232944 (MRC\,2125-237) on the right. The velocities are relative to $z=0.9540$. The grey band in the spectra shows the rms spectral-line noise \citep{Yoon_2025}. The favoured model is peaked, with a Bayes factor of of 4.4.}
    \label{fig:spectra_sed_J212805}
\end{figure*}

\begin{figure*}[t!]
    \centering
    \begin{subfigure}[t]{0.5\linewidth}
        \centering
        \includegraphics[height=7cm]{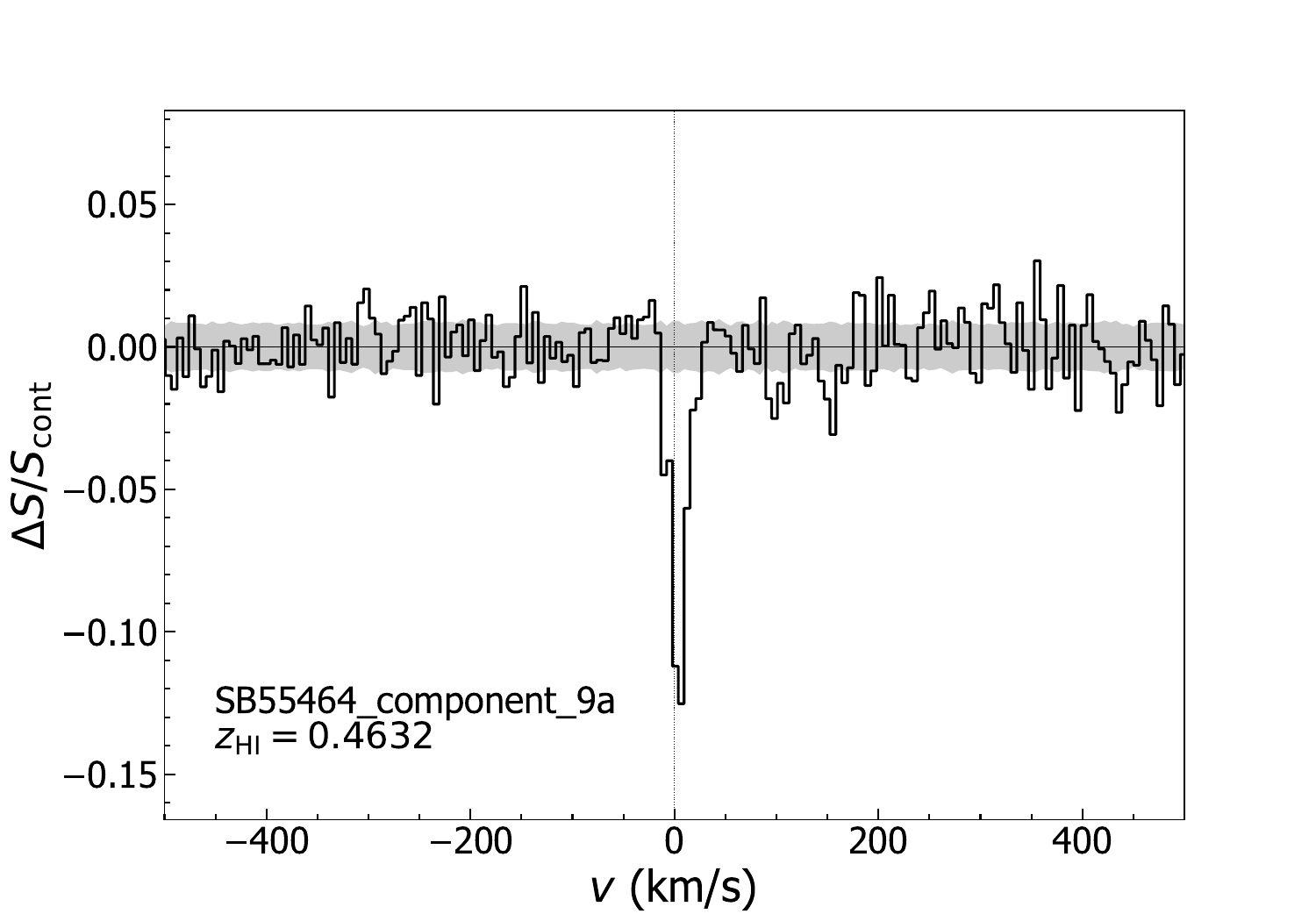}
    \end{subfigure}%
    ~ 
    \begin{subfigure}[t]{0.5\linewidth}
        \centering
        \includegraphics[height=6cm]{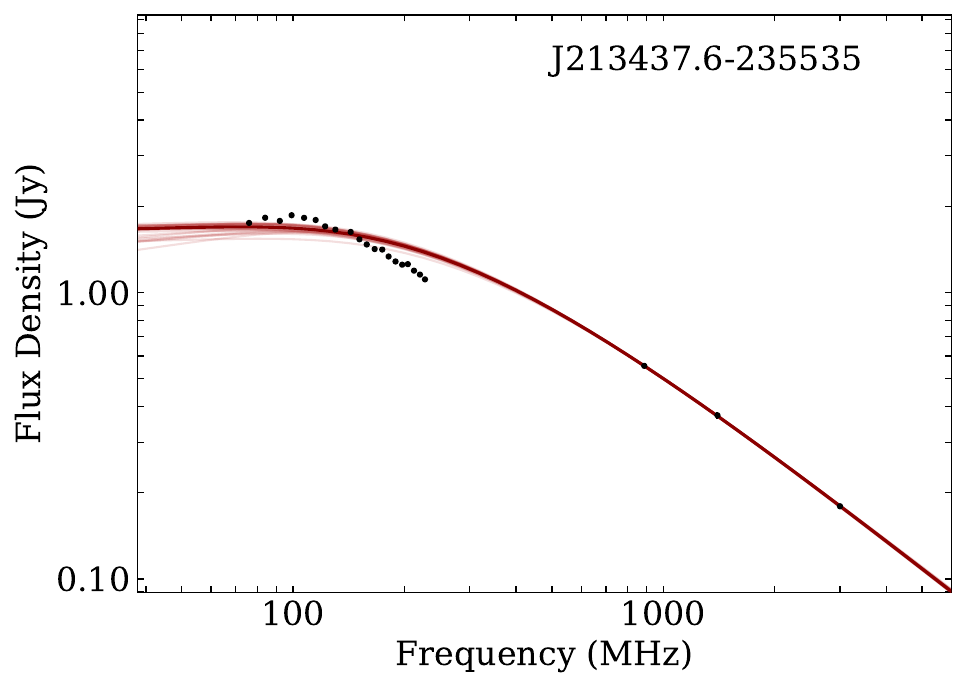}
    \end{subfigure}
    \caption{The \hi absorption spectral-line profile on the left, and radio SED fit for J213437.6-235535 (MRC\,2131-241) on the right. The velocities are relative to $z=0.4632$. The grey band in the spectra shows the rms spectral-line noise \citep{Yoon_2025}. The favoured model is peaked, with a Bayes factor of 1.8. This source demonstrates that the model does not always provide a perfect fit, reflecting the increased complexity or variability in some cases.}
    \label{fig:spectra_sed_J213437}
\end{figure*}

The \hi spectra of these two detections are shown in Figures \ref{fig:spectra_sed_J212805} and \ref{fig:spectra_sed_J213437}. The parameters fitted by the FLASHfinder are listed in Table \ref{tab:flash}.
Both of the sources with \hi detections were identified as PS radio sources by 
\cite{Callingham2017}. 

\begin{table*}[]
    \centering
    \begin{tabular}{lrrrrrrr>{\centering\arraybackslash}m{1.2cm}|c|>{\centering\arraybackslash}m{2.5cm}|}
 \hline
    Source name & FLASH field & $z_{\hi}$ & $\tau_{\text{peak}}$ & $\tau_{\text{int}}$ & Width & $\ln\,(B)$ & FLASH integrated flux \\
    & & & & (km s$^{-1})$ & (km s$^{-1})$ & & density (mJy) \\
    (1) & (2) & (3) & (4) & (5) & (6) & (7) & (8)\\
\hline
    MRC\,2125-237 & 350 & 0.9540 & $0.075\pm0.006$ & $16.08_{-1.13}^{+1.15}$ & $215.8_{-16.9}^{+18.5}$ & 125.7 & 1145.64\\
    MRC\,2131-241 & 350 & 0.4632 & $0.123\pm0.009$ & $2.25_{-0.14}^{+0.13}$ & $18.2_{-1.4}^{+1.4}$ & 198.0 & 535.85\\
    \hline \\
    \end{tabular}
    \caption{The two \hi absorption detections with FLASHfinder results for the spectra; (1) Source name; (2) FLASH field name; (3) \hi redshift; (4) Peak optical depth upper limit (except for detections) and uncertainty; (5) Integrated optical depth and uncertainty; (6) Linewidth and uncertainty; (7) $\ln\,(B)$; (8) FLASH integrated flux density (mJy) at 855.5 MHz. The uncertainties in columns (4) to (6) represent the 68\% interval around the median. The table is based on Tables 5 and 6 in \cite{Yoon_2025}. }
    \label{tab:flash}
\end{table*}

We also investigated 10 tentative detections of weaker lines with ln\,B values as low as 10. Only one of these was from SBID 55464 (Field 350), while the other nine were from SBID 62794 (Field 405), which is known to have some regions of poorer-quality data. None of these tentative lines were accepted as a robust detection of an \hi line. Four of them lay in a narrow spectral region between 832 and 835 MHz (\hi z = 0.701 to 0.707) where the spectral-line noise in SBID 62794 is higher than usual, which is likely due to the tropospheric ducting of RFI, (see \cite{Yoon_2025} for a description of how ducting affects the spectra). Two are at the position of known correlator artefacts, and the remaining four have ln B $<20$, and are statistically likely to be noise peaks.

\subsection{Upper limits for non-detections}
\begin{figure}[ht]
    \centering
    \includegraphics[width=1.0\linewidth]{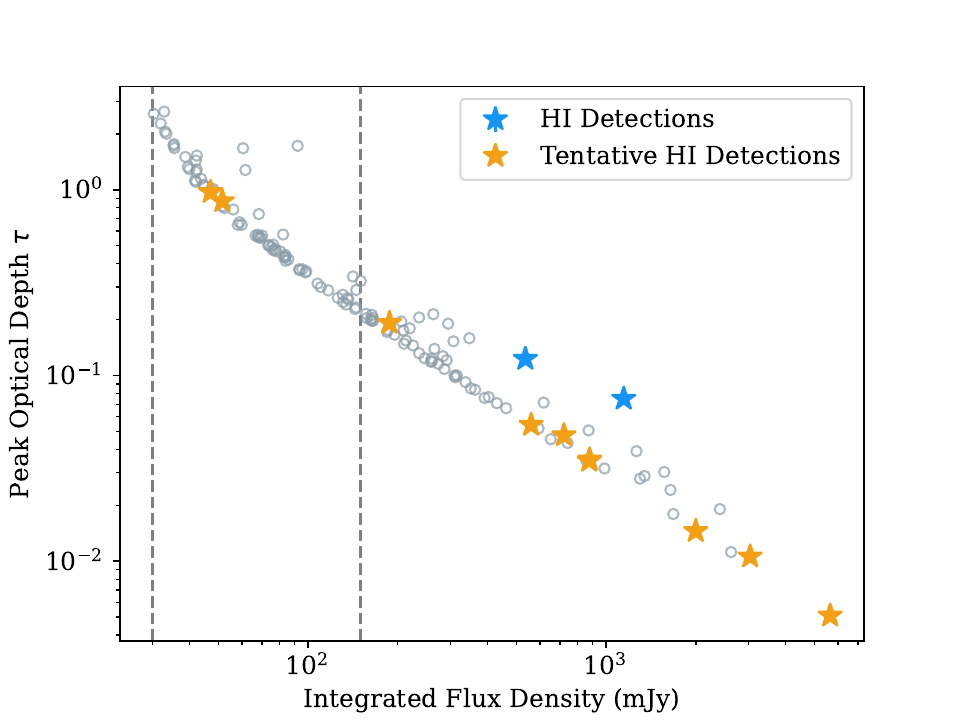}
    \caption{The upper limits of the peak optical depth for non-detections (shown by the grey circles) and the peak optical depths of the \hi lines detected (shown by the blue stars) against the integrated flux density. Sources with tentative detections have their upper limits shown by the orange stars. The vertical dashed lines show the 30 mJy limit used for the \hi line search, as well as the 150 mJy limit for the bright-source sample.}
    \label{fig:peak_opt_depth}
\end{figure}

For sources with no \hi detection, the upper limit of the peak optical depth was calculated using the equation for peak optical depth $\tau$ given by: $$I/I_0=e^{-\tau}$$ where $I_0$ is the peak flux density of the source, $I$ is the observed flux density at the peak of the line, and $(I_0-I)$ is the size of the `dip' below the continuum level. For an upper limit, where $\sigma$ is the rms noise in a single spectra channel, and we assume a detection limit of $5\sigma$, the equation becomes: $$\tau=-\ln[(I_0-5\sigma)/I_0]$$ We assume a value of $\sigma \sim 5.5$ mJy.beam$^{-1}$ per spectral channel. Tables \ref{tab:main_bright} and \ref{tab:main_add} show the upper limits of the peak optical depth for the sources with no \hi detections, and we present the upper limits of the peak optical depth for all the sources with non-detections, sources with tentative detections, as well as the peak optical depth for the two \hi detections against the integrated flux density in Figure \ref{fig:peak_opt_depth}.

\section{PROPERTIES OF THE \hi DETECTIONS}
\label{sec:hi_properties}
The LS10 images at the positions of the two sources (shown in Table \ref{tab:mrc}) with detections of \hi absorption are shown in Figure \ref{fig:legacy_images}.
Both of these objects, MRC 2125-237 (J212805.5-232944) and MRC 2131-241 (J213437.6-235535) are bright radio sources included in the MRC-1Jy catalogue \citep{McCarthy_1996}. In each figure, the red circles are centred on the MRC-1Jy radio position with a radius of 5\,arcsec on the sky.

The MRC-1Jy catalogue contains sources with a flux density above  1\,Jy at 408\,MHz, that lie within a strip of the southern sky at declinations between about $-20$ and $-30$\,degrees. The additional data available for the MRC-1Jy sample includes high-resolution radio images, deep optical images, and (for about half the sample) optical spectra and redshift measurements. 

Table \ref{tab:mrc} lists the radio positions measured for these two sources by \cite{kapahi1998} using high-resolution continuum images from the VLA at 5\,GHz. These positions are expected to be accurate to better than 0.1\,arcsec, while the LS10 optical positions are tied to the Gaia reference frame and should generally be accurate to within 0.2\,arcsec. The MRC-1Jy radio positions are more accurate compared to the ASKAP positions listed in the main tables (Tables \ref{tab:main_bright} and \ref{tab:main_add}) which have a typical uncertainty of $\sim 1$ arcsec. Since MRC-1Jy positions are not available for all sources, the ASKAP positions were used for the optical crossmatching in Section \ref{sec:optical}.

\begin{table*}[ht]
\begin{tabular}{lllllllclll}
\hline
Source & \multicolumn{2}{c}{MRC-1Jy radio position} & \multicolumn{2}{c}{MRC-1Jy optical position} & \multicolumn{2}{c}{LS10 optical position} & Radio-optical offset \\
  & \multicolumn{6}{c}{(J2000) }  & (MRC-1Jy-LS10, arcsec) \\
\hline
MRC\,2125-237 & 322.023127 & -23.496003 & 322.023334 & -23.495697 & 322.023141 & -23.495966 & 0.14 \\
MRC\,2131-241 (LS10 PSF) & 323.657001 & -23.926776 & .. & ..& 323.656892	& -23.926791 & 0.36 \\
MRC\,2131-241 (LS10 galaxy) & 323.657001 & -23.926776 & 323.657959 & -23.927246 & 323.657357 & -23.927154 & 1.80 \\
\hline
\end{tabular}
\caption{Radio and optical positions for the two radio sources with detected \hi absorption, and their potential optical counterparts.}
\label{tab:mrc}
\end{table*}

\subsection{MRC 2125-237 (J212805.5-232944) }  
The optical counterpart of the radio source MRC 2125-237 (shown in Figure \ref{fig:legacy_images} on the top), is a galaxy whose optical redshift of $z=0.95\pm0.05$ \citep{McCarthy_1996}, matches the \hi absorption redshift of $z=0.9540\pm0.0001$. This galaxy has an optical $g-i$ colour of 1.69 mag, placing it in the region occupied by high-excitation radio galaxies \citep[HERGs;][]{heckman2014} in this redshift range \citep{ching2017}. The galaxy has a fainter potential companion 2.5 arcsec away, corresponding to a projected separation of 20\,kpc on the sky at the \hi redshift. 

The broad \hi absorption FWHM linewidth of 216 km s$^{-1}$ suggests that this is an associated \hi system where the \hi gas lies in or around the same galaxy that hosts the radio source \citep[e.g.][]{Curran_2021}. This is also supported by the close match between the optical and \hi redshifts. As shown in \cite{Curran_2021}, the probability of the \hi absorber being intervening depends upon the full width at zero intensity (FWZI) of the profile. They find that for values of FWZI below approximately 120 km s$^{-1}$, there is a high probability (>70\%) that the absorber is intervening. For values greater than 120, the probability sharply drops, and for FWZI $\gtrsim 200$ km s$^{-1}$, the probability of the absorber being intervening is around 20\%. The FWZI for both detections was found by eye from the spectral-line profiles, since these values are not provided in the FLASH linefinder data. For MRC 2125-237 we find the FWZI to be approximately 400 km s$^{-1}$.

The ASKAP NSI value of $0.98\pm0.06$ implies that almost all the radio emission arises from a region smaller than 800\,pc in size, i.e. well within the size range expected for PS radio sources \citep{odea2021}. 

\subsection{MRC 2131-241 (J213437.6-235535) } 

\begin{figure}[t!]
    \centering
    \begin{subfigure}[t]{\linewidth}
        \centering
        \includegraphics[width=0.7\linewidth]{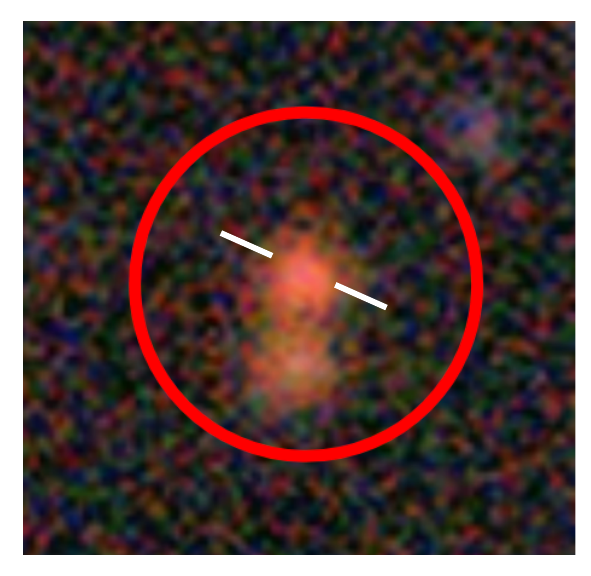}
    \end{subfigure}
    ~
    \begin{subfigure}[t]{\linewidth}
        \centering
        \includegraphics[width=0.7\linewidth]{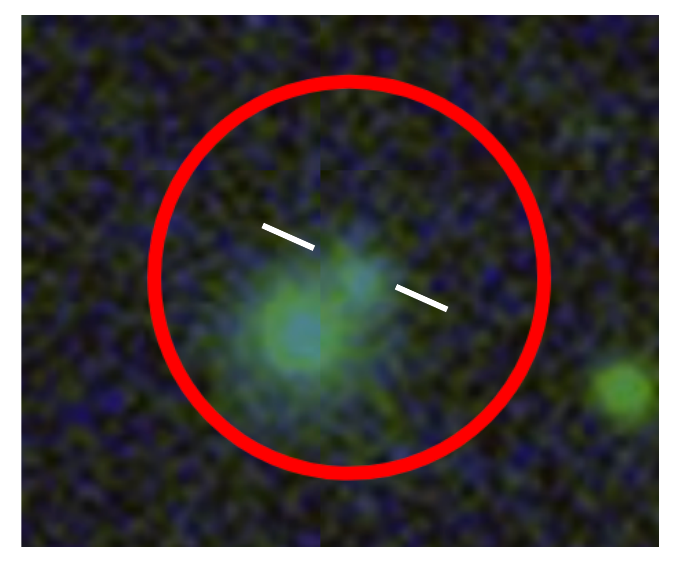}
    \end{subfigure}
%legacy_image_with_radio_pos_2125.pdf}
    \caption{Legacy Survey DR 10 images centred on the MRC-1Jy radio position of MRC\,2125-237 (J212805.5-232946) (on the top) and 
    %CHANGE HERE
    MRC\,2131-241 (J213437.6-235537) (on the bottom). The red circle on the two sources has a radius of 5\,arcsec, which corresponds to a projected distance of 40.1\,kpc on the sky at the \hi redshift of $z=0.9540$, and 29.6\,kpc on the sky at the \hi redshift of $z=0.4632$ respectively. The two white lines indicate the galaxy identified as the radio-source host.}
    \label{fig:legacy_images}
\end{figure}

\cite{McCarthy_1996} provided a finding chart for MRC\,2131-241, and identified the optical counterpart as the faint object near the centre of the circle in the lower image in Figure \ref{fig:legacy_images}, noting that ``The identification is a galaxy 1\,arcsec northwest of a star.''
Only $g$ and $i$ band LS10 images are available for this region of sky. Confusingly, the faint object near the centre of the circle, which \cite{McCarthy_1996} listed as a galaxy, is classified as a point source (PSF) in the LS10 catalogue; but the brighter object to the south-east, which \cite{McCarthy_1996} referred to as a star, is classified by LS10 as a galaxy. Additional high resolution optical imaging would be useful to clarify the nature of these two objects, but we can still draw some conclusions from the data available now. 

The narrow \hi FWHM linewidth of 18.2\,km\,s$^{-1}$ and FWZI of $\sim$50 km s$^{-1}$ suggests that this is an intervening line. The faint object closest to the MRC-1Jy radio position also has an $i$-band magnitude of 22.4, which is far fainter than expected for a radio galaxy at the \hi redshift of $z=0.4632$, again implying that the radio-source host is much more distant than the \hi gas seen in absorption. The brighter galaxy offset from the radio position in Figure \ref{fig:legacy_images} has a photometric redshift estimate of $z\sim0.66$, and a $g-i$ colour of 1.85 mag, consistent with the general population of star-forming galaxies at this redshift \citep{ching2017}. It therefore seems likely that this is an intervening line, with a potential \hi host galaxy at an impact parameter of around 10\,kpc from the background radio source. 

The ASKAP NSI value of $0.84\pm0.04$, as with the previous source MRC\,2125-237, implies that most of the radio emission arises from a compact region smaller than $\sim800$\,pc in diameter. 

\section{DISCUSSION}
\label{sec:discussion}

\subsection{\hi detection rate}
The \hi lines detected in this work from sources MRC 2131-241 and MRC 2125-237 have redshifts of 0.4632 and 0.9540, and FWHM velocity linewidths of 18 km s$^{-1}$ and 216 km s$^{-1}$ respectively. 
As discussed in \cite{Yoon_2025}, data validation is completed with a checklist, which then supplies a numerical score which the FLASH team uses to classify each SBID as either GOOD, UNCERTAIN, or BAD. This checklist includes, but is not limited to, measurements of the rms noise, visual inspection of the images, and a sample of spectra. 
Both of the detections in this work were from the FLASH field 350 (SBID 55464) which had GOOD quality spectral-line data. The other FLASH field 405 (SBID 62794) was validated with UNCERTAIN data quality due to a few glitches and weak artefacts present in a small part of the band. 

\cite{Yoon_2025} found that, as expected for an absorption-line search, the peak optical depth increases at lower flux densities due to the detection limit. We see this with the two detections as well as with the peak optical depth upper limits in Figure \ref{fig:peak_opt_depth}. The two detections in this figure are visually well above the detection limit, and the 150 mJy bright-source sample limit. Additionally, the mean value of the peak optical depth for the two detections is 0.099, which is greater than the median value of $\tau_{\text{pk}}=0.030$ obtained in the sample of 136 lines compiled by \cite{Curran_2021}.
\\\\
%\cite{Allison_2022} estimated that for an assumed \hi spin temperature $T_S=300$K, the rate of \hi detections in the FLASH survey is around 3.1 lines per ASKAP field. However, in the FLASH pilot survey, they found a significantly lower \hi detection rate \citep{Yoon_2025}. In the first pilot survey they obtained a rate of 0.46 lines per field, and in the second a rate of 0.32 lines per field. They believe that this is mostly due to the spectral-line artefacts such as glitches at the edges of beam-forming intervals, correlator dropouts, wobbles due to using incorrect parameters in fitting the spectral bandpass, and ducting. Ducting appears to be the only ongoing spectral-line artefact affecting the spectra, meaning that we now expect a higher detection rate since most issues have been resolved.

\subsection{Radio spectral index}
\label{sec:spec_index}
\begin{figure}
    \centering
    \begin{subfigure}[t]{\linewidth}
    \centering
        \includegraphics[height=6cm]{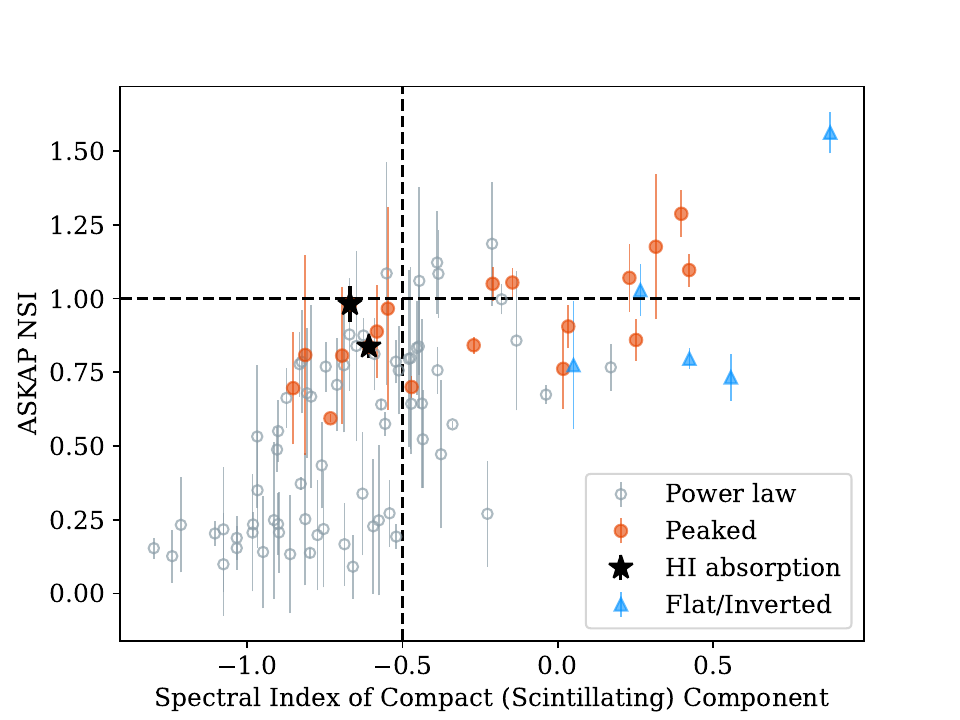}
    \end{subfigure}
    \begin{subfigure}[t]{\linewidth}
    \centering
        \includegraphics[height=6cm]{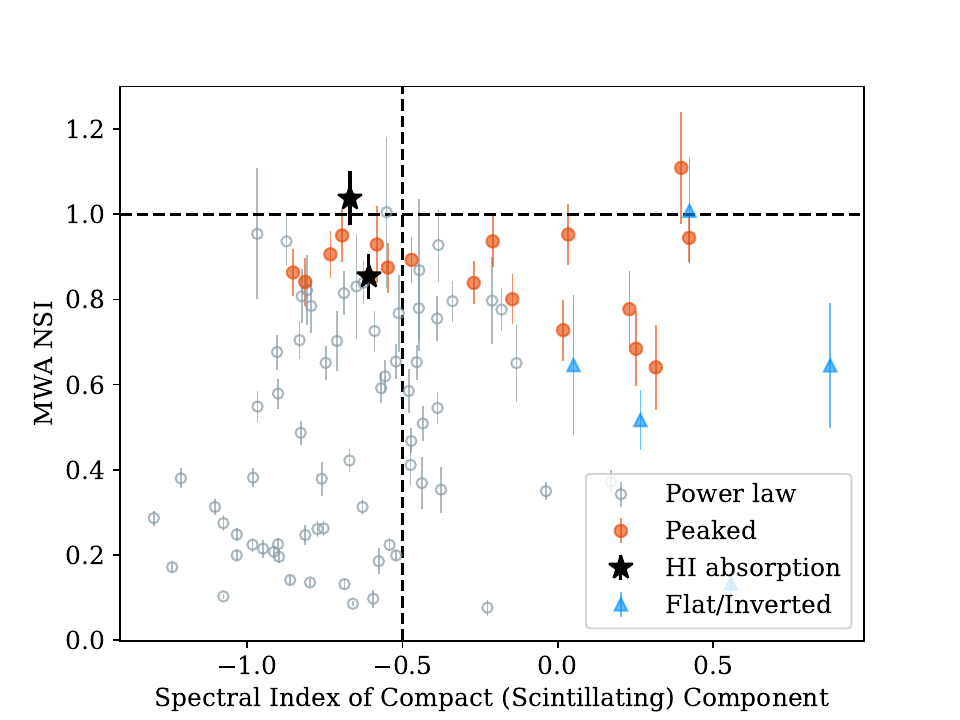}
    \end{subfigure}
    \caption{On the top: the ASKAP NSI vs. the spectral index of the compact scintillating component for all of the sources in the final catalogue shown by the grey circles. The peaked sources are shown by the orange triangles, and the two sources that had \hi absorption observed in their spectra are shown by the black stars. On the bottom: the MWA NSI vs. spectral index. The spectral index has been calculated between MWA 162 MHz and ASKAP 820 MHz.}
    \label{fig:spec_index}
\end{figure}

%Variability data from the VAST (ASKAP Variables and Slow Transients) survey which has good time resolution could be used to gain a larger sample of scintillating, compact sources.
Most extragalactic radio sources found in surveys at ASKAP frequencies have well-defined power law spectra, with a mean spectral index of -0.75 and dispersion of 0.2 \citep[e.g.][]{Kellermann_1988}. Sources with a spectral index $\gtrsim-0.5$ are usually part of the flat-spectrum population. This work focuses on the subset of sources with compact components identified by their IPS.
%The value of the spectral index of a source is characteristic of objects such as hot-spots in lobes which have a spectral index between $\sim-1.3- -0.5$. 
Sources dominated by compact structure have much more complex SEDs, and a figure of the spectral index and NSI of sources can help classify different populations of sources. 
Radio sources with compact hot spots in their lobes which are only a small fraction of the total flux density, have a low NSI with power law spectral index $\alpha$ $\sim-0.7$. 
%CSOs have a high NSI with power law $\sim-0.7$, 
%CSS has been defined in the introduction
CSS sources have a high NSI with a similar steep power law spectral index, while PS and flat-spectrum sources all have a mid to high NSI as seen in Figure \ref{fig:spec_index} and \cite{Chhetri_2026}.

The two-point radio spectral index $\alpha$ of a source, where the flux density is $S_1$ at frequency $\nu_1$ and $S_2$ at $\nu_2$ is: $$\alpha = \frac{\log(S_1/S_2)}{\log(\nu_1/\nu_2)}$$

The sources in the ASKAP IPS catalogue were observed at 820 MHz and also by MWA IPS at 162 MHz. The compact flux densities from both catalogues were used to characterise the spectral index of the compact component between 162 and 820 MHz. The compact flux density was obtained by multiplying the peak flux density by the NSI. The flux density measurements however were taken from RACS-low, which was made at a slightly different frequency, so there is a flux density error of a few percent at 820 MHz. We emphasise that the spectral index in this work is for the compact scintillating component of the source, while the SED model fitting uses the flux densities for the total source. However, for sources with a high NSI (compact sources), the compact flux density and the total flux density are the same.

Figure \ref{fig:spec_index} shows the ASKAP NSI and MWA NSI vs the two-point spectral index of the compact scintillating component for all of the sources in the sample. There were 92 sources which had both an ASKAP NSI and an MWA NSI, and therefore a measurement of the spectral index. The sources with only NSI upper limits have no detectable compact component, and so they were not included here. The sources classified as peaked from their radio SED fits are shown by the orange circles, and the two sources found to have \hi absorption in their spectra are shown by the black stars. The \hi absorption sources are part of the PS sources group. Power law sources are shown by the grey circles, and flat or inverted sources are shown by the blue triangles.

\begin{figure}
    \centering
    \begin{subfigure}[t]{\linewidth}
    \centering
        \includegraphics[height=6cm]{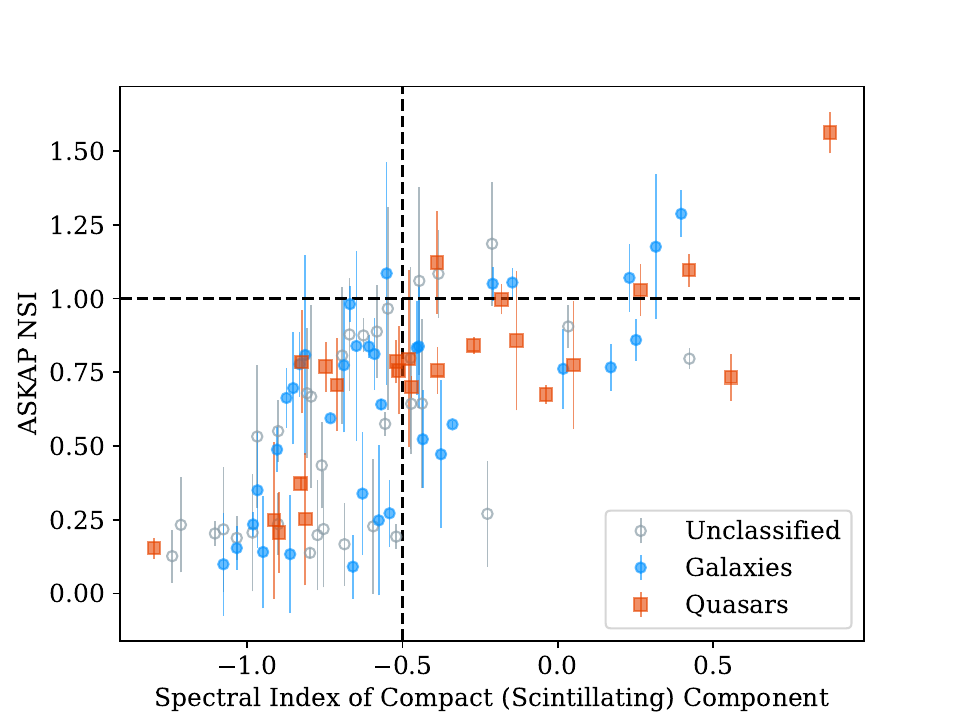}
    \end{subfigure}
    \begin{subfigure}[t]{\linewidth}
    \centering
        \includegraphics[height=6cm]{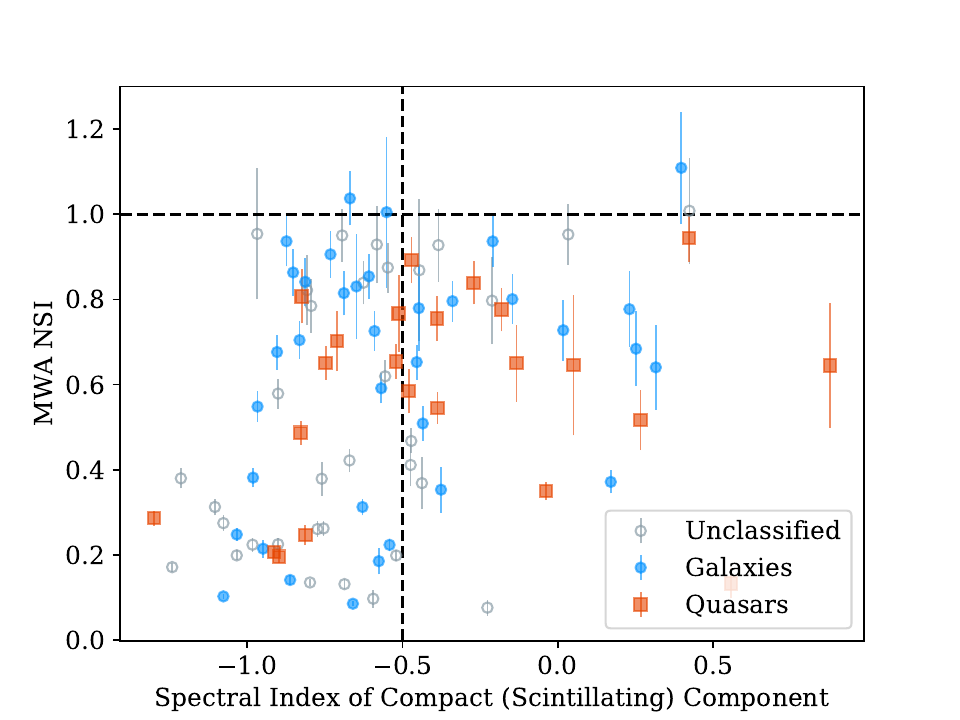}
    \end{subfigure}
    \caption{On the top: the ASKAP NSI vs. the spectral index of the compact scintillating component for all of the sources in the final catalogue, including sources classified as either galaxies or quasars using a photometric classification and/or a spectroscopic classification. On the bottom: the MWA NSI vs. spectral index. Galaxies are shown by blue circles and quasars are shown by orange squares. As in Figure \ref{fig:spec_index}, the spectral index has been calculated between MWA 162 MHz and ASKAP 820 MHz.}
    \label{fig:spec_index_gal_quasar}
\end{figure}

\subsection{Median NSI values for galaxies and quasars}
\label{sec:median_nsi}

Since the optical identifications for the bright-source sample are incomplete, and quasars are likely to be over-represented among the identified sources, we need to be cautious in making general statements about the statistics of the quasar and galaxy populations. 

However, we can compare the NSI values of confirmed galaxies and quasars in Tables \ref{tab:main_bright} and \ref{tab:main_add}, noting that the ASKAP NSI value represents the fraction of flux density arising from a component less than 0.1 arcsec in size. Table \ref{tab:nsi2} shows the results. 

\begin{table}[]
    \centering
    \begin{tabular}{llll}
    \hline
Type & Number & Redshift & NSI median \\
     &        & median & \\
    \hline
Galaxies & 33 & $1.03\pm0.16$ & $0.30\pm0.07$  \\    
Quasars  & 20 & $1.42\pm0.20$ & $0.69\pm0.11$ \\
Unclassified & 26 & .. & $0.22\pm0.07$ \\
\hline
    \end{tabular}
    \caption{Median values of NSI for optically-identified galaxies and quasars in the bright-source ($>150$\,mJy) sample. The listed median redshifts for the galaxies and quasars are based mainly on photometric redshift estimates. Most of the `Unclassified' objects are expected to be galaxies with extended radio emission.  }
    \label{tab:nsi2}
\end{table}

As can be seen from Table \ref{tab:nsi2}, there is a significant difference between the NSI medians for the galaxies and quasars. This could be because there are some blazars in our sample such as the uncommon source discussed below, which most likely have SEDs classified as flat/inverted. Blazars differ from quasars because their jet axis is closely aligned to our line of sight, causing relativistic beaming \citep{Giroletti_2016}. \cite{Chhetri_2018} find that the high-frequency (20 GHz) radio emission of blazars is often dominated by a compact radio core, and at lower frequencies (162 MHz), the emission is dominated by the extended structure which has been built up by the AGN activity over a long period of time.
From Table \ref{tab:nsi2} and Figure \ref{fig:spec_index_gal_quasar}, we see that the low NSI values are very similar for galaxies and unclassified sources. This provides strong evidence that the unclassified sources are more distant galaxies, and being more distant means higher radio luminosity, which is consistent with the suggestion by \cite{Chhetri_2026} that these are FRII sources with a small fraction of their flux density in scintillating hotspots.

\subsection{The uncommon source NVSS J215031-223200}
\label{sec:unusual_src}

\begin{figure}
    \centering
    \includegraphics[angle=0,width=1.0\hsize]{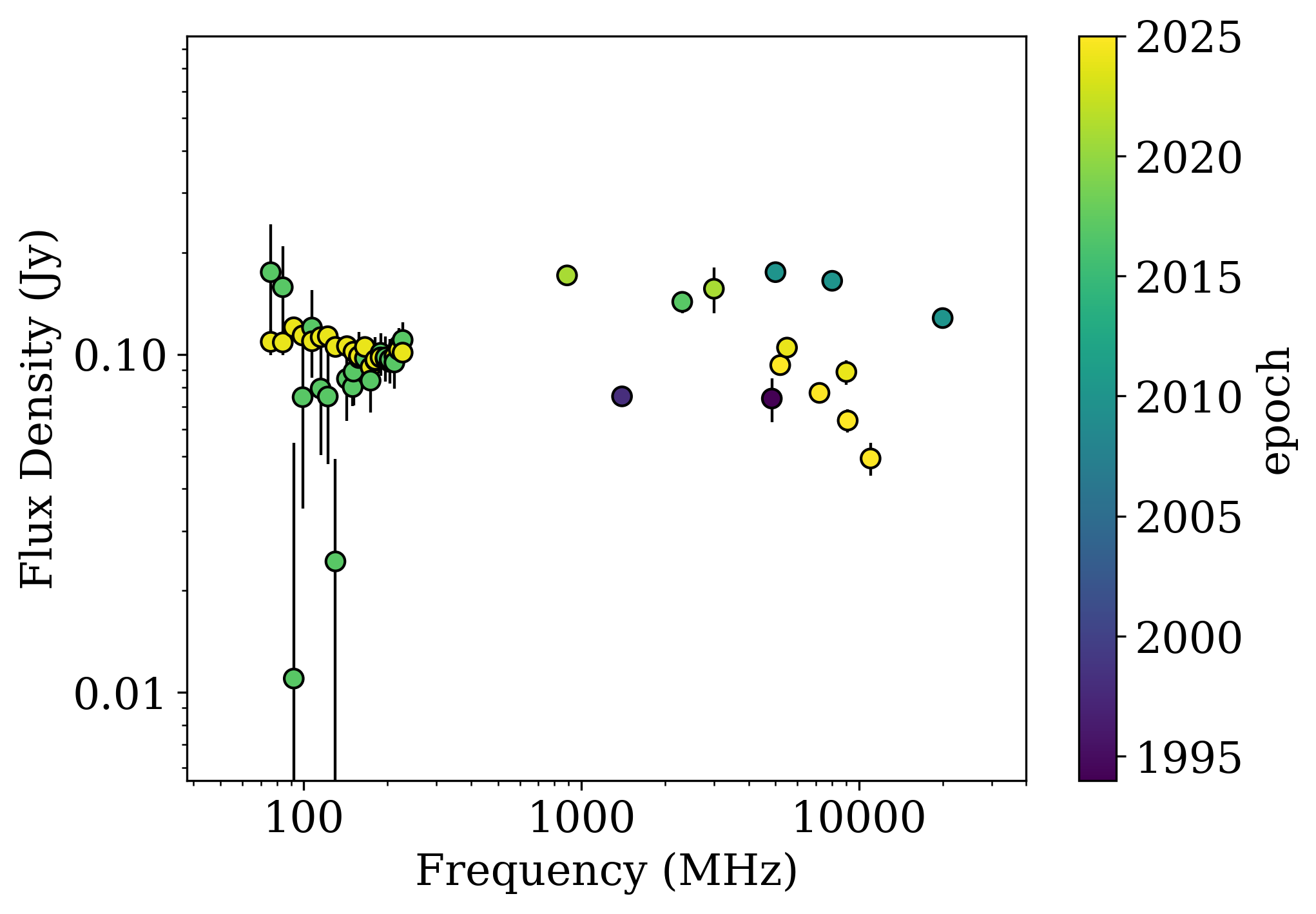}
    \caption{Radio SED of NVSS J215031-223200, showing high-frequency variability. }
    \label{fig:high_nsi}
\end{figure}

NVSS J215031-223200 was identified by eye as an outlier in terms of both compact-component spectral index and ASKAP NSI in Figure~\ref{fig:spec_index}, with an ASKAP NSI of $1.56\pm 0.07$. It is classified as a flat-spectrum radio source by \citealt{Healey2007}, and its SED is presented in Figure \ref{fig:high_nsi}. \textsc{RadioSED} prefers a PS model, and this preference is due to the large scatter in high frequency flux densities, which is the result of variability over the $\sim20$\,yr span of observations. However, even this best-fit model is a poor characterisation of the broadband SED, since \textsc{RadioSED} does not account for source variability in fitting, and we stress that this is not a true PS source. 

NVSS\,J215031-223200 is classified as a quasar, with a spectroscopic redshift of $2.113\pm0.638$ from the Gaia-unWISE quasar catalogue \citep[Quaia;][]{quaia2024}, but the radio variability of the source suggests that it is likely a blazar. The SEDs of blazars are usually strongly affected by relativistic beaming, while PS sources are not expected to show strong variability. 

Recent observations with the Australia Telescope Compact Array (ATCA) at $4-12\,$GHz, shown in Figure~\ref{fig:high_nsi}, reveal that this source has varied by a factor of $\sim3$ over the course of 12 months between initial monitoring in December 2024 (ATCA project C3664), and subsequent follow up (ATCA project  C3550) in December 2025. This high degree of variability is likely present down to at least ASKAP frequencies, based on the factor of $\sim2.5$ difference between RACS-low (888\,MHz) and NVSS (1.4\,GHz), although repeat observations by the MWA between 2013 and 2020 show the radio flux density is stable below $300\,$MHz to within the instrument uncertainties. 

This absence of radio variability at low frequencies is characteristic of flat-spectrum radio quasars \citep{Ciaramella2004}. It is interesting to note that the NSI of this source also decreases dramatically at MWA wavelengths (NSI(MWA) $=0.65\pm0.15$), which is most likely due to an extended steep spectrum structure becoming increasingly dominant at lower frequencies. This also causes the upturn in the SED in Figure \ref{fig:high_nsi} below 100\,MHz.
This source has no observed \hi absorption, and it also did not have a tentative detection. Its integrated flux density is just above the 150 mJy bright-source sample limit, and it is located along the trend in the integrated flux density-peak optical depth diagram in Figure \ref{fig:peak_opt_depth}.

\section{SUMMARY}
\label{sec:conclusion}
A search in the first ASKAP interplanetary scintillation field for \hi absorption using FLASH has resulted in two detections in J212805.5-232946 (MRC 2125-237) and J213437.6-23553 (MRC 2131-241) without optical preselection. Both of these detections were strong lines with high Bayesian evidence values $\ln\,(B)$. 
The short 2.5 minute observation of this field demonstrates the extraordinary efficiency of the IPS method for finding many thousands of compact sources with matches in the full FLASH survey. With these IPS observations at a similar frequency to FLASH, we were able to investigate the sub-arcsec structure of sources in the field. SED fitting in this field has affirmed that most peaked-spectrum (PS) sources are compact sources, and has shown that both \hi absorption detections have peaked SEDs. In future, SEDs could be created using data spanning different frequencies, particularly in the $\sim 300-700$ MHz range, which could be achieved with the upcoming SKA, allowing for better constrained fits.
%The available spectroscopic and photometric optical counterparts for the FLASH-IPS sources were found to have significantly different ASKAP NSIs, likely the result of beaming from blazars in the sample. 
The spectroscopic redshift for the source J212805.5-232946 matched the \hi absorption line redshift (0.95). J212805.5-232946 also has a broad \hi absorption line which, in combination with its optical redshift, points to it being an associated source. On the other hand, J213437.6-235537 has a narrow absorption line which suggests it is an intervening source. Both of these sources were classified as galaxies. There was an uncommon source identified as being an outlier in compact spectral index, with a very high ASKAP NSI, as well as having large scatter in flux density at higher frequencies.
We anticipate that since it is feasible to carry out many more IPS observations using ASKAP and MWA, this technique can be used to assist with identifying candidates for \hi absorption in large spectral-line datasets such as the full FLASH survey. In these large datasets, compact sources can be easily identified, and PS sources and \hi absorption can be identified from a much smaller sample. Our results also indicate that IPS may not only be linked to compact sources but to PS sources as well, demonstrating that IPS is a promising tool for current and future HI absorption studies.

\begin{acknowledgement}
The authors would like to thank Roberto Soria and Stas Shabala for providing useful comments which improved the clarity of this paper.
This scientific work uses data obtained from Inyarrimanha Ilgari Bundara / the Murchison Radio-astronomy Observatory. We acknowledge the Wajarri Yamaji People as the Traditional Owners and native title holders of the Observatory site. CSIRO’s ASKAP radio telescope is part of the Australia Telescope National Facility (https://ror.org/05qajvd42). Operation of ASKAP is funded by the Australian Government with support from the National Collaborative Research Infrastructure Strategy. ASKAP uses the resources of the Pawsey Supercomputing Research Centre. Establishment of ASKAP,
Inyarrimanha Ilgari Bundara, the CSIRO Murchison Radioastronomy Observatory and the Pawsey Supercomputing Research Centre are initiatives of the Australian Government, with support from the Government of Western Australia and the Science and Industry Endowment Fund.

The Legacy Surveys consist of three individual and complementary projects: the Dark Energy Camera Legacy Survey (DECaLS; Proposal ID \#2014B-0404; PIs: David Schlegel and Arjun Dey), the Beijing-Arizona Sky Survey (BASS; NOAO Prop. ID \#2015A-0801; PIs: Zhou Xu and Xiaohui Fan), and the Mayall z-band Legacy Survey (MzLS; Prop. ID \#2016A-0453; PI: Arjun Dey). DECaLS, BASS and MzLS together include data obtained, respectively, at the Blanco telescope, Cerro Tololo Inter-American Observatory, NSF’s NOIRLab; the Bok telescope, Steward Observatory, University of Arizona; and the Mayall telescope, Kitt Peak National Observatory, NOIRLab. Pipeline processing and analyses of the data were supported by NOIRLab and the Lawrence Berkeley National Laboratory (LBNL). The Legacy Surveys project is honored to be permitted to conduct astronomical research on Iolkam Du’ag (Kitt Peak), a mountain with particular significance to the Tohono O’odham Nation.

NOIRLab is operated by the Association of Universities for Research in Astronomy (AURA) under a cooperative agreement with the National Science Foundation. LBNL is managed by the Regents of the University of California under contract to the U.S. Department of Energy.

This project used data obtained with the Dark Energy Camera (DECam), which was constructed by the Dark Energy Survey (DES) collaboration. Funding for the DES Projects has been provided by the U.S. Department of Energy, the U.S. National Science Foundation, the Ministry of Science and Education of Spain, the Science and Technology Facilities Council of the United Kingdom, the Higher Education Funding Council for England, the National Center for Supercomputing Applications at the University of Illinois at Urbana-Champaign, the Kavli Institute of Cosmological Physics at the University of Chicago, Center for Cosmology and Astro-Particle Physics at the Ohio State University, the Mitchell Institute for Fundamental Physics and Astronomy at Texas A\&M University, Financiadora de Estudos e Projetos, Fundacao Carlos Chagas Filho de Amparo, Financiadora de Estudos e Projetos, Fundacao Carlos Chagas Filho de Amparo a Pesquisa do Estado do Rio de Janeiro, Conselho Nacional de Desenvolvimento Cientifico e Tecnologico and the Ministerio da Ciencia, Tecnologia e Inovacao, the Deutsche Forschungsgemeinschaft and the Collaborating Institutions in the Dark Energy Survey. The Collaborating Institutions are Argonne National Laboratory, the University of California at Santa Cruz, the University of Cambridge, Centro de Investigaciones Energeticas, Medioambientales y Tecnologicas-Madrid, the University of Chicago, University College London, the DES-Brazil Consortium, the University of Edinburgh, the Eidgenossische Technische Hochschule (ETH) Zurich, Fermi National Accelerator Laboratory, the University of Illinois at Urbana-Champaign, the Institut de Ciencies de l’Espai (IEEC/CSIC), the Institut de Fisica d’Altes Energies, Lawrence Berkeley National Laboratory, the Ludwig Maximilians Universitat Munchen and the associated Excellence Cluster Universe, the University of Michigan, NSF’s NOIRLab, the University of Nottingham, the Ohio State University, the University of Pennsylvania, the University of Portsmouth, SLAC National Accelerator Laboratory, Stanford University, the University of Sussex, and Texas A\&M University.

BASS is a key project of the Telescope Access Program (TAP), which has been funded by the National Astronomical Observatories of China, the Chinese Academy of Sciences (the Strategic Priority Research Program “The Emergence of Cosmological Structures” Grant \# XDB09000000), and the Special Fund for Astronomy from the Ministry of Finance. The BASS is also supported by the External Cooperation Program of Chinese Academy of Sciences (Grant \# 114A11KYSB20160057), and Chinese National Natural Science Foundation (Grant \# 12120101003, \# 11433005).

The Legacy Survey team makes use of data products from the Near-Earth Object Wide-field Infrared Survey Explorer (NEOWISE), which is a project of the Jet Propulsion Laboratory/California Institute of Technology. NEOWISE is funded by the National Aeronautics and Space Administration.

The Legacy Surveys imaging of the DESI footprint is supported by the Director, Office of Science, Office of High Energy Physics of the U.S. Department of Energy under Contract No. DE-AC02-05CH1123, by the National Energy Research Scientific Computing Center, a DOE Office of Science User Facility under the same contract; and by the U.S. National Science Foundation, Division of Astronomical Sciences under Contract No. AST-0950945 to NOAO.

The Photometric Redshifts for the Legacy Surveys (PRLS) catalog used in this paper was produced thanks to funding from the U.S. Department of Energy Office of Science, Office of High Energy Physics via grant DE-SC0007914.

This research was supported by an Australian Government Research Training Program (RTP) Scholarship.
HY is supported by the National Research Foundation of Korea (NRF) grant funded by the Korea government (MSIT, RS-2025-00516062), and by funding from Korean government (Korea AeroSpace Administration (KASA), grant number RS-2026-25587698).
This research has made use of NASA's Astrophysics Data System Bibliographic Services, the NASA/IPAC Extragalactic Database (NED) which is funded by the National Aeronautics and Space Administration and operated by the California Institute of Technology and the crossmatch service provided by CDS, Strasbourg.

% Python modules and packages
We made use of Astropy \citep{astropy:2013, astropy:2018, astropy:2022}, Matplotlib \citep{Hunter_2007}, NumPy \citep{Harris_2020}, and pandas \citep{McKinney_2010}. For the \texttt{RadioSED} code, a full list of modules and packages used is available on \href{https://github.com/ekerrison/RadioSED}{GitHub}.

\end{acknowledgement}

\section{DATA AVAILABILITY}
Archived data from ASKAP surveys, such as FLASH, can be obtained through the CSIRO ASKAP Science Data Archive, CASDA (http://data.csiro.au/). Other survey data used in SED construction and multiwavelength analysis is available through the CDS VizieR catalogue service. Measurements derived from the ASKAP-MWA IPS catalogue are available in the companion paper \cite{Chhetri_2026} and in the Appendix of this paper. Any additional data is available upon reasonable request to the corresponding author.

% PASA uses footnotes, not endnotes. \endnote in this template will behave like \footnote; and \printendnotes will not output anything.
% \printendnotes

\bibliography{bibtemplate}

\begin{thebibliography}{}
\providecommand\natexlab[1]{#1}
\providecommand\JournalTitle[1]{#1}

\bibitem[{{Aditya} {et~al.}(2024){Aditya}, {Yoon}, {Allison}, {An}, {Chhetri}, {Curran}, {Darling}, {Emig}, {Glowacki}, {Kerrison}, \& et~al.}]{Aditya_2024}
{Aditya}, J.~N.~H.~S., {Yoon}, H., {Allison}, J.~R., {et~al.} 2024, \href{http://dx.doi.org/10.1093/mnras/stad3722}{\JournalTitle{\mnras}, 527, 8511}

\bibitem[{{Allison} {et~al.}(2021){Allison}, {Sadler}, {Mahony}, {Moss}, \& {Yoon}}]{Allison_2021}
{Allison}, J.~R., {Sadler}, E.~M., {Mahony}, E.~K., {Moss}, V.~A., \& {Yoon}, H. 2021, \href{http://dx.doi.org/10.1002/asna.20210040}{\JournalTitle{Astronomische Nachrichten}, 342, 1062}

\bibitem[{{Allison} {et~al.}(2012){Allison}, {Sadler}, \& {Whiting}}]{Allison_2012}
{Allison}, J.~R., {Sadler}, E.~M., \& {Whiting}, M.~T. 2012, \href{http://dx.doi.org/10.1071/AS11040}{\JournalTitle{\pasa}, 29, 221}

\bibitem[{{Allison} {et~al.}(2020){Allison}, {Sadler}, {Bellstedt}, {Davies}, {Driver}, {Ellison}, {Huynh}, {Kapi{\'n}ska}, {Mahony}, {Moss}, {Robotham}, {Whiting}, {Curran}, {Darling}, {Hotan}, {Hunstead}, {Koribalski}, {Lagos}, {Pettini}, {Pimbblet}, \& {Voronkov}}]{Allison_2020}
{Allison}, J.~R., {Sadler}, E.~M., {Bellstedt}, S., {et~al.} 2020, \href{http://dx.doi.org/10.1093/mnras/staa949}{\JournalTitle{\mnras}, 494, 3627}

\bibitem[{{Allison} {et~al.}(2022){Allison}, {Sadler}, {Amaral}, {An}, {Curran}, {Darling}, {Edge}, {Ellison}, {Emig}, {Gaensler}, {Garratt-Smithson}, {Glowacki}, {Grasha}, {Koribalski}, {Lagos}, {Lah}, {Mahony}, {Mao}, {Morganti}, {Moss}, {Pettini}, {Pimbblet}, {Power}, {Salas}, {Staveley-Smith}, {Whiting}, {Wong}, {Yoon}, {Zheng}, \& {Zwaan}}]{Allison_2022}
{Allison}, J.~R., {Sadler}, E.~M., {Amaral}, A.~D., {et~al.} 2022, \href{http://dx.doi.org/10.1017/pasa.2022.3}{\JournalTitle{\pasa}, 39, e010}

\bibitem[{{Astropy Collaboration} {et~al.}(2013){Astropy Collaboration}, {Robitaille}, {Tollerud}, {Greenfield}, {Droettboom}, {Bray}, {Aldcroft}, {Davis}, {Ginsburg}, {Price-Whelan}, {Kerzendorf}, {Conley}, {Crighton}, {Barbary}, {Muna}, {Ferguson}, {Grollier}, {Parikh}, {Nair}, {Unther}, {Deil}, {Woillez}, {Conseil}, {Kramer}, {Turner}, {Singer}, {Fox}, {Weaver}, {Zabalza}, {Edwards}, {Azalee Bostroem}, {Burke}, {Casey}, {Crawford}, {Dencheva}, {Ely}, {Jenness}, {Labrie}, {Lim}, {Pierfederici}, {Pontzen}, {Ptak}, {Refsdal}, {Servillat}, \& {Streicher}}]{astropy:2013}
{Astropy Collaboration}, {Robitaille}, T.~P., {Tollerud}, E.~J., {et~al.} 2013, \href{http://dx.doi.org/10.1051/0004-6361/201322068}{\JournalTitle{\aap}, 558, A33}

\bibitem[{{Astropy Collaboration} {et~al.}(2018){Astropy Collaboration}, {Price-Whelan}, {Sip{\H{o}}cz}, {G{\"u}nther}, {Lim}, {Crawford}, {Conseil}, {Shupe}, {Craig}, {Dencheva}, {Ginsburg}, {Vand erPlas}, {Bradley}, {P{\'e}rez-Su{\'a}rez}, {de Val-Borro}, {Aldcroft}, {Cruz}, {Robitaille}, {Tollerud}, {Ardelean}, {Babej}, {Bach}, {Bachetti}, {Bakanov}, {Bamford}, {Barentsen}, {Barmby}, {Baumbach}, {Berry}, {Biscani}, {Boquien}, {Bostroem}, {Bouma}, {Brammer}, {Bray}, {Breytenbach}, {Buddelmeijer}, {Burke}, {Calderone}, {Cano Rodr{\'\i}guez}, {Cara}, {Cardoso}, {Cheedella}, {Copin}, {Corrales}, {Crichton}, {D'Avella}, {Deil}, {Depagne}, {Dietrich}, {Donath}, {Droettboom}, {Earl}, {Erben}, {Fabbro}, {Ferreira}, {Finethy}, {Fox}, {Garrison}, {Gibbons}, {Goldstein}, {Gommers}, {Greco}, {Greenfield}, {Groener}, {Grollier}, {Hagen}, {Hirst}, {Homeier}, {Horton}, {Hosseinzadeh}, {Hu}, {Hunkeler}, {Ivezi{\'c}}, {Jain}, {Jenness}, {Kanarek}, {Kendrew}, {Kern}, {Kerzendorf}, {Khvalko}, {King}, {Kirkby}, {Kulkarni},
  {Kumar}, {Lee}, {Lenz}, {Littlefair}, {Ma}, {Macleod}, {Mastropietro}, {McCully}, {Montagnac}, {Morris}, {Mueller}, {Mumford}, {Muna}, {Murphy}, {Nelson}, {Nguyen}, {Ninan}, {N{\"o}the}, {Ogaz}, {Oh}, {Parejko}, {Parley}, {Pascual}, {Patil}, {Patil}, {Plunkett}, {Prochaska}, {Rastogi}, {Reddy Janga}, {Sabater}, {Sakurikar}, {Seifert}, {Sherbert}, {Sherwood-Taylor}, {Shih}, {Sick}, {Silbiger}, {Singanamalla}, {Singer}, {Sladen}, {Sooley}, {Sornarajah}, {Streicher}, {Teuben}, {Thomas}, {Tremblay}, {Turner}, {Terr{\'o}n}, {van Kerkwijk}, {de la Vega}, {Watkins}, {Weaver}, {Whitmore}, {Woillez}, {Zabalza}, \& {Astropy Contributors}}]{astropy:2018}
{Astropy Collaboration}, {Price-Whelan}, A.~M., {Sip{\H{o}}cz}, B.~M., {et~al.} 2018, \href{http://dx.doi.org/10.3847/1538-3881/aabc4f}{\JournalTitle{\aj}, 156, 123}

\bibitem[{{Astropy Collaboration} {et~al.}(2022){Astropy Collaboration}, {Price-Whelan}, {Lim}, {Earl}, {Starkman}, {Bradley}, {Shupe}, {Patil}, {Corrales}, {Brasseur}, {N{"o}the}, {Donath}, {Tollerud}, {Morris}, {Ginsburg}, {Vaher}, {Weaver}, {Tocknell}, {Jamieson}, {van Kerkwijk}, {Robitaille}, {Merry}, {Bachetti}, {G{"u}nther}, {Aldcroft}, {Alvarado-Montes}, {Archibald}, {B{'o}di}, {Bapat}, {Barentsen}, {Baz{'a}n}, {Biswas}, {Boquien}, {Burke}, {Cara}, {Cara}, {Conroy}, {Conseil}, {Craig}, {Cross}, {Cruz}, {D'Eugenio}, {Dencheva}, {Devillepoix}, {Dietrich}, {Eigenbrot}, {Erben}, {Ferreira}, {Foreman-Mackey}, {Fox}, {Freij}, {Garg}, {Geda}, {Glattly}, {Gondhalekar}, {Gordon}, {Grant}, {Greenfield}, {Groener}, {Guest}, {Gurovich}, {Handberg}, {Hart}, {Hatfield-Dodds}, {Homeier}, {Hosseinzadeh}, {Jenness}, {Jones}, {Joseph}, {Kalmbach}, {Karamehmetoglu}, {Ka{l}uszy{'n}ski}, {Kelley}, {Kern}, {Kerzendorf}, {Koch}, {Kulumani}, {Lee}, {Ly}, {Ma}, {MacBride}, {Maljaars}, {Muna}, {Murphy}, {Norman}, {O'Steen},
  {Oman}, {Pacifici}, {Pascual}, {Pascual-Granado}, {Patil}, {Perren}, {Pickering}, {Rastogi}, {Roulston}, {Ryan}, {Rykoff}, {Sabater}, {Sakurikar}, {Salgado}, {Sanghi}, {Saunders}, {Savchenko}, {Schwardt}, {Seifert-Eckert}, {Shih}, {Jain}, {Shukla}, {Sick}, {Simpson}, {Singanamalla}, {Singer}, {Singhal}, {Sinha}, {Sip{H{o}}cz}, {Spitler}, {Stansby}, {Streicher}, {{{S}}umak}, {Swinbank}, {Taranu}, {Tewary}, {Tremblay}, {Val-Borro}, {Van Kooten}, {Vasovi{'c}}, {Verma}, {de Miranda Cardoso}, {Williams}, {Wilson}, {Winkel}, {Wood-Vasey}, {Xue}, {Yoachim}, {Zhang}, {Zonca}, \& {Astropy Project Contributors}}]{astropy:2022}
{Astropy Collaboration}, {Price-Whelan}, A.~M., {Lim}, P.~L., {et~al.} 2022, \href{http://dx.doi.org/10.3847/1538-4357/ac7c74}{\JournalTitle{\apj}, 935, 167}

\bibitem[{{Baker} {et~al.}(1999){Baker}, {Hunstead}, {Kapahi}, \& {Subrahmanya}}]{baker1999}
{Baker}, J.~C., {Hunstead}, R.~W., {Kapahi}, V.~K., \& {Subrahmanya}, C.~R. 1999, \href{http://dx.doi.org/10.1086/313209}{\JournalTitle{\apjs}, 122, 29}

\bibitem[{{Best} {et~al.}(2005){Best}, {Kauffmann}, {Heckman}, \& {Ivezi{\'c}}}]{best2005}
{Best}, P.~N., {Kauffmann}, G., {Heckman}, T.~M., \& {Ivezi{\'c}}, {\v Z}. 2005, \JournalTitle{\mnras}, 362, 9

\bibitem[{{Burbidge}(1970)}]{burbidge1970}
{Burbidge}, E.~M. 1970, \href{http://dx.doi.org/10.1086/180518}{\JournalTitle{\apjl}, 160, L33}

\bibitem[{Callingham {et~al.}(2017)Callingham, Ekers, Gaensler, Line, Hurley-Walker, Sadler, Tingay, Hancock, Bell, Dwarakanath, For, Franzen, Hindson, Johnston-Hollitt, Kapi{\'{n}}ska, Lenc, McKinley, Morgan, Offringa, Procopio, Staveley-Smith, Wayth, Wu, Zheng, Callingham, Ekers, Gaensler, Line, Hurley-Walker, Sadler, Tingay, Hancock, Bell, Dwarakanath, For, Franzen, Hindson, Johnston-Hollitt, Kapi{\'{n}}ska, Lenc, McKinley, Morgan, Offringa, Procopio, Staveley-Smith, Wayth, Wu, \& Zheng}]{Callingham2017}
Callingham, J.~R., Ekers, R.~D., Gaensler, B.~M., {et~al.} 2017, \href{http://dx.doi.org/10.3847/1538-4357/836/2/174}{\JournalTitle{ApJ}, 836, 174}

\bibitem[{{Chandola} {et~al.}(2011){Chandola}, {Sirothia}, \& {Saikia}}]{Chandola_2011}
{Chandola}, Y., {Sirothia}, S.~K., \& {Saikia}, D.~J. 2011, \href{http://dx.doi.org/10.1111/j.1365-2966.2011.19607.x}{\JournalTitle{\mnras}, 418, 1787}

\bibitem[{{Chhetri} {et~al.}(2018){Chhetri}, {Morgan}, {Ekers}, {Macquart}, {Sadler}, {Giroletti}, {Callingham}, \& {Tingay}}]{Chhetri_2018}
{Chhetri}, R., {Morgan}, J., {Ekers}, R.~D., {et~al.} 2018, \href{http://dx.doi.org/10.1093/mnras/stx2864}{\JournalTitle{\mnras}, 474, 4937}

\bibitem[{Chhetri {et~al.}(2026)Chhetri, Morgan, Ekers, Sadler, Moss, Waszewski, \& Gordon-Hall}]{Chhetri_2026}
Chhetri, R., Morgan, J.~S., Ekers, R.~D., {et~al.} 2026, First results of sub-arcsecond scale objects identified with ASKAP using interplanetary scintillation, \href{http://arxiv.org/abs/2606.28302}{{\sffamily arXiv:2606.28302 [astro-ph.CO]}}

\bibitem[{{Chhetri} {et~al.}(2023){Chhetri}, {Morgan}, {Moss}, {Ekers}, {Scott}, {Bannister}, {Day}, {Deller}, \& {Shannon}}]{Chhetri_2023}
{Chhetri}, R., {Morgan}, J., {Moss}, V., {et~al.} 2023, \href{http://dx.doi.org/10.1016/j.asr.2022.08.012}{\JournalTitle{Advances in Space Research}, 72, 5361}

\bibitem[{{Ching} {et~al.}(2017){Ching}, {Sadler}, {Croom}, {Johnston}, {Pracy}, {Couch}, {Hopkins}, {Jurek}, \& {Pimbblet}}]{ching2017}
{Ching}, J.~H.~Y., {Sadler}, E.~M., {Croom}, S.~M., {et~al.} 2017, \href{http://dx.doi.org/10.1093/mnras/stw2396}{\JournalTitle{\mnras}, 464, 1306}

\bibitem[{{Ciaramella} {et~al.}(2004){Ciaramella}, {Bongardo}, {Aller}, {Aller}, {De Zotti}, {L{\"a}hteenmaki}, {Longo}, {Milano}, {Tagliaferri}, {Ter{\"a}sranta}, {Tornikoski}, \& {Urpo}}]{Ciaramella2004}
{Ciaramella}, A., {Bongardo}, C., {Aller}, H.~D., {et~al.} 2004, \href{http://dx.doi.org/10.1051/0004-6361:20035771}{\JournalTitle{\aap}, 419, 485}

\bibitem[{{Clarke}(1964)}]{Clarke_1964}
{Clarke}, M. 1964, \JournalTitle{{PhD Thesis, Cambridge University}}

\bibitem[{{Condon} {et~al.}(1998){Condon}, {Cotton}, {Greisen}, {Yin}, {Perley}, {Taylor}, \& {Broderick}}]{condon1998}
{Condon}, J.~J., {Cotton}, W.~D., {Greisen}, E.~W., {et~al.} 1998, \href{http://dx.doi.org/10.1086/300337}{\JournalTitle{\aj}, 115, 1693}

\bibitem[{{Curran}(2021)}]{Curran_2021}
{Curran}, S.~J. 2021, \href{http://dx.doi.org/10.1093/mnras/stab1865}{\JournalTitle{\mnras}, 506, 1548}

\bibitem[{{Curran} {et~al.}(2013){Curran}, {Whiting}, {Sadler}, \& {Bignell}}]{Curran_2013}
{Curran}, S.~J., {Whiting}, M.~T., {Sadler}, E.~M., \& {Bignell}, C. 2013, \href{http://dx.doi.org/10.1093/mnras/sts171}{\JournalTitle{\mnras}, 428, 2053}

\bibitem[{{Dallacasa} {et~al.}(2000){Dallacasa}, {Stanghellini}, {Centonza}, \& {Fanti}}]{Dallacasa_2000}
{Dallacasa}, D., {Stanghellini}, C., {Centonza}, M., \& {Fanti}, R. 2000, \href{http://dx.doi.org/10.48550/arXiv.astro-ph/0012428}{\JournalTitle{\aap}, 363, 887}

\bibitem[{{Dey} {et~al.}(2019{\natexlab{a}}){Dey}, {Schlegel}, {Lang}, {Blum}, {Burleigh}, {Fan}, {Findlay}, {Finkbeiner}, {Herrera}, {Juneau}, {Landriau}, {Levi}, {McGreer}, {Meisner}, {Myers}, {Moustakas}, {Nugent}, {Patej}, {Schlafly}, {Walker}, {Valdes}, {Weaver}, {Y{\`e}che}, {Zou}, {Zhou}, {Abareshi}, {Abbott}, {Abolfathi}, {Aguilera}, {Alam}, {Allen}, {Alvarez}, {Annis}, {Ansarinejad}, {Aubert}, {Beechert}, {Bell}, {BenZvi}, {Beutler}, {Bielby}, {Bolton}, {Brice{\~n}o}, {Buckley-Geer}, {Butler}, {Calamida}, {Carlberg}, {Carter}, {Casas}, {Castander}, {Choi}, {Comparat}, {Cukanovaite}, {Delubac}, {DeVries}, {Dey}, {Dhungana}, {Dickinson}, {Ding}, {Donaldson}, {Duan}, {Duckworth}, {Eftekharzadeh}, {Eisenstein}, {Etourneau}, {Fagrelius}, {Farihi}, {Fitzpatrick}, {Font-Ribera}, {Fulmer}, {G{\"a}nsicke}, {Gaztanaga}, {George}, {Gerdes}, {Gontcho}, {Gorgoni}, {Green}, {Guy}, {Harmer}, {Hernandez}, {Honscheid}, {Huang}, {James}, {Jannuzi}, {Jiang}, {Joyce}, {Karcher}, {Karkar}, {Kehoe}, {Kneib},
  {Kueter-Young}, {Lan}, {Lauer}, {Le Guillou}, {Le Van Suu}, {Lee}, {Lesser}, {Perreault Levasseur}, {Li}, {Mann}, {Marshall}, {Mart{\'\i}nez-V{\'a}zquez}, {Martini}, {du Mas des Bourboux}, {McManus}, {Meier}, {M{\'e}nard}, {Metcalfe}, {Mu{\~n}oz-Guti{\'e}rrez}, {Najita}, {Napier}, {Narayan}, {Newman}, {Nie}, {Nord}, {Norman}, {Olsen}, {Paat}, {Palanque-Delabrouille}, {Peng}, {Poppett}, {Poremba}, {Prakash}, {Rabinowitz}, {Raichoor}, {Rezaie}, {Robertson}, {Roe}, {Ross}, {Ross}, {Rudnick}, {Safonova}, {Saha}, {S{\'a}nchez}, {Savary}, {Schweiker}, {Scott}, {Seo}, {Shan}, {Silva}, {Slepian}, {Soto}, {Sprayberry}, {Staten}, {Stillman}, {Stupak}, {Summers}, {Sien Tie}, {Tirado}, {Vargas-Maga{\~n}a}, {Vivas}, {Wechsler}, {Williams}, {Yang}, {Yang}, {Yapici}, {Zaritsky}, {Zenteno}, {Zhang}, {Zhang}, {Zhou}, \& {Zhou}}]{Legacy_paper}
{Dey}, A., {Schlegel}, D.~J., {Lang}, D., {et~al.} 2019{\natexlab{a}}, \href{http://dx.doi.org/10.3847/1538-3881/ab089d}{\JournalTitle{\aj}, 157, 168}

\bibitem[{{Dey} {et~al.}(2019{\natexlab{b}}){Dey}, {Schlegel}, {Lang}, {Blum}, {Burleigh}, {Fan}, {Findlay}, {Finkbeiner}, {Herrera}, {Juneau}, {Landriau}, {Levi}, {McGreer}, {Meisner}, {Myers}, {Moustakas}, {Nugent}, {Patej}, {Schlafly}, {Walker}, {Valdes}, {Weaver}, {Y{\`e}che}, {Zou}, {Zhou}, {Abareshi}, {Abbott}, {Abolfathi}, {Aguilera}, {Alam}, {Allen}, {Alvarez}, {Annis}, {Ansarinejad}, {Aubert}, {Beechert}, {Bell}, {BenZvi}, {Beutler}, {Bielby}, {Bolton}, {Brice{\~n}o}, {Buckley-Geer}, {Butler}, {Calamida}, {Carlberg}, {Carter}, {Casas}, {Castander}, {Choi}, {Comparat}, {Cukanovaite}, {Delubac}, {DeVries}, {Dey}, {Dhungana}, {Dickinson}, {Ding}, {Donaldson}, {Duan}, {Duckworth}, {Eftekharzadeh}, {Eisenstein}, {Etourneau}, {Fagrelius}, {Farihi}, {Fitzpatrick}, {Font-Ribera}, {Fulmer}, {G{\"a}nsicke}, {Gaztanaga}, {George}, {Gerdes}, {Gontcho}, {Gorgoni}, {Green}, {Guy}, {Harmer}, {Hernandez}, {Honscheid}, {Huang}, {James}, {Jannuzi}, {Jiang}, {Joyce}, {Karcher}, {Karkar}, {Kehoe}, {Kneib},
  {Kueter-Young}, {Lan}, {Lauer}, {Le Guillou}, {Le Van Suu}, {Lee}, {Lesser}, {Perreault Levasseur}, {Li}, {Mann}, {Marshall}, {Mart{\'\i}nez-V{\'a}zquez}, {Martini}, {du Mas des Bourboux}, {McManus}, {Meier}, {M{\'e}nard}, {Metcalfe}, {Mu{\~n}oz-Guti{\'e}rrez}, {Najita}, {Napier}, {Narayan}, {Newman}, {Nie}, {Nord}, {Norman}, {Olsen}, {Paat}, {Palanque-Delabrouille}, {Peng}, {Poppett}, {Poremba}, {Prakash}, {Rabinowitz}, {Raichoor}, {Rezaie}, {Robertson}, {Roe}, {Ross}, {Ross}, {Rudnick}, {Safonova}, {Saha}, {S{\'a}nchez}, {Savary}, {Schweiker}, {Scott}, {Seo}, {Shan}, {Silva}, {Slepian}, {Soto}, {Sprayberry}, {Staten}, {Stillman}, {Stupak}, {Summers}, {Sien Tie}, {Tirado}, {Vargas-Maga{\~n}a}, {Vivas}, {Wechsler}, {Williams}, {Yang}, {Yang}, {Yapici}, {Zaritsky}, {Zenteno}, {Zhang}, {Zhang}, {Zhou}, \& {Zhou}}]{dey2019}
---. 2019{\natexlab{b}}, \href{http://dx.doi.org/10.3847/1538-3881/ab089d}{\JournalTitle{\aj}, 157, 168}

\bibitem[{{Drinkwater} {et~al.}(1997){Drinkwater}, {Webster}, {Francis}, {Condon}, {Ellison}, {Jauncey}, {Lovell}, {Peterson}, \& {Savage}}]{drinkwater1997}
{Drinkwater}, M.~J., {Webster}, R.~L., {Francis}, P.~J., {et~al.} 1997, \href{http://dx.doi.org/10.1093/mnras/284.1.85}{\JournalTitle{\mnras}, 284, 85}

\bibitem[{{Dunlop} {et~al.}(1989){Dunlop}, {Peacock}, {Savage}, {Lilly}, {Heasley}, \& {Simon}}]{dunlop1989}
{Dunlop}, J.~S., {Peacock}, J.~A., {Savage}, A., {et~al.} 1989, \href{http://dx.doi.org/10.1093/mnras/238.4.1171}{\JournalTitle{\mnras}, 238, 1171}

\bibitem[{{Ger{\'e}b} {et~al.}(2015){Ger{\'e}b}, {Maccagni}, {Morganti}, \& {Oosterloo}}]{Gereb_2015}
{Ger{\'e}b}, K., {Maccagni}, F.~M., {Morganti}, R., \& {Oosterloo}, T.~A. 2015, \href{http://dx.doi.org/10.1051/0004-6361/201424655}{\JournalTitle{\aap}, 575, A44}

\bibitem[{Giroletti {et~al.}(2016)Giroletti, D’Ammando, Orienti, Lico, \& Collaboration}]{Giroletti_2016}
Giroletti, M., D’Ammando, F., Orienti, M., Lico, R., \& Collaboration, T. F.-L. 2016, \href{http://dx.doi.org/10.3390/galaxies4030030}{\JournalTitle{Galaxies}, 4}

\bibitem[{{Glowacki} {et~al.}(2019){Glowacki}, {Allison}, {Moss}, {Mahony}, {Sadler}, {Callingham}, {Ellison}, {Whiting}, {Bunton}, {Chippendale}, \& et~al.}]{Glowacki_2019}
{Glowacki}, M., {Allison}, J.~R., {Moss}, V.~A., {et~al.} 2019, \href{http://dx.doi.org/10.1093/mnras/stz2452}{\JournalTitle{\mnras}, 489, 4926}

\bibitem[{{Gupta} {et~al.}(2006){Gupta}, {Salter}, {Saikia}, {Ghosh}, \& {Jeyakumar}}]{Gupta_2006}
{Gupta}, N., {Salter}, C.~J., {Saikia}, D.~J., {Ghosh}, T., \& {Jeyakumar}, S. 2006, \href{http://dx.doi.org/10.1111/j.1365-2966.2006.11064.x}{\JournalTitle{\mnras}, 373, 972}

\bibitem[{{Hale} {et~al.}(2021){Hale}, {McConnell}, {Thomson}, {Lenc}, {Heald}, {Hotan}, {Leung}, {Moss}, {Murphy}, {Pritchard}, {Sadler}, {Stewart}, \& {Whiting}}]{hale2021}
{Hale}, C.~L., {McConnell}, D., {Thomson}, A.~J.~M., {et~al.} 2021, \href{http://dx.doi.org/10.1017/pasa.2021.47}{\JournalTitle{\pasa}, 38, e058}

\bibitem[{{Hancock} {et~al.}(2018){Hancock}, {Trott}, \& {Hurley-Walker}}]{Hancock2018}
{Hancock}, P.~J., {Trott}, C.~M., \& {Hurley-Walker}, N. 2018, \href{http://dx.doi.org/10.1017/pasa.2018.3}{\JournalTitle{\pasa}, 35, e011}

\bibitem[{Harris {et~al.}(2020)Harris, Millman, van~der Walt, Gommers, Virtanen, Cournapeau, Wieser, Taylor, Berg, Smith, Kern, Picus, Hoyer, van Kerkwijk, Brett, Haldane, del R{\'{i}}o, Wiebe, Peterson, G{\'{e}}rard-Marchant, Sheppard, Reddy, Weckesser, Abbasi, Gohlke, \& Oliphant}]{Harris_2020}
Harris, C.~R., Millman, K.~J., van~der Walt, S.~J., {et~al.} 2020, \href{http://dx.doi.org/10.1038/s41586-020-2649-2}{\JournalTitle{Nature}, 585, 357}

\bibitem[{{Healey} {et~al.}(2007){Healey}, {Romani}, {Taylor}, {Sadler}, {Ricci}, {Murphy}, {Ulvestad}, \& {Winn}}]{Healey2007}
{Healey}, S.~E., {Romani}, R.~W., {Taylor}, G.~B., {et~al.} 2007, \href{http://dx.doi.org/10.1086/513742}{\JournalTitle{\apjs}, 171, 61}

\bibitem[{{Heckman} \& {Best}(2014)}]{heckman2014}
{Heckman}, T.~M., \& {Best}, P.~N. 2014, \href{http://dx.doi.org/10.1146/annurev-astro-081913-035722}{\JournalTitle{\araa}, 52, 589}

\bibitem[{{Hewish} {et~al.}(1964){Hewish}, {Scott}, \& {Wills}}]{Hewish_1964}
{Hewish}, A., {Scott}, P.~F., \& {Wills}, D. 1964, \href{http://dx.doi.org/10.1038/2031214a0}{\JournalTitle{\nat}, 203, 1214}

\bibitem[{{Hewish} \& {Symonds}(1969)}]{Hewish_1969}
{Hewish}, A., \& {Symonds}, M.~D. 1969, \href{http://dx.doi.org/10.1016/0032-0633(69)90064-6}{\JournalTitle{\planss}, 17, 313}

\bibitem[{{Holt} {et~al.}(2008){Holt}, {Tadhunter}, \& {Morganti}}]{holt2008}
{Holt}, J., {Tadhunter}, C.~N., \& {Morganti}, R. 2008, \href{http://dx.doi.org/10.1111/j.1365-2966.2008.13089.x}{\JournalTitle{\mnras}, 387, 639}

\bibitem[{{Hotan} {et~al.}(2021){Hotan}, {Bunton}, {Chippendale}, {Whiting}, {Tuthill}, {Moss}, {McConnell}, {Amy}, {Huynh}, {Allison}, {Anderson}, {Bannister}, {Bastholm}, {Beresford}, {Bock}, {Bolton}, {Chapman}, {Chow}, {Collier}, {Cooray}, {Cornwell}, {Diamond}, {Edwards}, {Feain}, {Franzen}, {George}, {Gupta}, {Hampson}, {Harvey-Smith}, {Hayman}, {Heywood}, {Jacka}, {Jackson}, {Jackson}, {Jeganathan}, {Johnston}, {Kesteven}, {Kleiner}, {Koribalski}, {Lee-Waddell}, {Lenc}, {Lensson}, {Mackay}, {Mahony}, {McClure-Griffiths}, {McConigley}, {Mirtschin}, {Ng}, {Norris}, {Pearce}, {Phillips}, {Pilawa}, {Raja}, {Reynolds}, {Roberts}, {Roxby}, {Sadler}, {Shields}, {Schinckel}, {Serra}, {Shaw}, {Sweetnam}, {Troup}, {Tzioumis}, {Voronkov}, \& {Westmeier}}]{Hotan_2021}
{Hotan}, A.~W., {Bunton}, J.~D., {Chippendale}, A.~P., {et~al.} 2021, \href{http://dx.doi.org/10.1017/pasa.2021.1}{\JournalTitle{\pasa}, 38, e009}

\bibitem[{Hunter(2007)}]{Hunter_2007}
Hunter, J.~D. 2007, \href{http://dx.doi.org/10.1109/MCSE.2007.55}{\JournalTitle{Computing in Science \& Engineering}, 9, 90}

\bibitem[{{Hurley-Walker} {et~al.}(2022){Hurley-Walker}, {Galvin}, {Duchesne}, {Zhang}, {Morgan}, {Hancock}, {An}, {Franzen}, {Heald}, {Ross}, {Vernstrom}, {Anderson}, {Gaensler}, {Johnston-Hollitt}, {Kaplan}, {Riseley}, {Tingay}, \& {Walker}}]{Hurley-Walker_2022}
{Hurley-Walker}, N., {Galvin}, T.~J., {Duchesne}, S.~W., {et~al.} 2022, \href{http://dx.doi.org/10.1017/pasa.2022.17}{\JournalTitle{\pasa}, 39, e035}

\bibitem[{{Kapahi} {et~al.}(1998){Kapahi}, {Athreya}, {van Breugel}, {McCarthy}, \& {Subrahmanya}}]{kapahi1998}
{Kapahi}, V.~K., {Athreya}, R.~M., {van Breugel}, W., {McCarthy}, P.~J., \& {Subrahmanya}, C.~R. 1998, \href{http://dx.doi.org/10.1086/313144}{\JournalTitle{\apjs}, 118, 275}

\bibitem[{{Kellermann} \& {Owen}(1988)}]{Kellermann_1988}
{Kellermann}, K.~I., \& {Owen}, F.~N. 1988, in Galactic and Extragalactic Radio Astronomy, ed. K.~I. {Kellermann} \& G.~L. {Verschuur}, 563

\bibitem[{{Kerrison} {et~al.}(2024){Kerrison}, {Allison}, {Moss}, {Sadler}, \& {Rees}}]{Kerrison2024}
{Kerrison}, E.~F., {Allison}, J.~R., {Moss}, V.~A., {Sadler}, E.~M., \& {Rees}, G.~A. 2024, \href{http://dx.doi.org/10.1093/mnras/stae1796}{\JournalTitle{\mnras}, 533, 4248}

\bibitem[{{Kerrison} {et~al.}(2025){Kerrison}, {Sadler}, {Moss}, {Mahony}, {Driessen}, {Ross}, {Rose}, {Dobie}, \& {Murphy}}]{Kerrison_2025}
{Kerrison}, E.~F., {Sadler}, E.~M., {Moss}, V.~A., {et~al.} 2025, \href{http://dx.doi.org/10.1093/mnras/staf1643}{\JournalTitle{\mnras}, 543, 3895}

\bibitem[{{Lacy} {et~al.}(2020){Lacy}, {Baum}, {Chandler}, {Chatterjee}, {Clarke}, {Deustua}, {English}, {Farnes}, {Gaensler}, {Gugliucci}, {Hallinan}, {Kent}, {Kimball}, {Law}, {Lazio}, {Marvil}, {Mao}, {Medlin}, {Mooley}, {Murphy}, {Myers}, {Osten}, {Richards}, {Rosolowsky}, {Rudnick}, {Schinzel}, {Sivakoff}, {Sjouwerman}, {Taylor}, {White}, {Wrobel}, {Andernach}, {Beasley}, {Berger}, {Bhatnager}, {Birkinshaw}, {Bower}, {Brandt}, {Brown}, {Burke-Spolaor}, {Butler}, {Comerford}, {Demorest}, {Fu}, {Giacintucci}, {Golap}, {G{\"u}th}, {Hales}, {Hiriart}, {Hodge}, {Horesh}, {Ivezi{\'c}}, {Jarvis}, {Kamble}, {Kassim}, {Liu}, {Loinard}, {Lyons}, {Masters}, {Mezcua}, {Moellenbrock}, {Mroczkowski}, {Nyland}, {O'Dea}, {O'Sullivan}, {Peters}, {Radford}, {Rao}, {Robnett}, {Salcido}, {Shen}, {Sobotka}, {Witz}, {Vaccari}, {van Weeren}, {Vargas}, {Williams}, \& {Yoon}}]{Lacy_2020}
{Lacy}, M., {Baum}, S.~A., {Chandler}, C.~J., {et~al.} 2020, \href{http://dx.doi.org/10.1088/1538-3873/ab63eb}{\JournalTitle{\pasp}, 132, 035001}

\bibitem[{{Maccagni} {et~al.}(2017){Maccagni}, {Morganti}, {Oosterloo}, {Ger{\'e}b}, \& {Maddox}}]{Maccagni_2017}
{Maccagni}, F.~M., {Morganti}, R., {Oosterloo}, T.~A., {Ger{\'e}b}, K., \& {Maddox}, N. 2017, \href{http://dx.doi.org/10.1051/0004-6361/201730563}{\JournalTitle{\aap}, 604, A43}

\bibitem[{{Mahony} {et~al.}(2022){Mahony}, {Allison}, {Sadler}, {Ellison}, {Mao}, {Morganti}, {Moss}, {Seta}, {Tadhunter}, {Weng}, {Whiting}, {Yoon}, {Bell}, {Bunton}, {Harvey-Smith}, {Kimball}, {Koribalski}, \& {Voronkov}}]{Mahony_2022}
{Mahony}, E.~K., {Allison}, J.~R., {Sadler}, E.~M., {et~al.} 2022, \href{http://dx.doi.org/10.1093/mnras/stab3041}{\JournalTitle{\mnras}, 509, 1690}

\bibitem[{{McCarthy} {et~al.}(1996){McCarthy}, {Kapahi}, {van Breugel}, {Persson}, {Athreya}, \& {Subrahmanya}}]{McCarthy_1996}
{McCarthy}, P.~J., {Kapahi}, V.~K., {van Breugel}, W., {et~al.} 1996, \href{http://dx.doi.org/10.1086/192353}{\JournalTitle{\apjs}, 107, 19}

\bibitem[{McConnell {et~al.}(2020)McConnell, Hale, Lenc, Banfield, Heald, Hotan, Leung, Moss, Murphy, O’Brien, \& et~al.}]{McConnell_2020}
McConnell, D., Hale, C.~L., Lenc, E., {et~al.} 2020, \href{http://dx.doi.org/10.1017/pasa.2020.41}{\JournalTitle{Publications of the Astronomical Society of Australia}, 37, e048}

\bibitem[{{Morgan} {et~al.}(2019){Morgan}, {Macquart}, {Chhetri}, {Ekers}, {Tingay}, \& {Sadler}}]{Morgan_2019}
{Morgan}, J.~S., {Macquart}, J.-P., {Chhetri}, R., {et~al.} 2019, \href{http://dx.doi.org/10.1017/pasa.2018.40}{\JournalTitle{\pasa}, 36, e002}

\bibitem[{{Morgan} {et~al.}(2018){Morgan}, {Macquart}, {Ekers}, {Chhetri}, {Tokumaru}, {Manoharan}, {Tremblay}, {Bisi}, \& {Jackson}}]{Morgan_2018}
{Morgan}, J.~S., {Macquart}, J.~P., {Ekers}, R., {et~al.} 2018, \href{http://dx.doi.org/10.1093/mnras/stx2284}{\JournalTitle{\mnras}, 473, 2965}

\bibitem[{{Morganti} \& {Oosterloo}(2018)}]{Morganti_2018}
{Morganti}, R., \& {Oosterloo}, T. 2018, \href{http://dx.doi.org/10.1007/s00159-018-0109-x}{\JournalTitle{\aapr}, 26, 4}

\bibitem[{{Murphy} {et~al.}(2010){Murphy}, {Sadler}, {Ekers}, {Massardi}, {Hancock}, {Mahony}, {Ricci}, {Burke-Spolaor}, {Calabretta}, {Chhetri}, {de Zotti}, {Edwards}, {Ekers}, {Jackson}, {Kesteven}, {Lindley}, {Newton-McGee}, {Phillips}, {Roberts}, {Sault}, {Staveley-Smith}, {Subrahmanyan}, {Walker}, \& {Wilson}}]{Murphy_2010}
{Murphy}, T., {Sadler}, E.~M., {Ekers}, R.~D., {et~al.} 2010, \href{http://dx.doi.org/10.1111/j.1365-2966.2009.15961.x}{\JournalTitle{\mnras}, 402, 2403}

\bibitem[{{O'Dea}(1998)}]{odea1998}
{O'Dea}, C.~P. 1998, \href{http://dx.doi.org/10.1086/316162}{\JournalTitle{\pasp}, 110, 493}

\bibitem[{{O'Dea} \& {Saikia}(2021)}]{odea2021}
{O'Dea}, C.~P., \& {Saikia}, D.~J. 2021, \href{http://dx.doi.org/10.1007/s00159-021-00131-w}{\JournalTitle{\aapr}, 29, 3}

\bibitem[{{Orienti} {et~al.}(2007){Orienti}, {Dallacasa}, \& {Stanghellini}}]{Orienti_2007}
{Orienti}, M., {Dallacasa}, D., \& {Stanghellini}, C. 2007, \href{http://dx.doi.org/10.1051/0004-6361:20066122}{\JournalTitle{\aap}, 461, 923}

\bibitem[{{Owen} {et~al.}(1995){Owen}, {Ledlow}, \& {Keel}}]{owen1995}
{Owen}, F.~N., {Ledlow}, M.~J., \& {Keel}, W.~C. 1995, \href{http://dx.doi.org/10.1086/117252}{\JournalTitle{\aj}, 109, 14}

\bibitem[{{P{\'e}roux} \& {Howk}(2020)}]{Peroux_2020}
{P{\'e}roux}, C., \& {Howk}, J.~C. 2020, \href{http://dx.doi.org/10.1146/annurev-astro-021820-120014}{\JournalTitle{\araa}, 58, 363}

\bibitem[{{Peterson} {et~al.}(1979){Peterson}, {Wright}, {Jauncey}, \& {Condon}}]{peterson1979}
{Peterson}, B.~A., {Wright}, A.~E., {Jauncey}, D.~L., \& {Condon}, J.~J. 1979, \href{http://dx.doi.org/10.1086/157299}{\JournalTitle{\apj}, 232, 400}

\bibitem[{{Pihlstr{\"o}m} {et~al.}(2003){Pihlstr{\"o}m}, {Conway}, \& {Vermeulen}}]{Philstrom_2003}
{Pihlstr{\"o}m}, Y.~M., {Conway}, J.~E., \& {Vermeulen}, R.~C. 2003, \href{http://dx.doi.org/10.1051/0004-6361:20030469}{\JournalTitle{\aap}, 404, 871}

\bibitem[{{Planck Collaboration} {et~al.}(2020){Planck Collaboration}, {Aghanim}, {Akrami}, {Ashdown}, {Aumont}, {Baccigalupi}, {Ballardini}, {Banday}, {Barreiro}, {Bartolo}, {Basak}, {Battye}, {Benabed}, {Bernard}, {Bersanelli}, {Bielewicz}, {Bock}, {Bond}, {Borrill}, {Bouchet}, {Boulanger}, {Bucher}, {Burigana}, {Butler}, {Calabrese}, {Cardoso}, {Carron}, {Challinor}, {Chiang}, {Chluba}, {Colombo}, {Combet}, {Contreras}, {Crill}, {Cuttaia}, {de Bernardis}, {de Zotti}, {Delabrouille}, {Delouis}, {Di Valentino}, {Diego}, {Dor{\'e}}, {Douspis}, {Ducout}, {Dupac}, {Dusini}, {Efstathiou}, {Elsner}, {En{\ss}lin}, {Eriksen}, {Fantaye}, {Farhang}, {Fergusson}, {Fernandez-Cobos}, {Finelli}, {Forastieri}, {Frailis}, {Fraisse}, {Franceschi}, {Frolov}, {Galeotta}, {Galli}, {Ganga}, {G{\'e}nova-Santos}, {Gerbino}, {Ghosh}, {Gonz{\'a}lez-Nuevo}, {G{\'o}rski}, {Gratton}, {Gruppuso}, {Gudmundsson}, {Hamann}, {Handley}, {Hansen}, {Herranz}, {Hildebrandt}, {Hivon}, {Huang}, {Jaffe}, {Jones}, {Karakci}, {Keih{\"a}nen},
  {Keskitalo}, {Kiiveri}, {Kim}, {Kisner}, {Knox}, {Krachmalnicoff}, {Kunz}, {Kurki-Suonio}, {Lagache}, {Lamarre}, {Lasenby}, {Lattanzi}, {Lawrence}, {Le Jeune}, {Lemos}, {Lesgourgues}, {Levrier}, {Lewis}, {Liguori}, {Lilje}, {Lilley}, {Lindholm}, {L{\'o}pez-Caniego}, {Lubin}, {Ma}, {Mac{\'\i}as-P{\'e}rez}, {Maggio}, {Maino}, {Mandolesi}, {Mangilli}, {Marcos-Caballero}, {Maris}, {Martin}, {Martinelli}, {Mart{\'\i}nez-Gonz{\'a}lez}, {Matarrese}, {Mauri}, {McEwen}, {Meinhold}, {Melchiorri}, {Mennella}, {Migliaccio}, {Millea}, {Mitra}, {Miville-Desch{\^e}nes}, {Molinari}, {Montier}, {Morgante}, {Moss}, {Natoli}, {N{\o}rgaard-Nielsen}, {Pagano}, {Paoletti}, {Partridge}, {Patanchon}, {Peiris}, {Perrotta}, {Pettorino}, {Piacentini}, {Polastri}, {Polenta}, {Puget}, {Rachen}, {Reinecke}, {Remazeilles}, {Renzi}, {Rocha}, {Rosset}, {Roudier}, {Rubi{\~n}o-Mart{\'\i}n}, {Ruiz-Granados}, {Salvati}, {Sandri}, {Savelainen}, {Scott}, {Shellard}, {Sirignano}, {Sirri}, {Spencer}, {Sunyaev}, {Suur-Uski}, {Tauber}, {Tavagnacco},
  {Tenti}, {Toffolatti}, {Tomasi}, {Trombetti}, {Valenziano}, {Valiviita}, {Van Tent}, {Vibert}, {Vielva}, {Villa}, {Vittorio}, {Wandelt}, {Wehus}, {White}, {White}, {Zacchei}, \& {Zonca}}]{Planck_paper}
{Planck Collaboration}, {Aghanim}, N., {Akrami}, Y., {et~al.} 2020, \href{http://dx.doi.org/10.1051/0004-6361/201833910}{\JournalTitle{\aap}, 641, A6}

\bibitem[{{Rickett}(1973)}]{Rickett_1973}
{Rickett}, B.~J. 1973, \href{http://dx.doi.org/10.1029/JA078i010p01543}{\JournalTitle{\jgr}, 78, 1543}

\bibitem[{{Roster} {et~al.}(2024){Roster}, {Salvato}, {Krippendorf}, {Saxena}, {Shirley}, {Buchner}, {Wolf}, {Dwelly}, {Bauer}, {Aird}, {Ricci}, {Assef}, {Anderson}, {Liu}, {Merloni}, {Weller}, \& {Nandra}}]{roster2024}
{Roster}, W., {Salvato}, M., {Krippendorf}, S., {et~al.} 2024, \href{http://dx.doi.org/10.1051/0004-6361/202452361}{\JournalTitle{\aap}, 692, A260}

\bibitem[{{Roster} {et~al.}(2026){Roster}, {Sadler}, {Mahony}, {Salvato}, {Yoon}, {Kluge}, {Shirley}, {Kerrison}, {Buchner}, {Igo}, {Davies}, {Allison}, {Shabala}, {Moss}, {Starck}, {Whiting}, {Nandra}, \& {Weller}}]{Roster_2026}
{Roster}, W., {Sadler}, E., {Mahony}, E., {et~al.} 2026, \href{http://dx.doi.org/10.1017/pasa.2026.10196}{\JournalTitle{\pasa}, 43, e065}

\bibitem[{{Sadler} {et~al.}(2019){Sadler}, {Chhetri}, {Morgan}, {Mahony}, {Jarrett}, \& {Tingay}}]{sadler2019}
{Sadler}, E.~M., {Chhetri}, R., {Morgan}, J., {et~al.} 2019, \href{http://dx.doi.org/10.1093/mnras/sty3033}{\JournalTitle{\mnras}, 483, 1354}

\bibitem[{{Sadler} {et~al.}(2020){Sadler}, {Moss}, {Allison}, {Mahony}, {Whiting}, {Johnston}, {Ellison}, {Lagos}, \& {Koribalski}}]{Sadler_2020}
{Sadler}, E.~M., {Moss}, V.~A., {Allison}, J.~R., {et~al.} 2020, \href{http://dx.doi.org/10.1093/mnras/staa2390}{\JournalTitle{\mnras}, 499, 4293}

\bibitem[{{Saxena} {et~al.}(2024){Saxena}, {Salvato}, {Roster}, {Shirley}, {Buchner}, {Wolf}, {Kohl}, {Starck}, {Dwelly}, {Comparat}, {Malyali}, {Krippendorf}, {Zenteno}, {Lang}, {Schlegel}, {Zhou}, {Dey}, {Valdes}, {Myers}, {Assef}, {Ricci}, {Temple}, {Merloni}, {Koekemoer}, {Anderson}, {Morrison}, {Liu}, \& {Nandra}}]{saxena2024}
{Saxena}, A., {Salvato}, M., {Roster}, W., {et~al.} 2024, \href{http://dx.doi.org/10.1051/0004-6361/202450886}{\JournalTitle{\aap}, 690, A365}

\bibitem[{{Skilling}(2004)}]{Skilling_2004}
{Skilling}, J. 2004, \href{http://dx.doi.org/10.1063/1.1835238}{in American Institute of Physics Conference Series, Vol. 735, Bayesian Inference and Maximum Entropy Methods in Science and Engineering: 24th International Workshop on Bayesian Inference and Maximum Entropy Methods in Science and Engineering, ed. R.~{Fischer}, R.~{Preuss}, \& U.~V. {Toussaint}}, 395

\bibitem[{Slob {et~al.}(2022)Slob, Callingham, R{\"{o}}ttgering, Williams, Duncan, de~Gasperin, Hardcastle, \& Miley}]{Slob2022}
Slob, M.~M., Callingham, J.~R., R{\"{o}}ttgering, H. J.~A., {et~al.} 2022, \href{http://dx.doi.org/10.1051/0004-6361/202244651}{\JournalTitle{Astronomy {\&} Astrophysics}, 668, A186}

\bibitem[{{Snellen} {et~al.}(1998){Snellen}, {Schilizzi}, {de Bruyn}, {Miley}, {Rengelink}, {Roettgering}, \& {Bremer}}]{Snellen_1998}
{Snellen}, I.~A.~G., {Schilizzi}, R.~T., {de Bruyn}, A.~G., {et~al.} 1998, \href{http://dx.doi.org/10.1051/aas:1998281}{\JournalTitle{\aaps}, 131, 435}

\bibitem[{{Storey-Fisher} {et~al.}(2024){Storey-Fisher}, {Hogg}, {Rix}, {Eilers}, {Fabbian}, {Blanton}, \& {Alonso}}]{quaia2024}
{Storey-Fisher}, K., {Hogg}, D.~W., {Rix}, H.-W., {et~al.} 2024, \href{http://dx.doi.org/10.3847/1538-4357/ad1328}{\JournalTitle{\apj}, 964, 69}

\bibitem[{{Su} {et~al.}(2022{\natexlab{a}}){Su}, {Sadler}, {Allison}, {Mahony}, {Moss}, {Whiting}, {Yoon}, {Aditya}, {Bellstedt}, {Robotham}, \& et~al.}]{Su_2022}
{Su}, R., {Sadler}, E.~M., {Allison}, J.~R., {et~al.} 2022{\natexlab{a}}, \href{http://dx.doi.org/10.1093/mnras/stac2257}{\JournalTitle{\mnras}, 516, 2947}

\bibitem[{{Su} {et~al.}(2022{\natexlab{b}}){Su}, {Sadler}, {Allison}, {Mahony}, {Moss}, {Whiting}, {Yoon}, {Aditya}, {Bellstedt}, {Robotham}, \& et~al.}]{Renzhi_2022}
---. 2022{\natexlab{b}}, \href{http://dx.doi.org/10.1093/mnras/stac2257}{\JournalTitle{\mnras}, 516, 2947}

\bibitem[{{Tingay} {et~al.}(2013){Tingay}, {Goeke}, {Bowman}, {Emrich}, {Ord}, {Mitchell}, {Morales}, {Booler}, {Crosse}, {Wayth}, {Lonsdale}, {Tremblay}, {Pallot}, {Colegate}, {Wicenec}, {Kudryavtseva}, {Arcus}, {Barnes}, {Bernardi}, {Briggs}, {Burns}, {Bunton}, {Cappallo}, {Corey}, {Deshpande}, {Desouza}, {Gaensler}, {Greenhill}, {Hall}, {Hazelton}, {Herne}, {Hewitt}, {Johnston-Hollitt}, {Kaplan}, {Kasper}, {Kincaid}, {Koenig}, {Kratzenberg}, {Lynch}, {Mckinley}, {Mcwhirter}, {Morgan}, {Oberoi}, {Pathikulangara}, {Prabu}, {Remillard}, {Rogers}, {Roshi}, {Salah}, {Sault}, {Udaya-Shankar}, {Schlagenhaufer}, {Srivani}, {Stevens}, {Subrahmanyan}, {Waterson}, {Webster}, {Whitney}, {Williams}, {Williams}, \& {Wyithe}}]{Tingay_2013}
{Tingay}, S.~J., {Goeke}, R., {Bowman}, J.~D., {et~al.} 2013, \href{http://dx.doi.org/10.1017/pasa.2012.007}{\JournalTitle{\pasa}, 30, e007}

\bibitem[{{Titov} {et~al.}(2011){Titov}, {Jauncey}, {Johnston}, {Hunstead}, \& {Christensen}}]{titov2011}
{Titov}, O., {Jauncey}, D.~L., {Johnston}, H.~M., {Hunstead}, R.~W., \& {Christensen}, L. 2011, \href{http://dx.doi.org/10.1088/0004-6256/142/5/165}{\JournalTitle{\aj}, 142, 165}

\bibitem[{{Vermeulen} {et~al.}(2003){Vermeulen}, {Pihlstr{\"o}m}, {Tschager}, {de Vries}, {Conway}, {Barthel}, {Baum}, {Braun}, {Bremer}, {Miley}, {O'Dea}, {R{\"o}ttgering}, {Schilizzi}, {Snellen}, \& {Taylor}}]{Vermeulen_2003}
{Vermeulen}, R.~C., {Pihlstr{\"o}m}, Y.~M., {Tschager}, W., {et~al.} 2003, \href{http://dx.doi.org/10.1051/0004-6361:20030468}{\JournalTitle{\aap}, 404, 861}

\bibitem[{{V{\'e}ron-Cetty} {et~al.}(2000){V{\'e}ron-Cetty}, {Woltjer}, {Staveley-Smith}, \& {Ekers}}]{Veron-Cetty_2000}
{V{\'e}ron-Cetty}, M.~P., {Woltjer}, L., {Staveley-Smith}, L., \& {Ekers}, R.~D. 2000, \JournalTitle{\aap}, 362, 426

\bibitem[{{Wang} {et~al.}(2025){Wang}, {Bannister}, {Gupta}, {Deng}, {Pilawa}, {Tuthill}, {Bunton}, {Flynn}, {Glowacki}, {Jaini}, {Lee}, {Lenc}, {Lucero}, {Paek}, {Radhakrishnan}, {Thyagarajan}, {Uttarkar}, {Wang}, {Bhat}, {James}, {Moss}, {Murphy}, {Reynolds}, {Shannon}, {Spitler}, {Tzioumis}, {Caleb}, {Deller}, {Gordon}, {Marnoch}, {Ryder}, {Simha}, {Anderson}, {Ball}, {Brodrick}, {Cooray}, {Gupta}, {Hayman}, {Ng}, {Pearce}, {Phillips}, {Voronkov}, \& {Westmeier}}]{wang2025}
{Wang}, Z., {Bannister}, K.~W., {Gupta}, V., {et~al.} 2025, \href{http://dx.doi.org/10.1017/pasa.2024.107}{\JournalTitle{\pasa}, 42, e005}

\bibitem[{{W}es {M}c{K}inney(2010)}]{McKinney_2010}
{W}es {M}c{K}inney. 2010, \href{http://dx.doi.org/10.25080/Majora-92bf1922-00a}{in {P}roceedings of the 9th {P}ython in {S}cience {C}onference, ed. {S}t\'efan van~der {W}alt \& {J}arrod {M}illman}, 56

\bibitem[{{Whiting} \& {Humphreys}(2012)}]{Whiting_2012}
{Whiting}, M., \& {Humphreys}, B. 2012, \href{http://dx.doi.org/10.1071/AS12028}{\JournalTitle{\pasa}, 29, 371}

\bibitem[{{Yoon} {et~al.}(2025){Yoon}, {Sadler}, {Mahony}, {Aditya}, {Allison}, {Glowacki}, {Kerrison}, {Moss}, {Su}, {Weng}, {Whiting}, {Wong}, {Callingham}, {Curran}, {Darling}, {Edge}, {Ellison}, {Emig}, {Garratt-Smithson}, {German}, {Grasha}, {Koribalski}, {Morganti}, {Oosterloo}, {P{\'e}roux}, {Pettini}, {Pimbblet}, {Zheng}, {Zwaan}, {Ball}, {Bock}, {Brodrick}, {Bunton}, {Cooray}, {Edwards}, {Hayman}, {Hotan}, {Lee-Waddell}, {McClure-Griffiths}, {Ng}, {Phillips}, {Raja}, {Voronkov}, \& {Westmeier}}]{Yoon_2025}
{Yoon}, H., {Sadler}, E.~M., {Mahony}, E.~K., {et~al.} 2025, \href{http://dx.doi.org/10.1017/pasa.2025.10046}{\JournalTitle{\pasa}, 42, e088}

\end{thebibliography}
%\section{\uppercase{Supporting Information}}
%Placeholder for link to supplementary data. Will include tables that are continuations of Tables 4 and 5.

%%%% Redshift refs for table below:
% Ba99 Baker+1999
% Bu70 Burbidge 1970
% D597 Drinkwater+ 1997
% Du89 Dunlop+1989
% Ho08 Holt+2008
% Mc96 McCarthy+1996
% Ow95 Owen+1995
% Pe79 Peterson+ 1979
% quaia Storey-Fisher+2024
% Ti11 Titov+2011

\appendix
\section{Data table}

\begin{landscape}
\begin{table*}[!ht]
\hspace*{-6.0cm}
    \begin{tabular}{lllllrrrllllrllll}
        Source &  NED name  & RAdeg & DEdeg &  Object Type & F\_tot & F\_pk & S/N & z\_phot & z\_spec & err & z\_ref &  NSI & err & NSI & err & SED \\  
        & & \multicolumn{2}{c}{J2000} & (LS 10) & (mJy) & (mJy) & Cont & PICZL & & z\_spec & & ASKAP & NSI & MWA & NSI & \\
    \hline
        J212805.5-232944 & MRC 2125-237   & 322.022934 & -23.495761 & Galaxy   & 1114.88 & 1093.66 & 49.36 & .. & 0.95 & 0.05 & Mc96 & 0.98 & 0.06 & 1.04 & 0.06 & PS \\ 
        J212857.5-224244 & NVSS  J212857-224245 & 322.239795 & -22.712374 & Galaxy & 239.37 & 232.78 & 26.86 & 1.5 & .. & .. & .. & < 0.18 & .. & 0.38 & 0.02 & PL \\ 
        J212937.1-224828 & MRC 2126-230   & 322.404881 & -22.807968 & .. & 1301.06 & 769.16 & 106.69 & .. & .. & .. & .. & 0.13 & 0.09 & 0.17 & 0.01 & PL \\ 
        J213153.3-221102 & NVSS J213153-221103  & 322.972376 & -22.184104 & .. & 164.29 & 150.42 & 51.72 & .. & .. & .. & .. & < 0.11 & .. & 0.15 & 0.04 & PL \\ 
        J213154.3-230827 & NVSS  J213154-230828  & 322.976505 & -23.140914 & Quasar & 162.07 & 153.57 & 43.72 & .. & 1.484 & 0.088 & Quaia & < 0.16 & .. & 0.20 & 0.02 & PL \\ 
        &&&& \\
        J213250.4-230232 & MRC 2129-232 & 323.210089 & -23.042455 & .. & 477.5 & 252.83 & 65.9 & .. & .. & .. & .. & < 0.07 & .. & < 0.00 & < 0.02 & PL \\ 
        J213334.2-213818 & NVSS J213334-213818   & 323.392862 & -21.638441 & Quasar   & 276.24 & 271.67 & 80.54 & 0.1 & .. & .. & .. & 0.7 & 0.04 & 0.89 & 0.05 & PS \\ 
        J213339.0-234916 & NVSS J213339-234917 & 323.412796 & -23.821358 & Galaxy & 207.1 & 198.8 & 45.27 & 1.18 & .. & .. & .. & < 0.18 & .. & 0.38 & 0.02 & PL \\ 
        J213411.6-173542 & PKS 2131-178   & 323.548689 & -17.595065 & Galaxy   & 332.38 & 323.23 & 22.15 & 0.56 & .. & .. & .. & 0.66 & 0.1 & 0.94 & 0.06 & PL \\ 
        J213429.3-244308 & NVSS J213429-244309 & 323.62245 & -24.719154 & Quasar & 303.28 & 241.04 & 25.34 & 1.15 & .. & .. & .. & < 0.55 & .. & 0.25 & 0.02 & PL \\ 
        &&&& \\
        J213437.6-235535 & MRC 2131-241   & 323.656723 & -23.926639 & Galaxy   & 554.65 & 533.23 & 113.29 & 0.66 & .. & .. & .. & 0.84 & 0.04 & 0.85 & 0.05 & PS \\ 
        J213516.8-173429 & PKS 2132-177  & 323.820021 & -17.574727 & .. & 1585.87 & 1142.84 & 104.13 & .. & .. & .. & .. & 0.2 & 0.04 & 0.31 & 0.02 & PL \\ 
        J213729.4-195327 & NVSS J213729-195326  & 324.37281 & -19.89097 & .. & 261.48 & 239.2 & 74.7 & .. & .. & .. & .. & 0.17 & 0.14 & 0.13 & 0.01 & PL \\ 
        J213749.9-204232 & PKS 2135-20 & 324.458263 & -20.709059 & Galaxy  & 5223.87 & 5076.27 & 552.18 & 0.55 & 0.635 & 0.001 & Ho08 & 0.59 & 0.01 & 0.91 & 0.05 & PS \\ 
        J213801.0-253148 & MRC 2135-257  & 324.504382 & -25.530162 & Galaxy  & 819 & 801.79 & 61.81 & .. & 1.31 & 0.05 & Mc96 & 0.83 & 0.16 & 0.65 & 0.04 & PL \\ 
        &&&& \\
        J213802.1-221823 & NVSS J213802-221824  & 324.508853 & -22.306561 & Quasar  & 180.82 & 169.62 & 98.22 & 0.2 & .. & .. & .. & 0.21 & 0.14 & 0.20 & 0.01 & PL \\ 
        J213805.0-184330 & PKS 2135-18  & 324.521035 & -18.725081 & .. & 2225.38 & 1894.2 & 351.16 & .. & .. & .. & .. & 0.14 & 0.02 & 0.14 & 0.01 & PL \\ 
        J213837.1-243954 & PKS 2135-248  & 324.65472 & -24.665006 & Quasar  & 405.41 & 394.64 & 135.97 & 0.62 & 0.819 & 0.001 & Pe79 & 1.14 & 0.04 & .. & .. & PS \\ 
        J213838.3-175040 & MRC 2135-180  & 324.659658 & -17.844705 & .. & 597.32 & 578.81 & 79.52 & .. & .. & .. & .. & 0.57 & 0.04 & 0.62 & 0.04 & PL \\ 
        J213841.9-181045 & PKS 2135-184  & 324.674769 & -18.179252 & Galaxy  & 2402.56 & 2353.56 & 293.25 & 0.1 & 0.1887 & 0.0001 & Ow95 & 0.57 & 0.01 & 0.80 & 0.05 & PL \\ 
        &&&& \\
        J213847.4-184931 & NVSS J213847-184930  & 324.697736 & -18.825485 & .. & 399.1 & 388.94 & 76.27 &  .. & .. & .. & .. & 0.8 & 0.04 & 1.01 & 0.12 & I \\ 
        J213857.0-195725 & NVSS J213857-195723  & 324.737765 & -19.957044 & .. & 226.11 & 219.2 & 59.26 & .. & .. & .. & .. & 0.87 & 0.06 & 0.84 & 0.05 & PL \\ 
        J213908.5-245718 & NVSS J213908-245718  & 324.785738 & -24.955083 & .. & 232.76 & 225.97 & 50.98 & .. & .. & .. & .. & 0.64 & 0.17 & 0.47 & 0.03 & PL \\ 
        J213913.1-245414 & MRC 2136-251  & 324.804976 & -24.904024 & Galaxy & 729.56 & 697.9 & 162.31 & 0.6 & 0.939 & .. & Ba99 & 0.27 & 0.11 & 0.22 & 0.01 & PL \\ 
        J213924.5-255643 & MRC 2136-261  & 324.852211 & -25.945518 & .. & 1540.24 & 888.59 & 55.53 & .. & .. & .. & .. & 0.88 & 0.19 & 0.42 & 0.03 & PL \\ 
        &&&& \\
        J213926.4-172352 & NVSS J213926-172351  & 324.860129 & -17.397855 & .. & 240.84 & 234.84 & 38.44 & .. & .. & .. & .. & 0.91 & 0.07 & 0.95 & 0.07 & PS \\ 
        J213939.5-183810 & MRC 2136-188  & 324.914888 & -18.636148 & .. & 823.17 & 796.25 & 142.44 & .. & .. & .. & .. & 0.19 & 0.04 & 0.20 & 0.01 & PL \\ 
        J213945.0-250114 & NVSS J213945-250115 & 324.93767 & -25.0208 & Galaxy & 214.09 & 197.02 & 44.38 & 1.58 & .. & .. & .. & < 0.30 & .. & 0.19 & 0.02 & PL \\ 
        J213950.9-191630 & NVSS J213951-191629 & 324.96239 & -19.275152 & .. & 274.56 & 173.92 & 32.94 & .. & .. & .. & .. & 0.27 & 0.18 & 0.08 & 0.02 & PL \\ 
        J213957.7-212034 & NVSS J213957-212034  & 324.99066 & -21.342938 & .. & 207.03 & 165 & 48.32 & .. & .. & .. & .. & 0.22 & 0.21 & 0.27 & 0.02 & PL \\ 
                 \hline
\end{tabular}
\caption{Bright-source sample}
\label{tab:main_bright}
\end{table*}

\begin{table*}[!ht]
\hspace*{-6.0cm}
    \begin{tabular}{lllllrrrllllrllll}
        Source &  NED name  & RAdeg & DEdeg &  Object Type & F\_tot & F\_pk & S/N & z\_phot & z\_spec & err & z\_ref &  NSI & err & NSI & err & SED \\  
        & & \multicolumn{2}{c}{J2000} & (LS 10) & (mJy) & (mJy) & Cont & PICZL & & z\_spec & & ASKAP & NSI & MWA & NSI & \\
    \hline
        J214010.7-161628 & MRC 2137-165  & 325.044656 & -16.274532 & .. & 1193.14 & 566.08 & 27.9 & .. & .. & .. & .. & < 0.21 & .. & 0.09 & 0.01 & PL \\ 
        J214011.6-182529 & NVSS J214011-182528 & 325.048503 & -18.424837 & .. & 194.34 & 187.38 & 38.17 & .. & .. & .. & .. & 0.23 & 0.16 & 0.38 & 0.02 & PL \\ 
        J214032.5-211559 & NVSS J214032-211559  & 325.135587 & -21.266663 & .. & 316.24 & 250.32 & 93.7 & .. & .. & .. & .. & 0.24 & 0.1 & 0.23 & 0.01 & PL \\ 
        J214119.6-204102 & NVSS J214119-204102  & 325.331842 & -20.684099 & .. & 197.22 & 182.79 & 63.3 & .. & .. & .. & .. & < 0.11 & .. & 0.14 & 0.02 & PL \\ 
        J214159.8-254757 & NVSS J214159-254758 & 325.499304 & -25.799395 & .. & 254.44 & 213.4 & 34.33 & ~ & .. & .. & .. & < 0.61 & .. & 0.13 & 0.01 & PL \\ 
        &&&& \\
        J214201.7-191401 & NVSS J214201-191400  & 325.507319 & -19.233788 & Quasar  & 362.74 & 352.25 & 112.49 & 1.82 & .. & .. & .. & 0.84 & 0.03 & 0.84 & 0.05 & PS \\ 
        J214204.5-232919 & NVSS J214204-232920  & 325.519116 & -23.488684 & Galaxy  & 280.62 & 265.93 & 167.41 & 1.14 & .. & .. & .. & 0.13 & 0.2 & 0.14 & 0.01 & PL \\ 
        J214212.9-180743 & NVSS J214212-180742 & 325.553896 & -18.128764 & Galaxy & 206.05 & 192.92 & 41.41 & 1.66 & .. & .. & .. & < 0.10 & .. & 0.21 & 0.02 & PL \\ 
        J214230.8-244438 & NVSS  J214230-244440  & 325.628434 & -24.744064 & Quasar  & 251.71 & 221.22 & 111.98 & 0.58 & 0.833 & .. & deWitt23 & 0.73 & 0.08 & 0.13 & 0.04 & I \\ 
        J214233.3-241231 & NVSS J214233-241232 & 325.638763 & -24.208714 & Galaxy & 259.09 & 220.28 & 114.65 & 0.63 & .. & .. & .. & < 0.10 & .. & < 0.05 & < 0.02 & PL \\ 
        &&&& \\
        J214236.9-180620 & NVSS J214236-180619  & 325.653883 & -18.105705 & .. & 169.43 & 160.9 & 32.5 & .. & .. & .. & .. & 0.21 & 0.2 & 0.22 & 0.02 & PL \\ 
        J214241.9-230337 & NVSS J214241-230340  & 325.674633 & -23.060547 & Quasar  & 175.3 & 171.6 & 104.72 & 0.31 & .. & .. & .. & 1 & 0.05 & 0.78 & 0.05 & PL \\ 
        J214304.2-164147 & NVSS J214304-164146  & 325.767726 & -16.696639 & Galaxy  & 255.26 & 250.01 & 24.7 & 1.09 & .. & .. & .. & 0.76 & 0.13 & 0.73 & 0.07 & PS \\ 
        J214307.4-183553 & MRC 2140-188  & 325.78102 & -18.598324 & Galaxy  & 551.49 & 533.79 & 167.47 & 0.15 & 0.1873 & 0.0002 & Ow95 & 0.23 & 0.04 & 0.38 & 0.02 & PL \\ 
        J214329.1-181100 & MRC 2140-184  & 325.871374 & -18.183392 & Quasar  & 1230.97 & 1104.38 & 256.2 & 1.64 & .. & .. & .. & 0.15 & 0.04 & 0.29 & 0.02 & PL \\ 
        &&&& \\
        J214408.8-244240 & NVSS J214408-244241  & 326.036783 & -24.711146 & Quasar & 172.67 & 163.73 & 72.05 & 1.33 & .. & .. & .. & < 0.17 & .. & < 0.04 & < 0.05 & PL \\ 
        J214431.7-240210 & NVSS J214431-240211  & 326.132174 & -24.036154 & .. & 315.54 & 245.53 & 137.65 & .. & .. & .. & .. & 0.2 & 0.19 & 0.26 & 0.02 & PL \\ 
        J214449.1-221552 & NVSS J214449-221552  & 326.204797 & -22.264564 & Galaxy & 200.36 & 197 & 127.22 & 0.54 & .. & .. & .. & 0.49 & 0.08 & 0.68 & 0.04 & PL \\ 
        J214525.5-194841 & NVSS J214525-194842 & 326.35641 & -19.811429 & Galaxy  & 225.08 & 172.67 & 100.3 & 1.55 & .. & .. & .. & 0.14 & 0.19 & 0.21 & 0.02 & PL \\ 
        J214548.9-182600 & NVSS J214549-182559  & 326.454069 & -18.43357 & Galaxy  & 183.58 & 179.71 & 59.89 & 1.19 & .. & .. & .. & 0.86 & 0.07 & 0.68 & 0.09 & PS \\ 
        &&&& \\
        J214641.5-242957 & NVSS J214641-242957  & 326.673226 & -24.499255 & .. & 249.59 & 241.99 & 120.51 & .. & .. & .. & .. & 0.55 & 0.11 & 0.58 & 0.04 & PL \\ 
        J214703.2-174012 & PKS 2144-17  & 326.763378 & -17.670175 & Quasar  & 1566.25 & 1527.43 & 284.52 & 0.67 & 0.684 & .. & Bu70 & 0.37 & 0.02 & 0.49 & 0.03 & PL \\ 
        J214717.4-194647 & NVSS J214717-194646 & 326.822803 & -19.77975 & .. & 246.4 & 222.78 & 90.59 & .. & .. & .. & .. & 0.23 & 0.23 & < 0.00 & < 0.06 & PL \\ 
        J214740.7-232341 & MRC 2144-236  & 326.919624 & -23.394978 & Galaxy  & 578.85 & 493.93 & 255.16 & 1.34 & .. & .. & .. & 0.1 & 0.17 & 0.10 & 0.01 & PL \\ 
        J214809.0-251611 & NVSS J214809-251612  & 327.03756 & -25.26982 & .. & 305.18 & 272.43 & 62.9 & .. & .. & .. & .. & < 0.34 & .. & 0.14 & 0.01 & PL \\ 
        &&&& \\
        J214827.6-194126 & MRC 2145-199  & 327.115112 & -19.690663 & Galaxy  & 668.15 & 610.56 & 196.18 & 0.8 & .. & .. & .. & 0.09 & 0.11 & 0.09 & 0.01 & PL \\ 
        J214836.7-172344 & PKS 2145-17  & 327.153103 & -17.395775 & Quasar  & 952.37 & 920.01 & 140.24 & 1.58 & 2.13 & .. & Dr97 & 0.67 & 0.03 & 0.35 & 0.02 & PL \\ 
        J214941.7-191226 & MRC 2146-194  & 327.423955 & -19.207341 & Galaxy  & 802.07 & 669.19 & 202.44 & 1.01 & .. & .. & .. & 0.15 & 0.07 & 0.25 & 0.01 & PL \\ 
        J215002.8-174537 & NVSS  J215002-174537  & 327.511845 & -17.760509 & Galaxy  & 331.48 & 321.23 & 70.15 & 0.64 & .. & .. & .. & 0.77 & 0.08 & 0.37 & 0.03 & PL \\ 
        J215031.1-220820 & NVSS J215031-220820  & 327.629643 & -22.138931 & Galaxy & 158.81 & 157.03 & 95.3 & 1.41 & .. & .. & .. & < 0.15 & .. & 0.27 & 0.02 & PL \\ 
                 \hline
\end{tabular}

\end{table*}

\begin{table*}[!ht]
\hspace*{-6.0cm}
    \begin{tabular}{lllllrrrllllrllll}
        Source &  NED name  & RAdeg & DEdeg &  Object Type & F\_tot & F\_pk & S/N & z\_phot & z\_spec & err & z\_ref &  NSI & err & NSI & err & SED \\  
        & & \multicolumn{2}{c}{J2000} & (LS 10) & (mJy) & (mJy) & Cont & PICZL & & z\_spec & & ASKAP & NSI & MWA & NSI & \\
    \hline
        J215031.1-223200 & NVSS J215031-223200  & 327.629708 & -22.533454 & Quasar  & 171.19 & 169.88 & 80.74 & 1.7 & 2.113 & 0.638 & Quaia & 1.56 & 0.07 & 0.65 & 0.15 & F \\ 
        J215116.5-223647 & MRC 2148-228 & 327.818997 & -22.613062 & Galaxy & 616.08 & 447.38 & 254.7 & .. & 0.85 & 0.05 & Mc96 & < 0.05 & .. & 0.06 & 0.00 & PL \\ 
        J215150.9-194606 & PKS 2149-20  & 327.962384 & -19.768507 & Galaxy  & 2830.53 & 2631.81 & 342.71 & 0.34 & 0.424 & .. & Ba99 & 0.64 & 0.02 & 0.59 & 0.04 & PL \\ 
        J215255.4-200103 & NVSS J215255-200102  & 328.23124 & -20.017599 & Quasar  & 242.44 & 189.64 & 45.79 & 1.51 & .. & .. & .. & 0.25 & 0.27 & 0.21 & 0.01 & PL \\ 
        J215343.1-200103 & PKS 2150-202     & 328.429811 & -20.01752 & .. & 1203.51 & 1003.95 & 202.68 & .. & .. & .. & .. & 0.19 & 0.07 & 0.20 & 0.01 & PL \\ 
        &&&& \\
        J215445.3-213357 & NVSS J215445-213356 & 328.689165 & -21.565843 & Quasar & 164.62 & 152.42 & 105.21 & 3.97 & 2.561 & 0.401 & Quaia & < 0.14  & .. & 0.26 & 0.02 & PL \\ 
        J215449.8-213551 & NVSS J215449-213550  & 328.707682 & -21.597631 & .. & 168.58 & 162.64 & 106.26 & .. & .. & .. & .. & 0.32 & 0.21 & .. & .. & PL \\ 
        J215513.7-213702 & PKS 2152-218  & 328.807304 & -21.617432 & Quasar  & 158.11 & 152.84 & 90.76 & 0.62 & 0.306 & .. & Du89 & 1.03 & 0.09 & 0.52 & 0.07 & F \\ 
        J215517.3-233805 & NVSS J215517-233806  & 328.822459 & -23.634893 & Galaxy & 168.43 & 158.07 & 50.64 & 1.06 & .. & .. & .. & < 0.36 & .. & 0.17 & 0.02 & PL \\ 
        J215607.3-183741 & PKS 2153-188  & 329.030705 & -18.62812 & Galaxy  & 283.08 & 278.3 & 60.74 & 0.49 & 0.31 & .. & Du89 & 0.34 & 0.21 & 0.31 & 0.02 & PL \\ 
        &&&& \\
        J215729.1-180703 & PKS 2154-183  & 329.371289 & -18.117601 & Quasar  & 1846.47 & 1803.57 & 99.99 & 1.08 & 1.423 & .. & Du89 & 1.1 & 0.06 & 0.94 & 0.06 & PS \\ 
        J215735.6-201746 & PKS 2154-205  & 329.398686 & -20.296199 & Quasar  & 287 & 279.2 & 116.72 & 1.96 & 2.009 & 0.075 & Quaia & 0.77 & 0.08 & 0.65 & 0.04 & PL \\ 
        J215816.4-191810 & PKS 2155-195  & 329.568462 & -19.303013 & .. & 197.14 & 155.18 & 35.56 & .. & .. & .. & .. & < 0.24 & .. & 0.12 & 0.02 & PL \\ 
        J215911.8-185830 & PKS 2156-192  & 329.799417 & -18.975096 & Galaxy & 274.24 & 244.2 & 30.84 & 0.49 & .. & .. & .. & < 0.31 & .. & 0.29 & 0.02 & PL \\ 
        J215938.3-212743 & NVSS  J215938-212743  & 329.909795 & -21.461953 & Galaxy & 163.54 & 155.6 & 86.5 & 1.27 & .. & .. & .. & < 0.23 & .. & 0.23 & 0.02 & PL \\ 
        &&&& \\
        J215957.0-211040 & PKS 2157-214    & 329.987885 & -21.178018 & .. & 441.7 & 423.58 & 183.96 & .. & .. & .. & .. & 0.22 & 0.2 & 0.26 & 0.02 & PL \\ 
        J220027.5-231117 & NVSS  J220027-231118 & 330.114667 & -23.18829 & Quasar & 181.48 & 178 & 22.9 & 0.96 & .. & .. & .. & < 1.21 & .. & 0.83 & 0.06 & PS \\ 
        J220040.2-230722 & NVSS J220040-230723 & 330.167736 & -23.122855 & Galaxy & 224.8 & 221.67 & 28.36 & 0.31 & .. & .. & .. & < 0.98 & .. & 0.82 & 0.05 & PL \\ 
        J220041.1-185540 & PKS 2157-191   & 330.171284 & -18.928032 & Galaxy & 371.27 & 354.47 & 28.63 & 1.54 & .. & .. & .. & < 0.41 & .. & 0.15 & 0.01 & PL \\ 
        J220051.2-220557 & NVSS J220051-220557  & 330.213559 & -22.099424 & Quasar & 255.72 & 183.57 & 68.3 & .. & 1.585 & 0.131 & Quaia & < 0.32 & .. & 0.10 & 0.01 & PL \\ 
        &&&& \\
        J220054.6-194724 & NVSS J220054-194723  & 330.227886 & -19.790154 & Galaxy & 174.52 & 168.76 & 39.22 & 4.79 & .. & .. & .. & < 0.47 & .. & 0.42 & 0.03 & PL \\ 
        J220127.0-202536 & PKS 2158-206  & 330.362846 & -20.426705 & Quasar  & 523.11 & 509.88 & 173.3 & 1.84 & 2.272 & .. & Ba99 & 0.79 & 0.07 & 0.65 & 0.04 & PL \\ 
        J220150.2-211805 & PKS 2159-215  & 330.459264 & -21.301556 & Galaxy  & 328.8 & 318.2 & 119.6 & 1.41 & .. & .. & .. & 0.81 & 0.12 & 0.73 & 0.05 & PL \\ 
        J220212.5-190253 & PKS 2159-192   & 330.552406 & -19.048233 & Galaxy & 466.14 & 447.89 & 23.61 & 1.52 & .. & .. & .. & < 0.50 & .. & 0.31 & 0.02 & PL \\ 
        J220217.1-201748 & PKS 2159-205  & 330.571573 & -20.296709 & Galaxy  & 154.15 & 151.44 & 41.04 & 0.63 & .. & .. & .. & 1.18 & 0.25 & 0.64 & 0.10 & PS \\ 
        &&&& \\
        J220241.8-195509 & PKS 2159-201 & 330.674251 & -19.919429 & Galaxy  & 808.49 & 780.1 & 135.89 & 1.48 & .. & .. & .. & 0.78 & 0.11 & 0.70 & 0.04 & PL \\ 
        J220245.7-214332 & NVSS J220245-21433 & 330.690603 & -21.725633 & Galaxy & 286.68 & 278.87 & 71.64 & 1.52 & .. & .. & .. & < 0.34 & .. & 0.45 & 0.03 & PL \\ 
        J220402.4-212653 & PKS 2201-217     & 331.010277 & -21.448263 & Galaxy & 327.98 & 278.75 & 62.47 & .. & 0.0732 & .. & Ow95 & < 0.41 & .. & < 0.04 & < 0.03 & PL \\ \hline
        \multicolumn{15}{l}{(1) RACS/IPS source name; (2) NED name; (3) IPS Right Ascension (deg); (4) IPS Declination (deg); (5) Legacy Survey DR10 Object type; (6) RACS total flux density (mJy); (7) RACS peak flux density (mJy/beam);}\\
        \multicolumn{15}{l}{(8) IPS Continuum signal to noise; (9) Photometric redshift;(10) Spectroscopic redshift; (11) Spectroscopic redshift uncertainty; (12) Reference catalogue for the spectroscopic redshift; (13) ASKAP NSI;}\\
        \multicolumn{15}{l}{(14) Uncertainty in ASKAP NSI; (15) MWA NSI; (16) Uncertainty in MWA NSI; (17) SED fit (from current work, see Section \ref{sec:SED}).}\\
        \multicolumn{15}{l}{{\it Notes. } Redshift references (z\_ref): Ba99=\cite{baker1999}; Bu70=\cite{burbidge1970}; Dr79=\cite{drinkwater1997}; Du89=\cite{dunlop1989}; Ho08=\cite{holt2008}; }\\
        \multicolumn{15}{l}{Mc96=\cite{McCarthy_1996}; Ow95=\cite{owen1995}; Pe79=\cite{peterson1979}; Quaia=Gaia-unWISE Quasar Catalog \cite{quaia2024}; Ti11=\cite{titov2011}.}\\
        \multicolumn{15}{l}{PL = Power law, PS = Peaked-spectrum, F = Flat, I = Inverted.}
    \end{tabular}

\end{table*}

\begin{table*}[!ht]
\hspace*{-6.0cm}
    \begin{tabular}{lllllrrrllllrllll}
        Source &  NED name  & RAdeg & DEdeg &  Object Type & F\_tot & F\_pk & S/N & z\_phot & z\_spec & err & z\_ref &  NSI & err & NSI & err & SED \\  
        & & \multicolumn{2}{c}{J2000} & (LS 10) & (mJy) & (mJy) & Cont & PICZL & & z\_spec & & ASKAP & NSI & MWA & NSI & \\
    \hline
        J213236.6-222229 & NVSS J213236-222229   & 323.152860 & -22.374862 & .. & 83.97 & 69.21 & 27.93 & .. & .. & .. & .. & 0.43 & 0.15 & 0.38 & 0.04 & PL \\ 
        J213254.3-210942 & NVSS J213254-210942   & 323.226367 & -21.161672 & .. & 59.69 & 57.63 & 8.90 & .. & .. & .. & .. & 0.64 & 0.29 & 0.37 & 0.06 & PL \\ 
        J213313.9-215402 & NVSS J213313-215402   & 323.308118 & -21.900614 & .. & 42.16 & 41.17 & 12.97 & .. & .. & .. & .. & 0.53 & 0.24 & 0.95 & 0.15 & PL \\ 
        J213319.8-224421 & NVSS J213319-224422   & 323.332683 & -22.739322 & Galaxy   & 124.58 & 123.04 & 34.83 & 0.72 & .. & .. & .. & 1.29 & 0.08 & 1.11 & 0.13 & PS \\ 
        J213320.2-181051 & NVSS J213320-181051   & 323.334515 & -18.181087 & Quasar   & 108.21 & 105.57 & 8.50 & 1.79 & .. & .. & .. & 0.86 & 0.24 & 0.65 & 0.09 & PL \\ 
        &&&& \\
        J213409.2-200840 & NVSS J213409-200839   & 323.538430 & -20.144485 & Galaxy   & 83.97 & 29.82 & 7.17 & 0.98 & .. & .. & .. & 0.84 & 0.32 & 0.83 & 0.12 & PL \\ 
        J213418.7-212949 & NVSS J213418-212949   & 323.577937 & -21.497180 & Galaxy   & 59.69 & 36.34 & 11.77 & 0.36 & .. & .. & .. & 0.82 & 0.23 & .. & .. & PS \\ 
        J213535.2-204545 & NVSS J213535-204546  & 323.896863 & -20.762590 & Galaxy & 42.16 & 36.24 & 10.75 & 1.18 & .. & .. & .. & 0.93 & 0.30 & .. & .. & PS \\ 
        J213539.8-200016 & NVSS J213539-200015  & 323.916195 & -20.004637 & .. & 124.58 & 37.78 & 11.42 & .. & .. & .. & .. & 0.63 & 0.23 & .. & .. & PL \\ 
        J213551.3-173710 & NVSS J213551-173708  & 323.963947 & -17.619564 & Galaxy & 108.21 & 63.18 & 8.29 & 1.56 & .. & .. & .. & 0.74 & 0.29 & .. & .. & PL \\ 
        &&&& \\
        J213628.1-230529 & NVSS J213628-230530  & 324.117481 & -23.091422 & Quasar  & 37.13 & 35.47 & 20.82 & 1.96 & .. & .. & .. & 0.97 & 0.19 & .. & .. & PL \\ 
        J213656.0-163730 & MRC 2134-168  & 324.233704 & -16.625202 & .. & 378.14 & 367.62 & 10.64 & .. & .. & .. & .. & 0.81 & 0.23 & 0.95 & 0.06 & PS \\ 
        J213700.3-224847 & NVSS J213700-224848  & 324.251592 & -22.813091 & Quasar  & 80.67 & 80.32 & 46.76 & 0.04 & .. & .. & .. & 0.93 & 0.09 & .. & .. & PS \\ 
        J213741.6-174516 & NVSS J213741-174515  & 324.423437 & -17.754516 & Galaxy  & 109.54 & 106.20 & 14.31 & 1.42 & .. & .. & .. & 0.70 & 0.19 & 0.86 & 0.05 & PS \\ 
        J213827.5-230634 & NVSS J213827-230635  & 324.614718 & -23.109656 & Galaxy  & 72.85 & 71.12 & 18.81 & 5.82 & .. & .. & .. & 1.17 & 0.24 & < 0.21 & < 0.15 & PL \\ 
        &&&& \\
        J213830.6-205022 & NVSS J213830-205022  & 324.627880 & -20.839507 & .. & 126.58 & 120.64 & 17.25 & .. & .. & .. & .. & 0.67 & 0.31 & 0.78 & 0.06 & PL \\ 
        J213914.6-221310 & NVSS J213914-221310 & 324.811219 & -22.219504 & Quasar  & 65.34 & 63.80 & 37.11 & 0.64 & 1.05 & 0.19 & Quaia & 0.73 & 0.14 & .. & .. & PL \\ 
        J213924.7-200359 & NVSS J213924-200358  & 324.852949 & -20.066533 & Galaxy  & 37.03 & 34.39 & 7.80 & 0.14 & .. & .. & .. & 1.16 & 0.40 & .. & .. & PL \\ 
        J213944.6-191225 & NVSS J213944-191224  & 324.936151 & -19.207174 & Quasar  & 73.09 & 70.04 & 19.02 & 1.62 & .. & .. & .. & 0.76 & 0.15 & 0.77 & 0.09 & PL \\ 
        J213958.2-195443 & NVSS J213958-195442  & 324.992716 & -19.912086 & .. & 37.62 & 36.96 & 8.98 & .. & .. & .. & .. & 1.06 & 0.32 & 0.87 & 0.17 & PL \\ 
        &&&& \\
        J214027.3-180904 & NVSS J214027-180904  & 325.113779 & -18.151279 & .. & 56.59 & 40.73 & 6.98 & .. & .. & .. & .. & 0.95 & 0.40 & < 0.00 & < 0.13 & PL \\ 
        J214032.6-213956 & NVSS J214032-213956  & 325.135991 & -21.665589 & Galaxy  & 127.98 & 125.57 & 74.28 & 3.67 & .. & .. & .. & 1.05 & 0.06 & 0.94 & 0.06 & PS \\ 
        J214046.8-192357 & NVSS  J214046-192356  & 325.195243 & -19.399238 & Quasar  & 92.02 & 90.04 & 26.45 & 1.33 & 1.01 & 0.07 & Quaia & 0.94 & 0.11 & .. & .. & I \\ 
        J214118.6-185859 & NVSS J214118-185858  & 325.327624 & -18.983245 & Galaxy  & 31.54 & 30.03 & 9.57 & 1.46 & .. & .. & .. & 0.88 & 0.31 & .. & .. & PS \\ 
        J214132.1-230213 & NVSS J214132-230212  & 325.384129 & -23.037204 & Quasar & 137.83 & 117.73 & 72.06 & 1.50 & .. & .. & .. & 0.25 & 0.22 & 0.25 & 0.02 & PL \\ 
        &&&& \\
        J214154.5-201837 & NVSS J214154-201836  & 325.477475 & -20.310343 & .. & 51.36 & 52.78 & 23.37 & .. & .. & .. & .. & 1.08 & 0.15 & 0.93 & 0.09 & PL \\ 
        J214158.4-194454 & NVSS J214158-194454  & 325.493341 & -19.748524 & Quasar & 46.23 & 39.16 & 17.90 & 1.88 & .. & .. & .. & 1.04 & 0.19 & .. & .. & PL \\ 
        J214215.8-255126 & NVSS J214215-255127  & 325.566127 & -25.857245 & Quasar  & 88.50 & 86.23 & 10.01 & 0.64 & .. & .. & .. & 3.80 & 0.49 & .. & .. & F \\ 
        J214250.0-195153 & NVSS J214249-195151  & 325.708564 & -19.864793 & Quasar  & 48.33 & 46.54 & 22.65 & 1.30 & .. & .. & .. & 0.79 & 0.18 & 0.81 & 0.06 & PL \\ 
        J214320.9-175339 & NVSS J214320-175339  & 325.837470 & -17.894259 & Quasar  & 79.52 & 76.76 & 15.65 & 1.56 & 1.44 & 0.09 & Quaia & 1.05 & 0.20 & .. & .. & F \\ 
        \end{tabular}
\caption{Additional sample}
\label{tab:main_add}

\end{table*}

\begin{table*}[!ht]
\hspace*{-6.0cm}
    \begin{tabular}{lllllrrrllllrllll}
        Source &  NED name  & RAdeg & DEdeg &  Object Type & F\_tot & F\_pk & S/N & z\_phot & z\_spec & err & z\_ref &  NSI & err & NSI & err & SED \\  
        & & \multicolumn{2}{c}{J2000} & (LS 10) & (mJy) & (mJy) & Cont & PICZL & & z\_spec & & ASKAP & NSI & MWA & NSI & \\
    \hline
        J214341.4-173705 & NVSS J214341-173704  & 325.922770 & -17.618244 & Quasar  & 72.29 & 70.38 & 11.90 & 1.72 & .. & .. & .. & 0.80 & 0.30 & 0.58 & 0.05 & PL \\ 
        J214355.8-201756 & NVSS J214355-201755  & 325.982907 & -20.298975 & Quasar & 42.25 & 40.40 & 19.63 & 1.47 & .. & .. & .. & 1.02 & 0.19 & .. & .. & PS \\ 
        J214403.1-221742 & NVSS J214403-221744  & 326.013107 & -22.295104 & Galaxy  & 49.30 & 47.75 & 29.84 & 0.62 & .. & .. & .. & 0.72 & 0.23 & .. & .. & PL \\ 
        J214409.6-195600 & NVSS J214409-195600  & 326.040327 & -19.933403 & .. & 45.92 & 32.20 & 16.67 & .. & .. & .. & .. & 0.71 & 0.26 & < 0.34 & < 0.13 & PL \\ 
        J214420.3-172202 & NVSS J214420-172201  & 326.084880 & -17.367228 & Galaxy  & 135.16 & 131.90 & 26.54 & 0.30 & .. & .. & .. & 0.52 & 0.17 & 0.51 & 0.04 & PL \\ 
        &&&& \\
        J214422.9-205741 & NVSS J214423-205740  & 326.095824 & -20.961471 & .. & 54.51 & 53.91 & 35.51 & .. & .. & .. & .. & 0.68 & 0.22 & 0.82 & 0.08 & PL \\ 
        J214436.3-183650 & NVSS J214436-183650  & 326.151383 & -18.614158 & Galaxy  & 65.54 & 64.20 & 23.16 & 1.66 & .. & .. & .. & 0.90 & 0.16 & .. & .. & PS \\ 
        J214443.9-225345 & NVSS J214444-225346  & 326.183297 & -22.895886 & Galaxy  & 147.78 & 144.78 & 106.37 & 1.42 & .. & .. & .. & 1.05 & 0.05 & 0.80 & 0.06 & PS \\ 
        J214516.0-224029 & NVSS J214515-224026  & 326.317059 & -22.674976 & .. & 58.54 & 40.92 & 11.52 & .. & .. & .. & .. & 1.87 & 0.39 & < 0.02 & < 0.06 & PL \\ 
        J214533.4-234633 & NVSS J214533-234634  & 326.389448 & -23.776107 & .. & 35.93 & 24.67 & 10.49 & .. & .. & .. & .. & 1.95 & 0.45 & < 0.46 & < 0.19 & PL \\ 
        &&&& \\
        J214539.5-231334 & NVSS J214539-231335  & 326.414938 & -23.226289 & Galaxy  & 101.98 & 100.49 & 70.44 & 1.34 & .. & .. & .. & 0.35 & 0.20 & 0.55 & 0.04 & PL \\ 
        J214557.7-184553 & NVSS J214557-184552  & 326.490533 & -18.764743 & Galaxy & 42.67 & 41.65 & 17.98 & 1.56 & .. & .. & .. & 0.69 & 0.26 & .. & .. & PS \\ 
        J214615.8-203846 & NVSS J214615-203843  & 326.565841 & -20.646186 & Quasar  & 66.00 & 54.04 & 42.25 & 0.66 & 0.69 & 0.07 & Quaia & 0.71 & 0.16 & 0.70 & 0.07 & PL \\ 
        J214618.5-241422 & NVSS J214618-241422  & 326.577187 & -24.239499 & Galaxy  & 115.72 & 113.30 & 60.90 & 1.20 & .. & .. & .. & 1.07 & 0.11 & 0.78 & 0.09 & PS \\ 
        J214622.4-203536 & NVSS J214622-203536  & 326.593491 & -20.593582 & Quasar  & 51.18 & 47.65 & 32.57 & 1.40 & .. & .. & .. & 1.26 & 0.14 & .. & .. & PS \\ 
        &&&& \\
        J214632.3-232633 & NVSS J214632-232633  & 326.634757 & -23.442640 & Galaxy  & 82.78 & 81.15 & 48.90 & 1.49 & .. & .. & .. & 1.04 & 0.13 & .. & .. & PL \\ 
        J214657.2-191339 & NVSS J214657-191339  & 326.738520 & -19.227702 & Quasar  & 146.13 & 138.46 & 52.13 & 0.66 & .. & .. & .. & 0.76 & 0.08 & 0.55 & 0.04 & PL \\ 
        J214716.7-222128 & NVSS J214716-222129   & 326.819745 & -22.357856 & .. & 134.58 & 111.87 & 70.80 & .. & .. & .. & .. & 0.23 & 0.23 & 0.10 & 0.02 & PL \\ 
        J214728.0-253503 & NVSS J214728-253503  & 326.866827 & -25.584234 & .. & 92.03 & 49.61 & 7.10 & .. & .. & .. & .. & 5.64 & 0.70 & < 0.00 & < 0.10 & PL \\ 
        J214733.7-225527 & NVSS J214733-225528  & 326.890742 & -22.924388 & .. & 63.78 & 62.78 & 47.24 & .. & .. & .. & .. & 0.89 & 0.16 & 0.93 & 0.09 & PS \\ 
        &&&& \\
        J214740.1-192730 & NVSS J214740-192729  & 326.917190 & -19.458476 & Galaxy  & 63.22 & 60.99 & 19.21 & 0.88 & .. & .. & .. & 0.84 & 0.21 & 0.78 & 0.10 & PL \\ 
        J214747.8-225315 & NVSS J214748-225316  & 326.949429 & -22.887612 & Galaxy & 42.16 & 36.08 & 23.32 & 0.28 & .. & .. & .. & 1.33 & 0.24 & .. & .. & PS \\ 
        J214811.4-201228 & NVSS J214811-201227  & 327.047864 & -20.207980 & Galaxy  & 29.91 & 28.84 & 17.52 & 0.51 & .. & .. & .. & 1.23 & 0.29 & .. & .. & PS \\ 
        J214854.6-190333 & NVSS J214854-190332  & 327.227709 & -19.059337 & Galaxy  & 33.55 & 31.28 & 9.04 & 1.45 & .. & .. & .. & 0.97 & 0.44 & .. & .. & PS \\ 
        J215001.7-182226 & NVSS J215001-182226  & 327.507085 & -18.374129 & Quasar  & 121.03 & 118.49 & 23.07 & 1.50 & 1.26 & 0.28 & Quaia & 0.77 & 0.22 & 0.65 & 0.17 & F \\ 
        &&&& \\
        J215018.8-215811 & NVSS J215018-215810  & 327.578340 & -21.969865 & Galaxy  & 125.84 & 121.53 & 75.04 & 0.80 & .. & .. & .. & 0.25 & 0.25 & 0.19 & 0.03 & PL \\ 
        J215036.2-172049 & NVSS J215036-172049  & 327.650846 & -17.347121 & .. & 85.64 & 85.08 & 14.16 & .. & .. & .. & .. & 0.97 & 0.34 & 0.87 & 0.06 & PS \\ 
        J215050.9-232908 & NVSS J215051-232907  & 327.712401 & -23.485575 & .. & 51.33 & 50.62 & 29.00 & .. & .. & .. & .. & 0.80 & 0.31 & 0.41 & 0.05 & PL \\ 
        J215103.1-231407 & NVSS J215103-231407  & 327.763299 & -23.235386 & Galaxy  & 87.85 & 86.32 & 46.51 & 1.10 & .. & .. & .. & 0.47 & 0.25 & 0.35 & 0.05 & PL \\ 
        J215156.4-215122 & NVSS J215156-215120  & 327.985017 & -21.856229 & Quasar & 45.98 & 41.70 & 30.94 & 1.84 & .. & .. & .. & 0.86 & 0.28 & < 0.00 & < 0.35 & PL \\ 
        \end{tabular}

\end{table*}

\begin{table*}[!ht]
\hspace*{-6.0cm}
    \begin{tabular}{lllllrrrllllrllll}
        Source &  NED name  & RAdeg & DEdeg &  Object Type & F\_tot & F\_pk & S/N & z\_phot & z\_spec & err & z\_ref &  NSI & err & NSI & err & SED \\  
        & & \multicolumn{2}{c}{J2000} & (LS 10) & (mJy) & (mJy) & Cont & PICZL & & z\_spec & & ASKAP & NSI & MWA & NSI & \\
    \hline
        J215207.9-194148 & NVSS J215208-194147  & 328.033155 & -19.696698 & .. & 41.09 & 21.36 & 6.72 & .. & .. & .. & .. & 1.90 & 0.62 & .. & .. & PL \\ 
        J215229.9-230947 &  NVSS J215230-230948  & 328.124892 & -23.163178 & Quasar  & 37.07 & 36.21 & 22.72 & 1.71 & 1.64 & 0.12 & Quaia & 1.04 & 0.33 & .. & .. & PS \\ 
        J215249.5-204422 & NVSS J215249-204420  & 328.206652 & -20.739479 & Galaxy & 39.31 & 37.99 & 13.87 & 5.60 & .. & .. & .. & 1.09 & 0.38 & 1.00 & 0.18 & PL \\ 
        J215336.3-215031 & NVSS J215336-215031  & 328.401436 & -21.842033 & .. & 61.52 & 60.75 & 37.65 & .. & .. & .. & .. & 0.83 & 0.24 & < 0.00 & < 0.39 & PL \\ 
        J215346.2-223859 & NVSS J215346-223900  & 328.442640 & -22.649876 & Galaxy  & 30.87 & 30.73 & 6.20 & 0.38 & .. & .. & .. & 4.49 & 0.79 & .. & .. & PL \\ 
        &&&& \\
        J215438.6-190220 & NVSS J215438-190219  & 328.660876 & -19.038911 & Galaxy  & 87.43 & 85.91 & 27.39 & 1.50 & .. & .. & .. & 0.77 & 0.23 & 0.81 & 0.05 & PL \\ 
        J215503.3-202612 & NVSS J215503-202612  & 328.764132 & -20.436822 & .. & 72.85 & 70.65 & 25.64 & .. & .. & .. & .. & 1.19 & 0.21 & 0.80 & 0.10 & PL \\ 
        J215601.7-223907 & NVSS J215601-223908  & 329.007413 & -22.652022 & Quasar  & 93.36 & 92.90 & 57.45 & 1.50 & .. & .. & .. & 1.12 & 0.17 & 0.75 & 0.05 & PL \\ 
        J215850.1-220831 & NVSS J215850-220833  & 329.709061 & -22.142214 & Galaxy  & 76.09 & 75.25 & 45.24 & 0.10 & .. & .. & .. & 0.81 & 0.34 & 0.84 & 0.06 & PS \\ 
        \hline
        \multicolumn{15}{l}{See Table \ref{tab:main_bright} for notes.}
    \end{tabular}
\end{table*}
\end{landscape}

\section{RADIO SEDS}
\subsection{Peaked and Compact Sources}
Figure \ref{fig:compact_peaked_seds} shows the radio SED fitting for every source in the final sample (as shown in Tables \ref{tab:main_bright} and \ref{tab:main_add}) that was classified as compact (an NSI > 0.8 from ASKAP), and had a peaked SED. This list does not include the two already shown in Figures \ref{fig:spectra_sed_J212805} and \ref{fig:spectra_sed_J213437}.

\begin{figure*}
    \centering
    \includegraphics[width=0.45\linewidth]{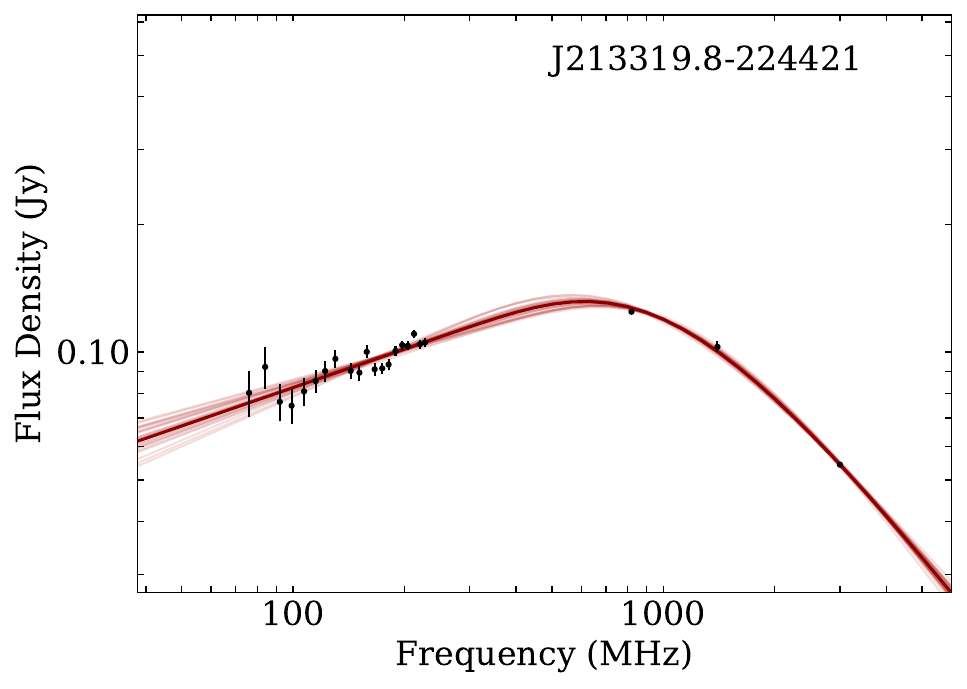}
    \includegraphics[width=0.45\linewidth]{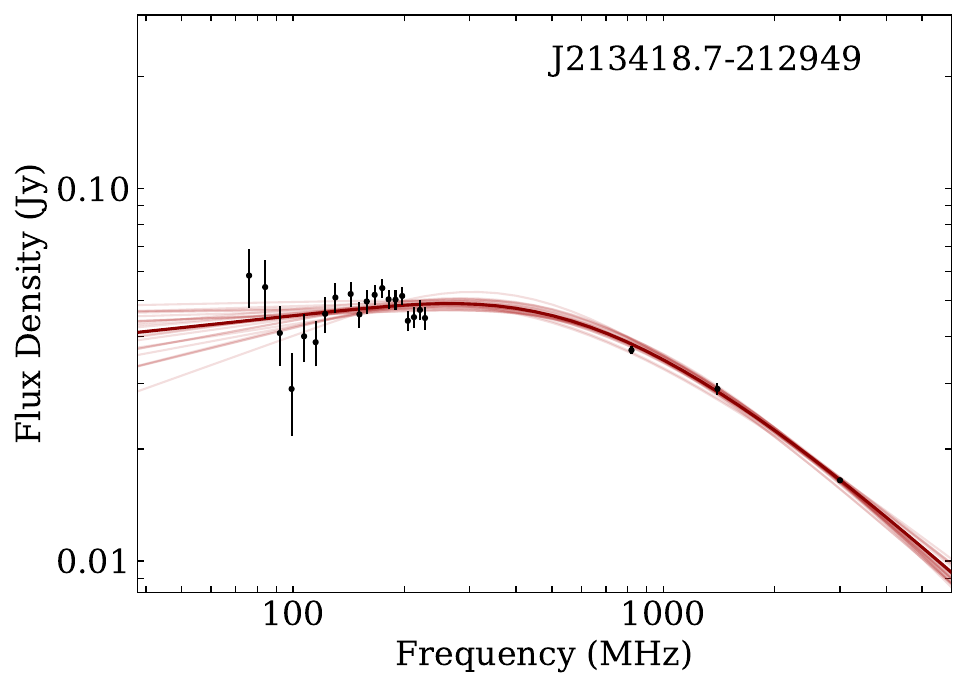}
    
    \includegraphics[width=0.45\linewidth]{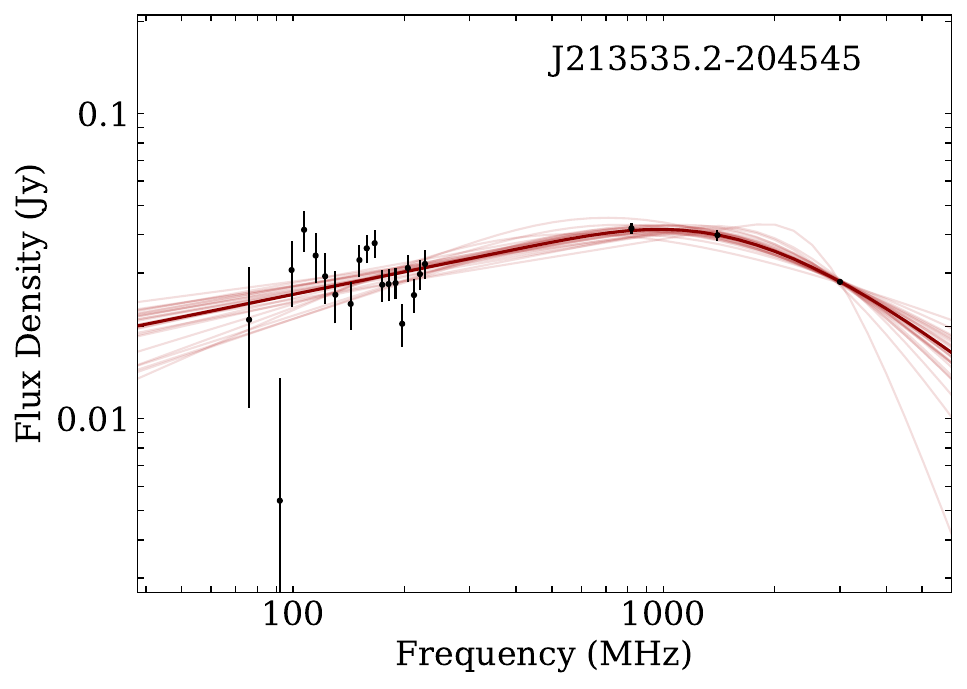}
    \includegraphics[width=0.45\linewidth]{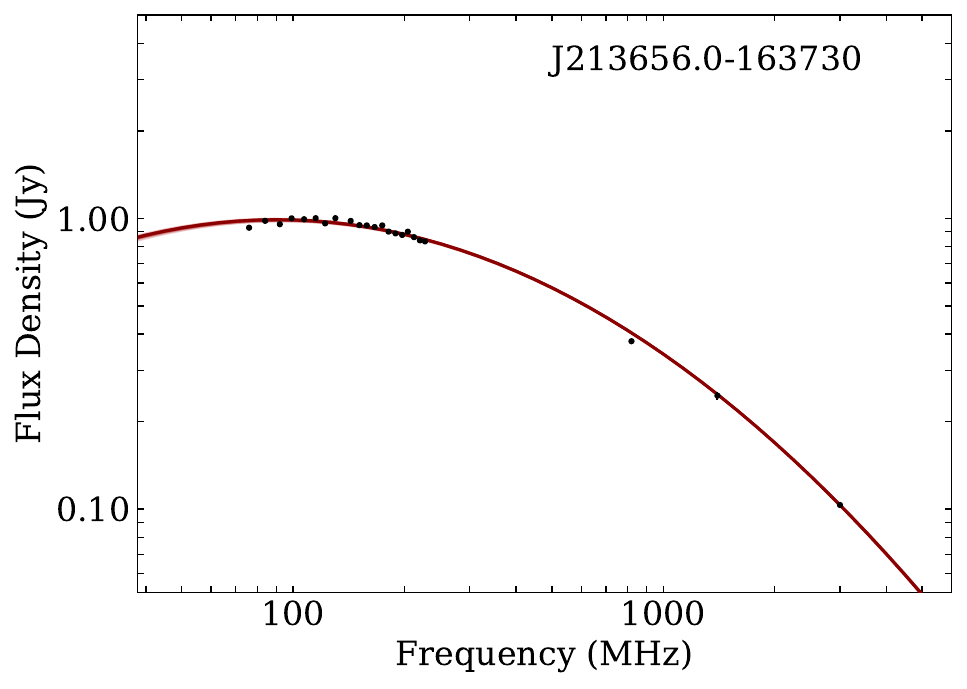}
    
    \includegraphics[width=0.45\linewidth]{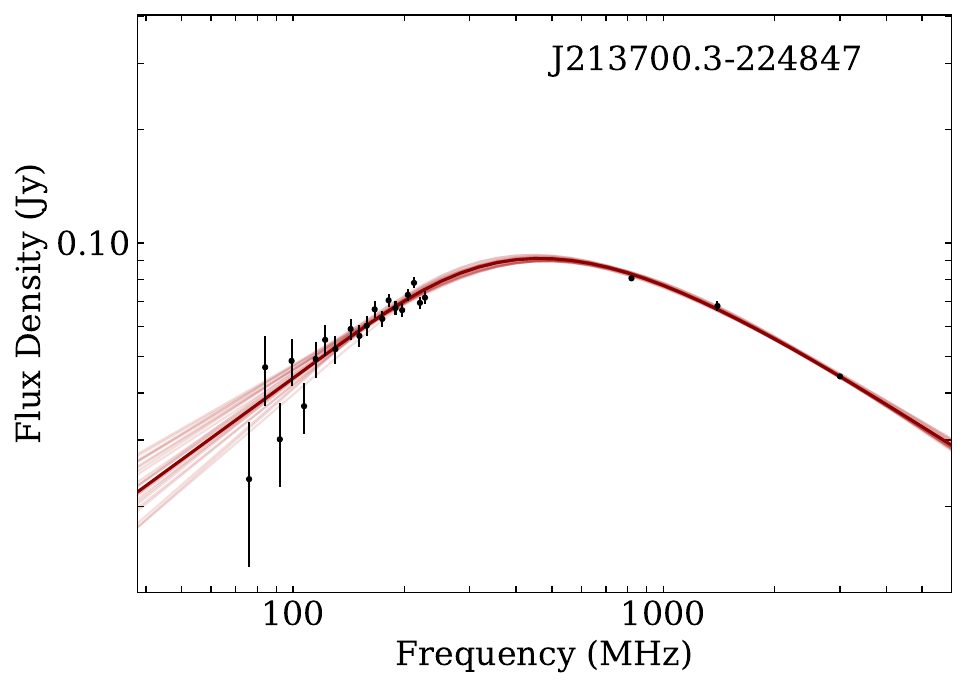}
    \includegraphics[width=0.45\linewidth]{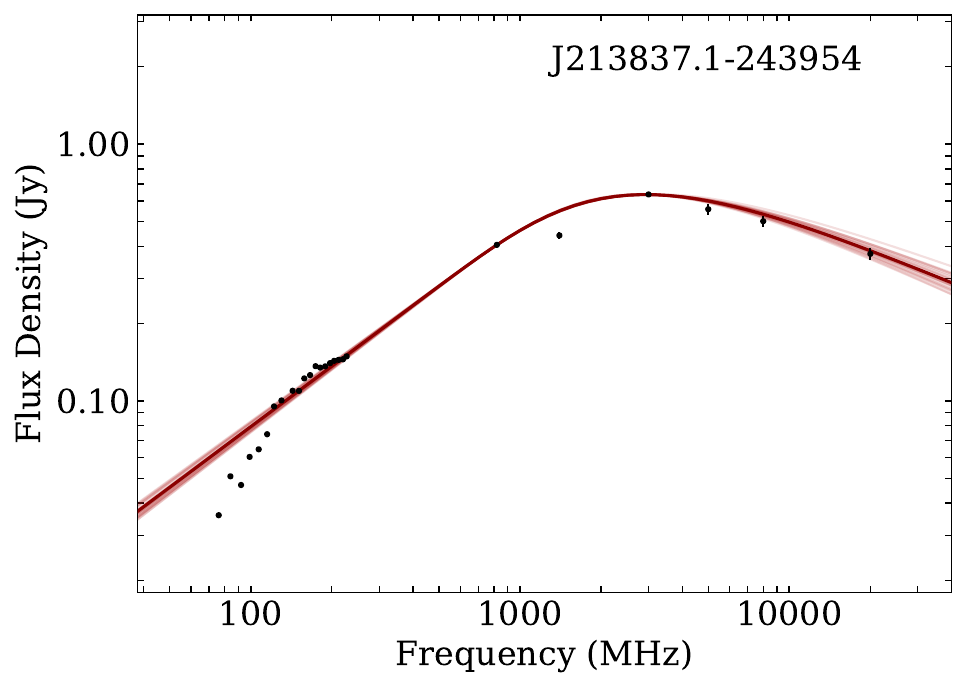}

    \caption{Radio SED fits for all of the compact sources in the final sample that had PS SEDs, not including those already shown in Figures \ref{fig:spectra_sed_J212805} and \ref{fig:spectra_sed_J213437}. This figure continues onto the next 3 pages.}
    \label{fig:compact_peaked_seds}
\end{figure*}

\begin{figure*}\ContinuedFloat
    \centering
    \includegraphics[width=0.45\linewidth]{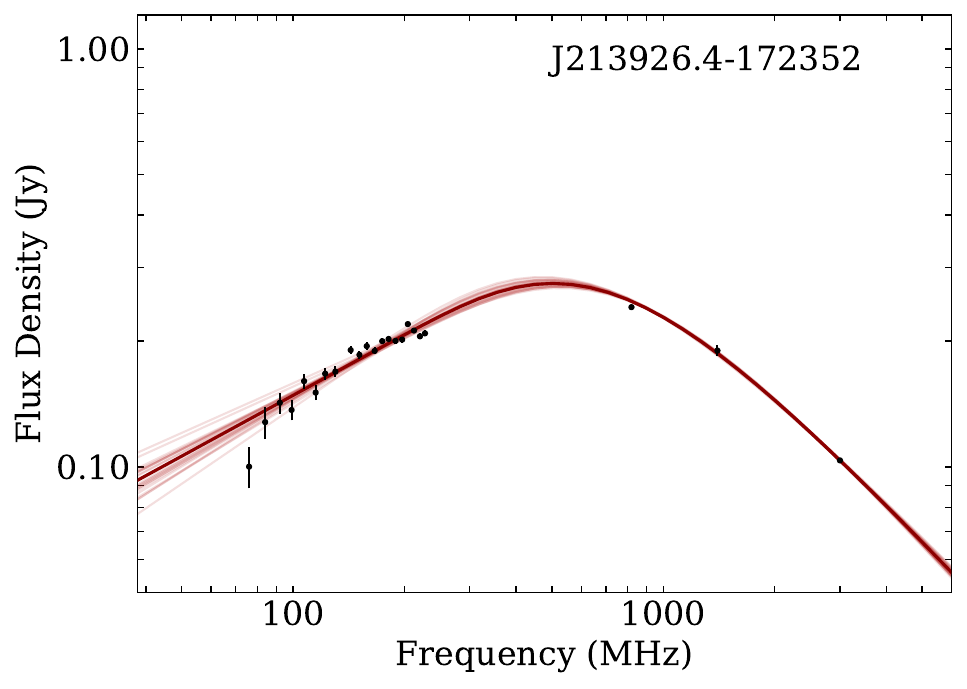}
    \includegraphics[width=0.45\linewidth]{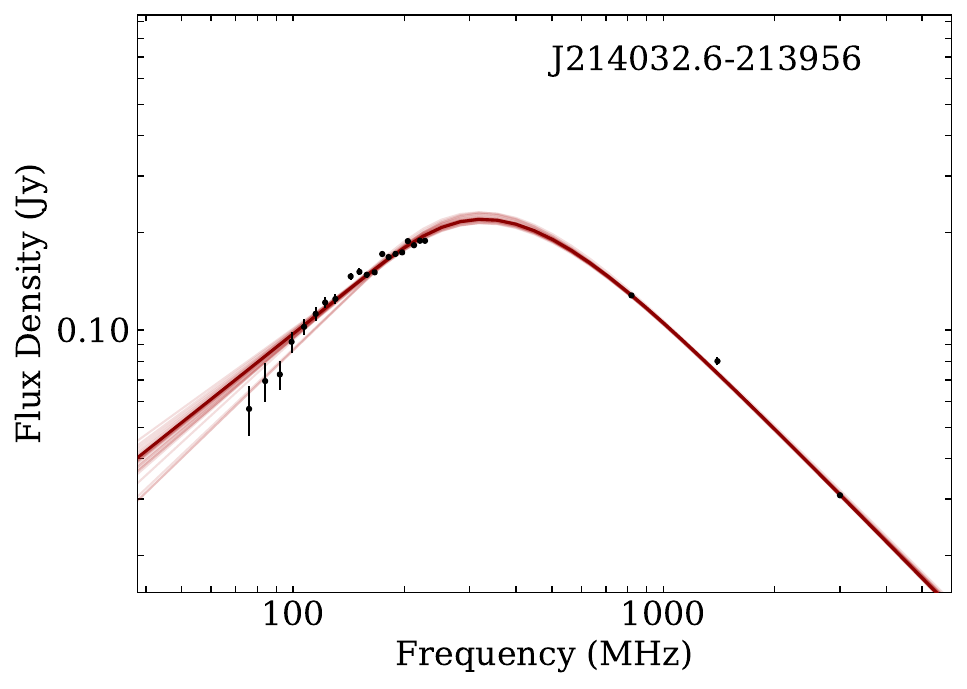}

    \includegraphics[width=0.45\linewidth]{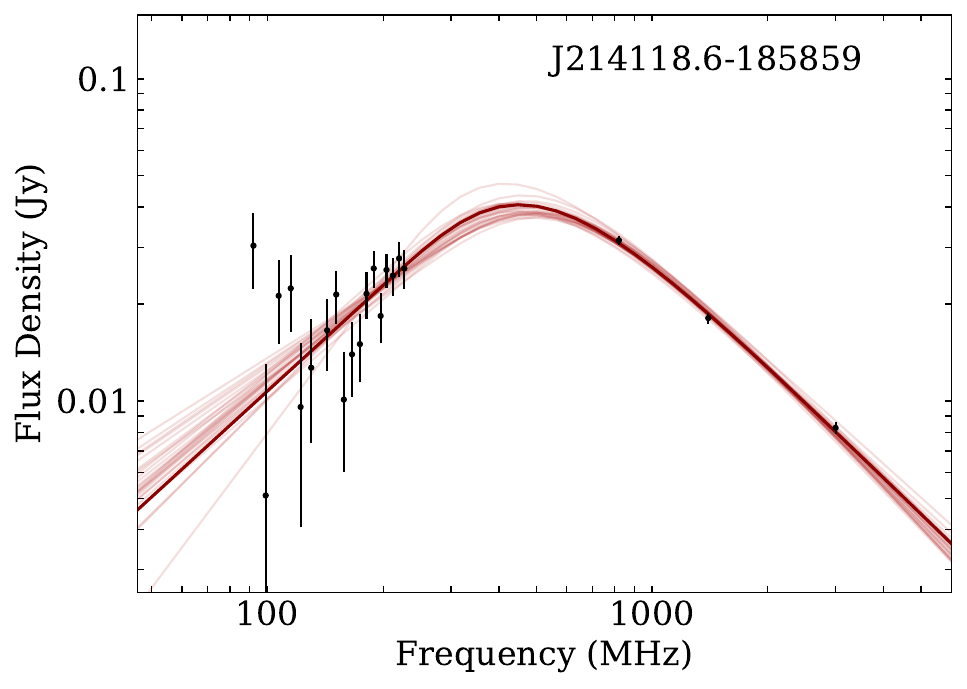}
    \includegraphics[width=0.45\linewidth]{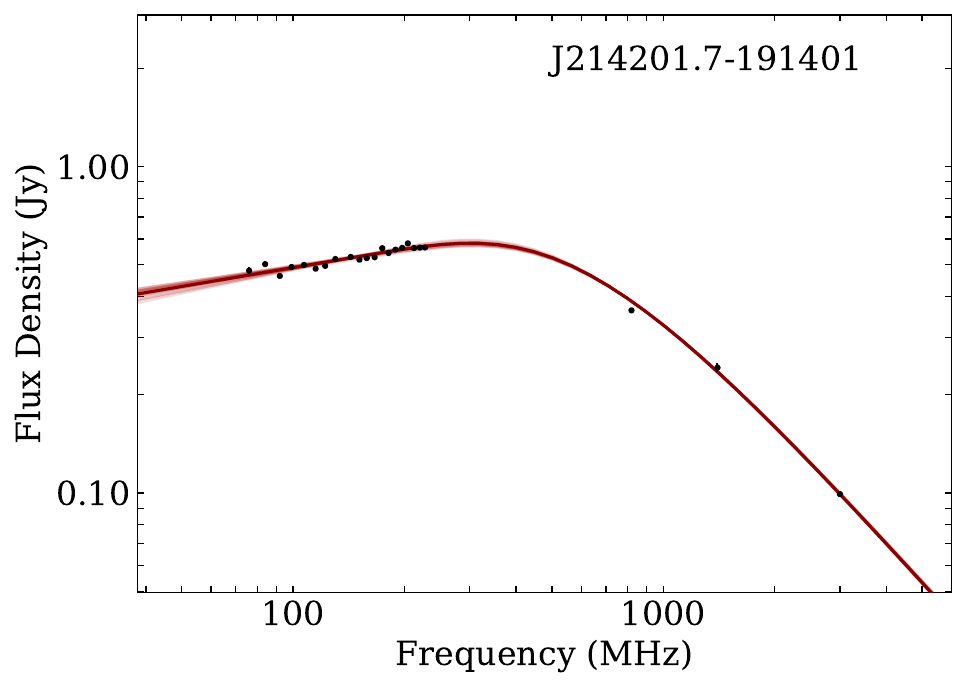}
    
    \includegraphics[width=0.45\linewidth]{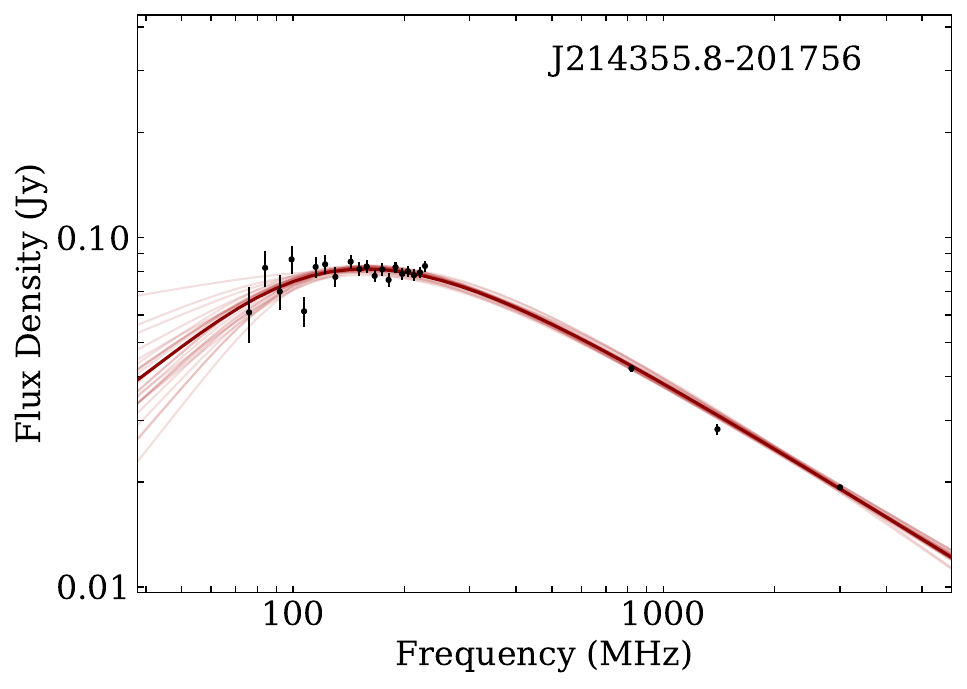}
    \includegraphics[width=0.45\linewidth]{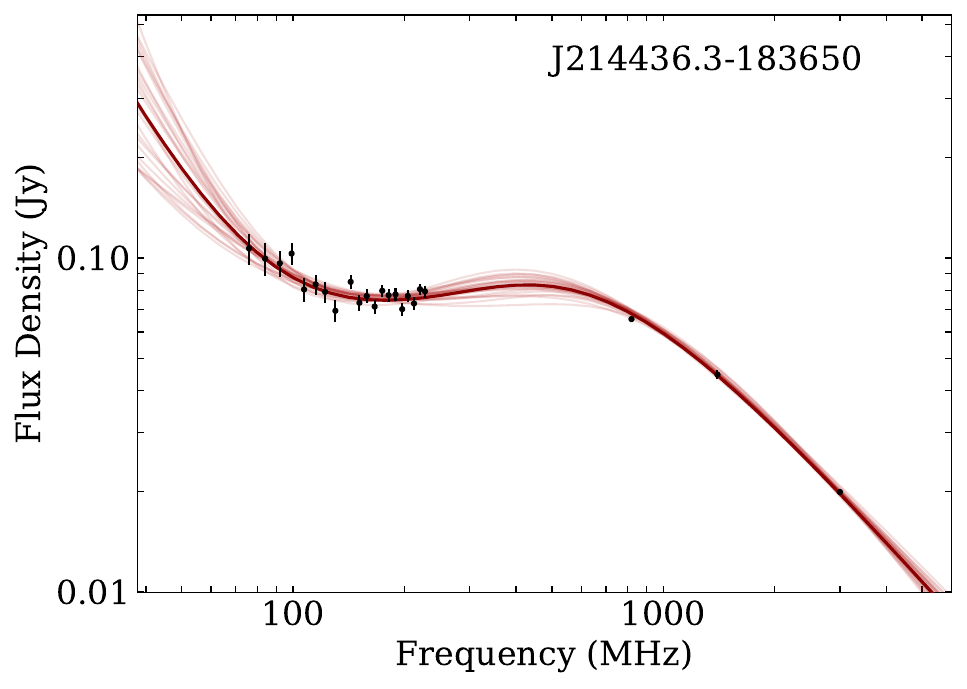}

    \includegraphics[width=0.45\linewidth]{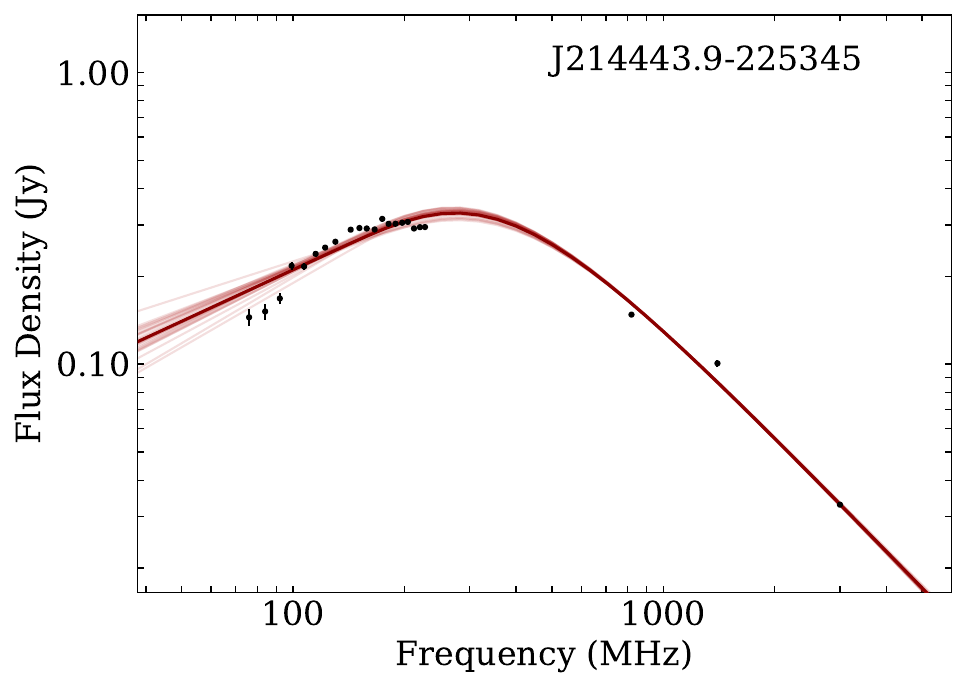}
    \includegraphics[width=0.45\linewidth]{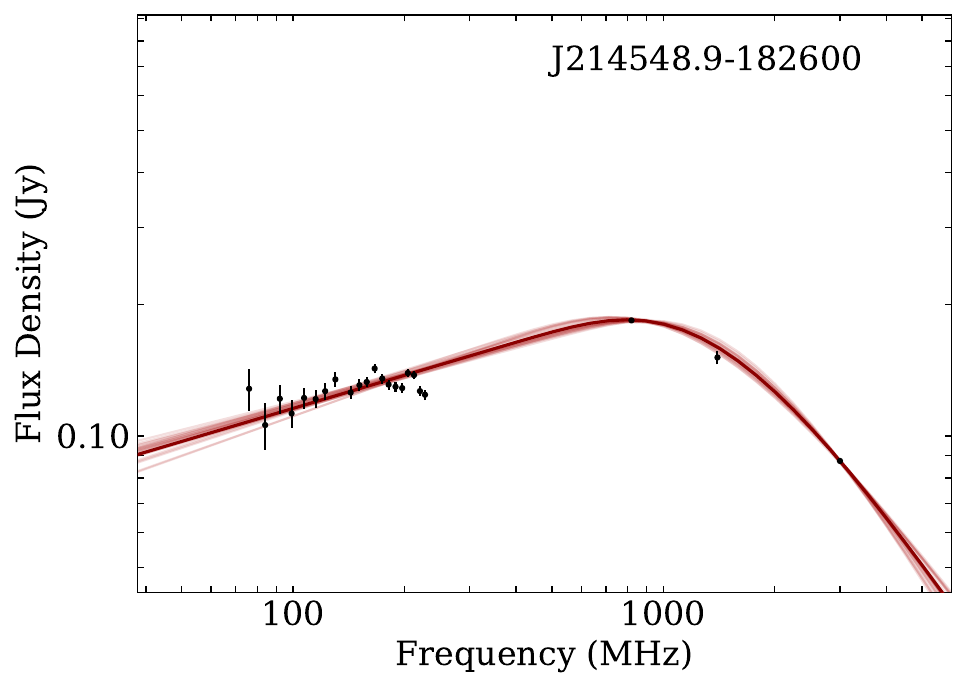}
    
\end{figure*}

\begin{figure*}\ContinuedFloat
    \centering
    \includegraphics[width=0.45\linewidth]{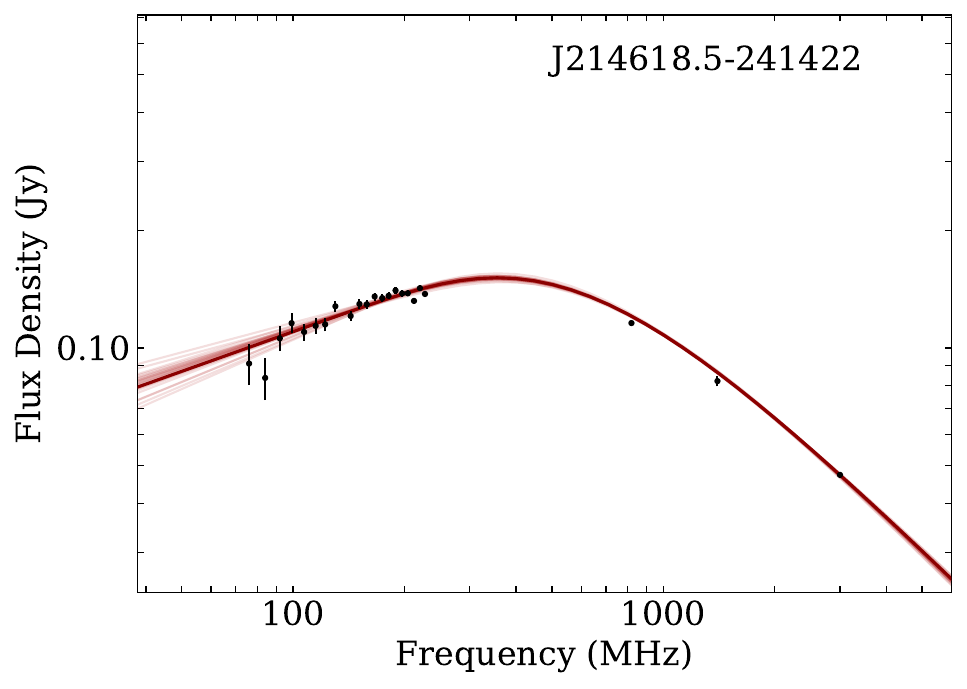}
    \includegraphics[width=0.45\linewidth]{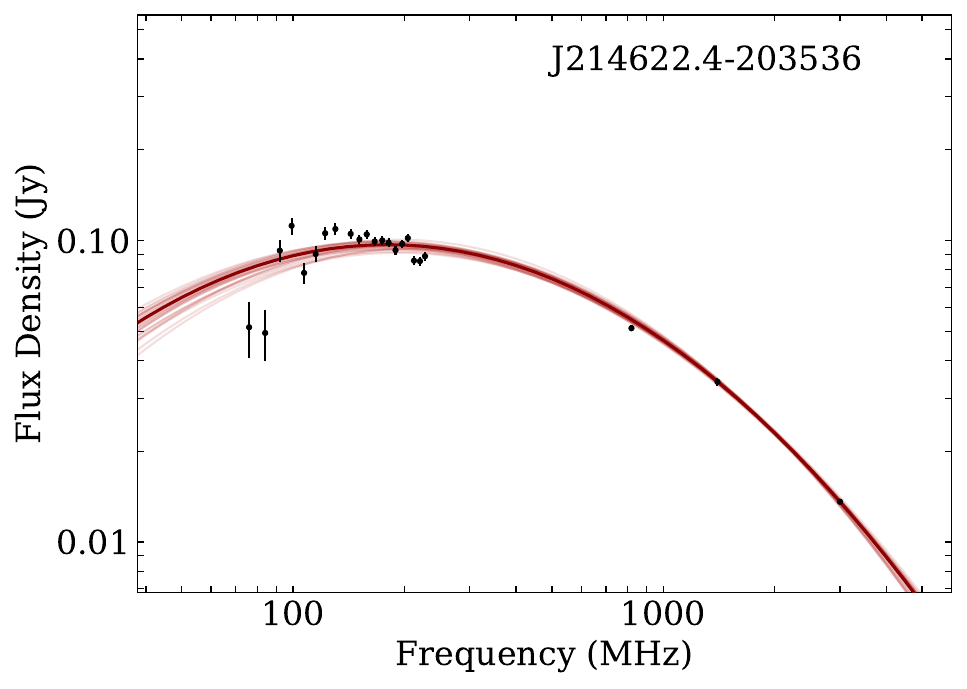}

    \includegraphics[width=0.45\linewidth]{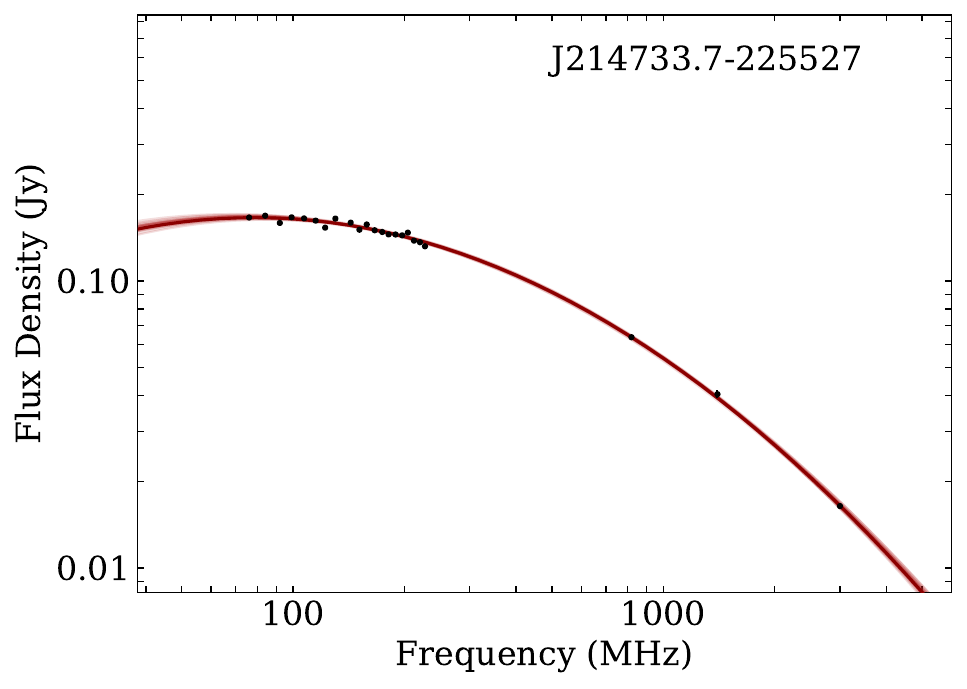}
    \includegraphics[width=0.45\linewidth]{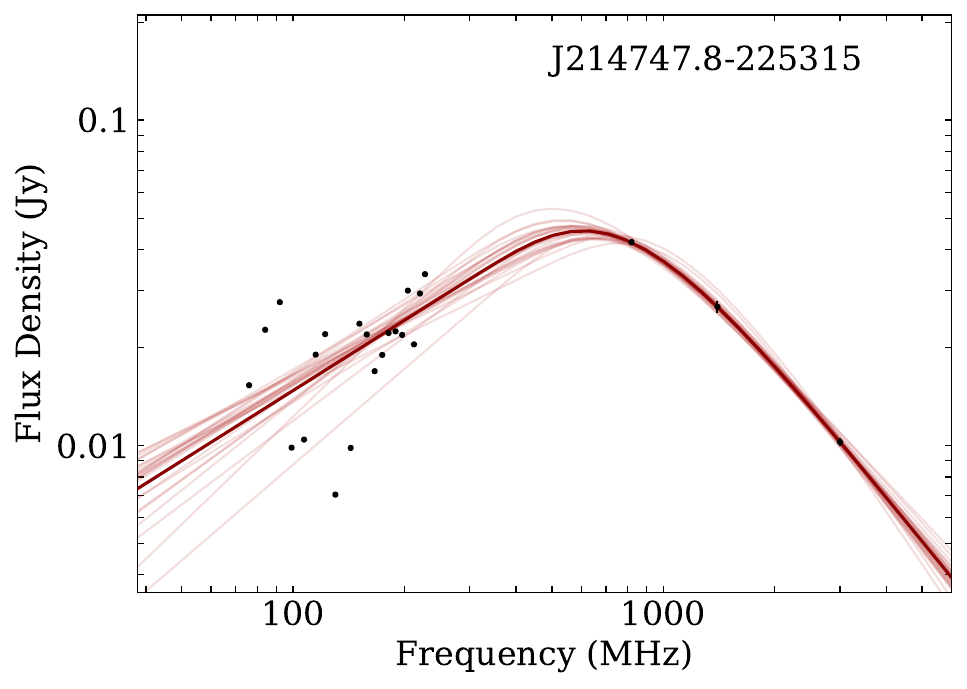}

    \includegraphics[width=0.45\linewidth]{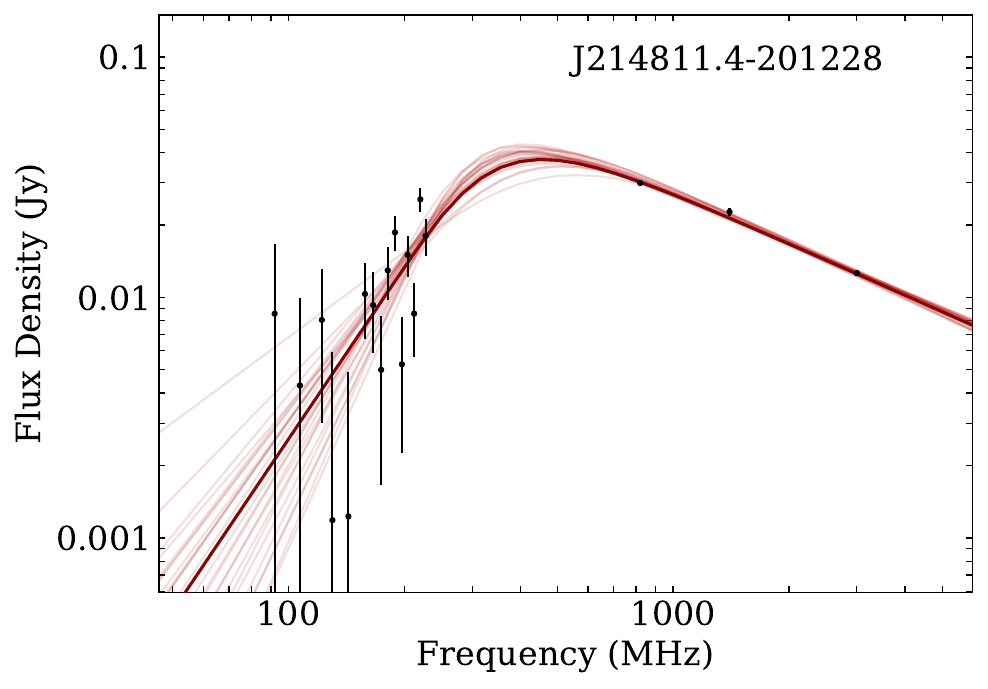}
    \includegraphics[width=0.45\linewidth]{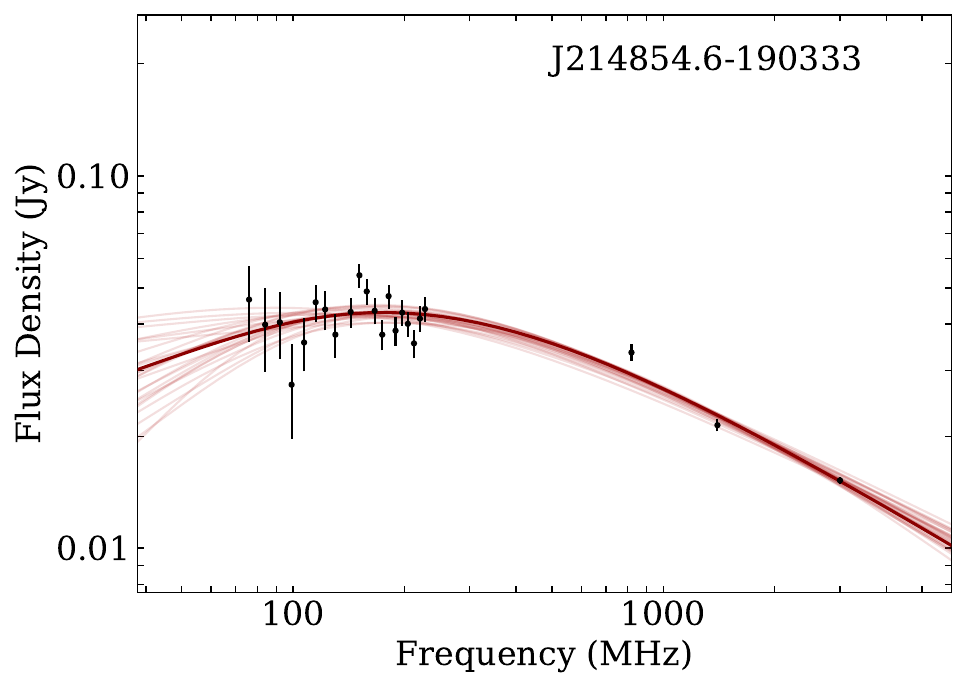}
    
    \includegraphics[width=0.45\linewidth]{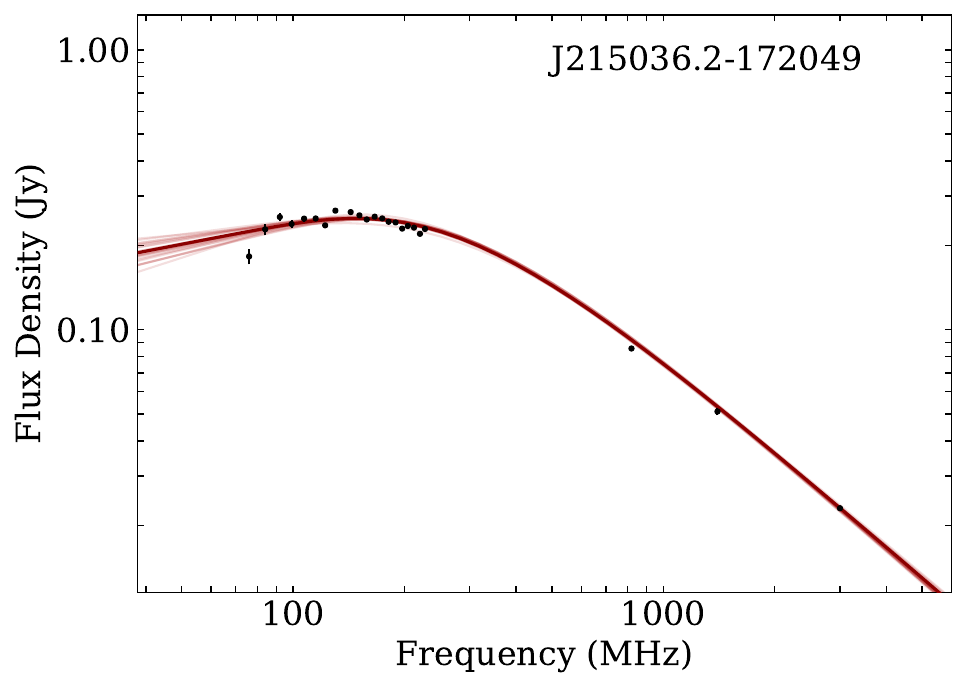}
    \includegraphics[width=0.45\linewidth]{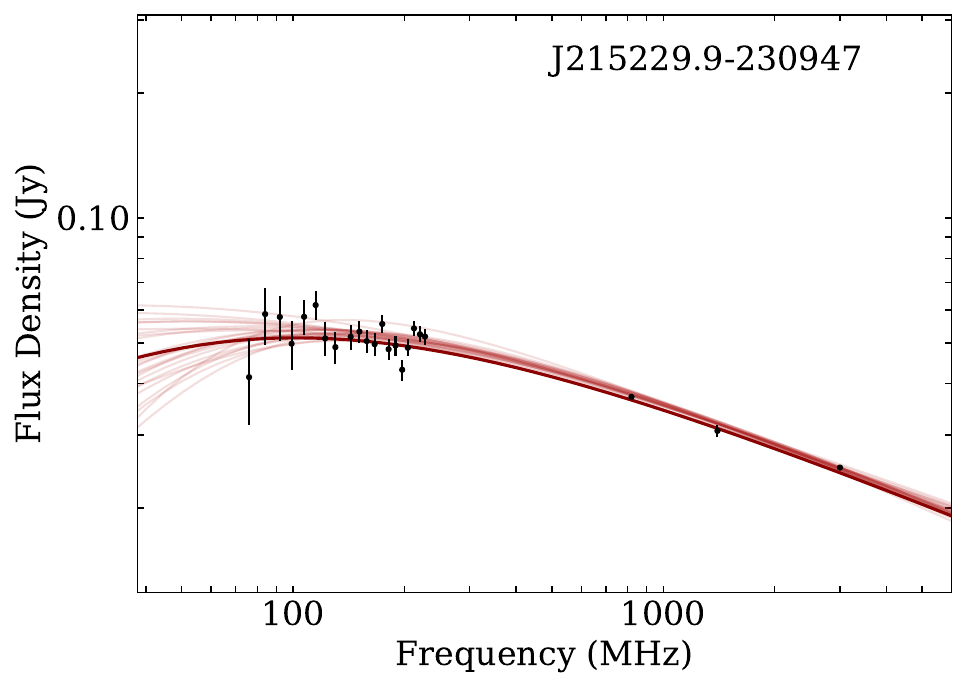}
    
\end{figure*}

\begin{figure*}\ContinuedFloat
    \centering
    \includegraphics[width=0.45\linewidth]{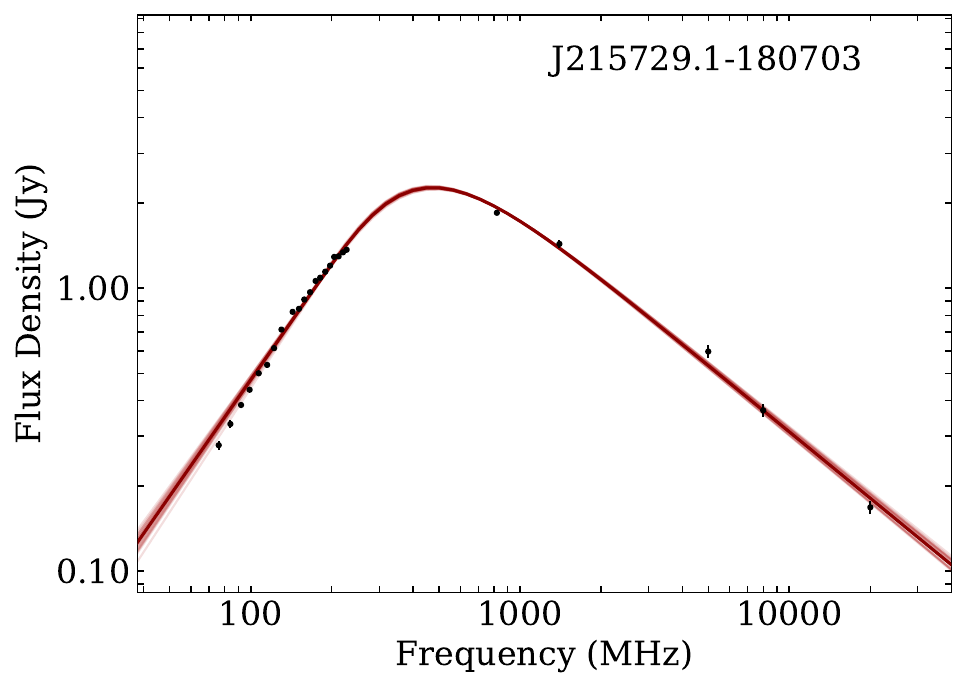}
    \includegraphics[width=0.45\linewidth]{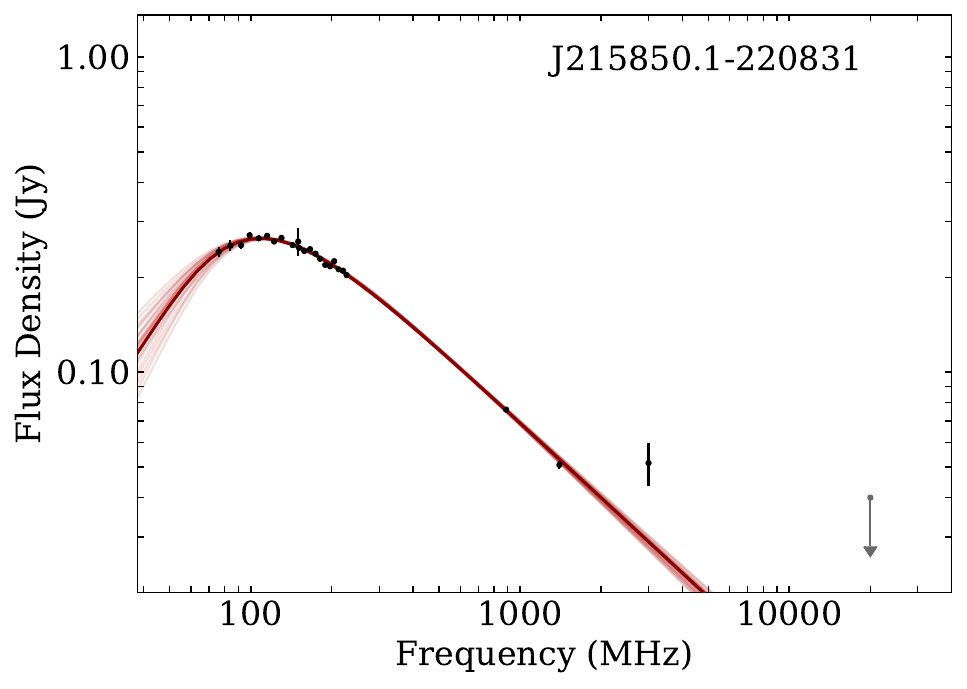}
    \includegraphics[width=0.45\linewidth]{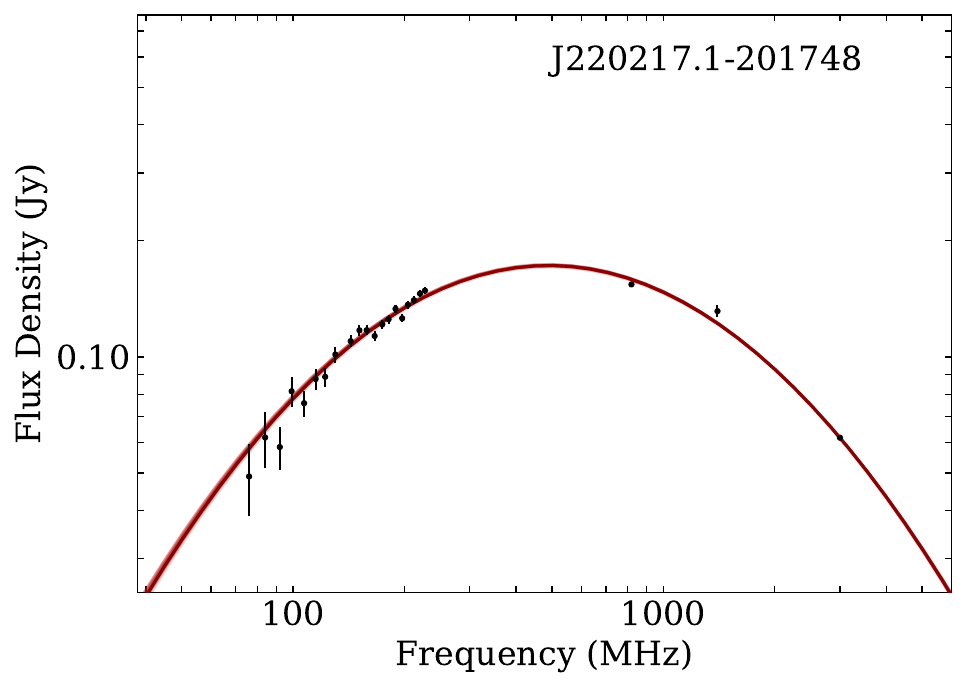}
\end{figure*}

\subsection{Peaked and Slightly Resolved Sources}
\begin{figure*}
    \centering
    \includegraphics[width=0.45\linewidth]{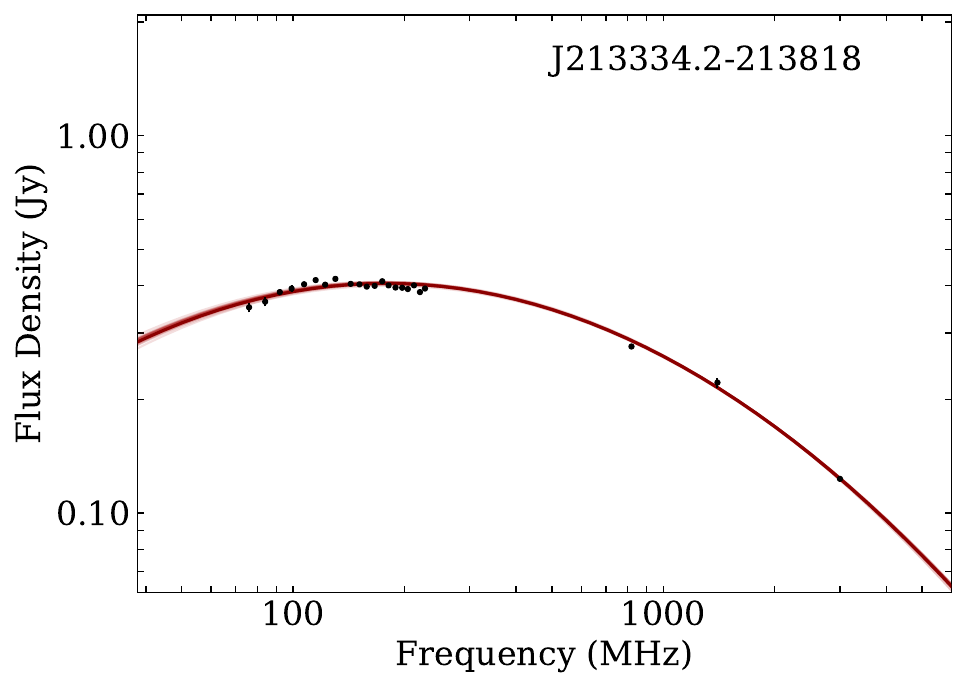}
    \includegraphics[width=0.45\linewidth]{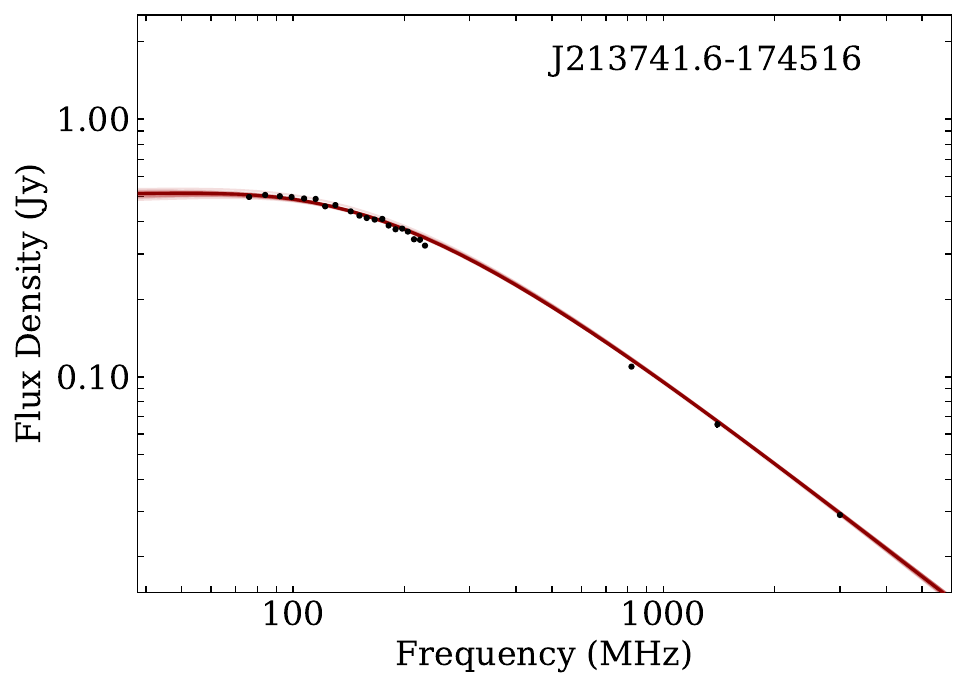}

    \includegraphics[width=0.45\linewidth]{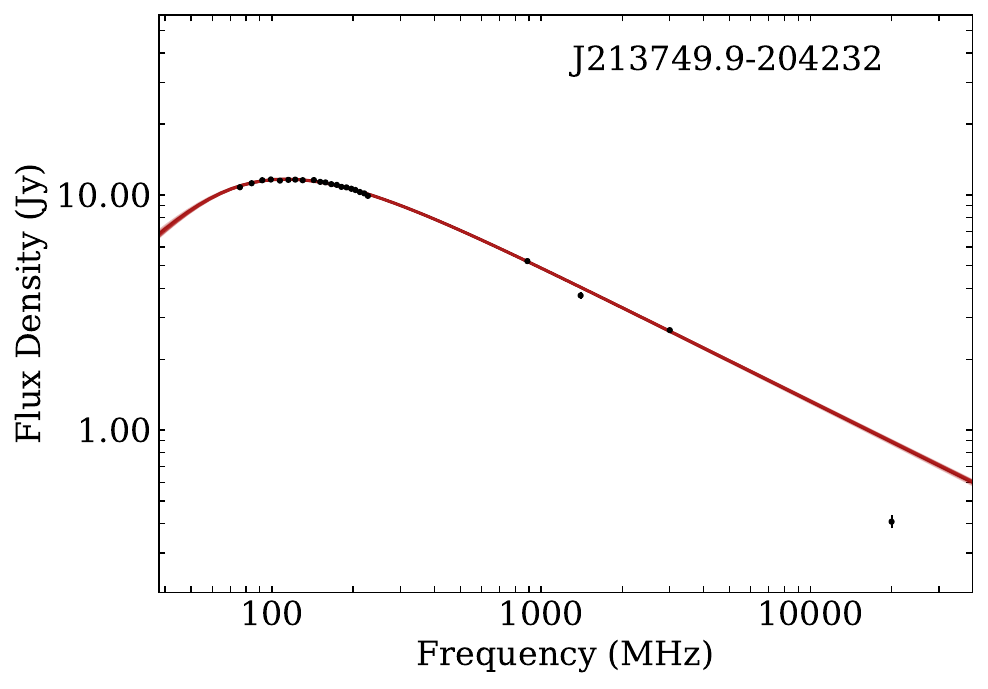}
    \includegraphics[width=0.45\linewidth]{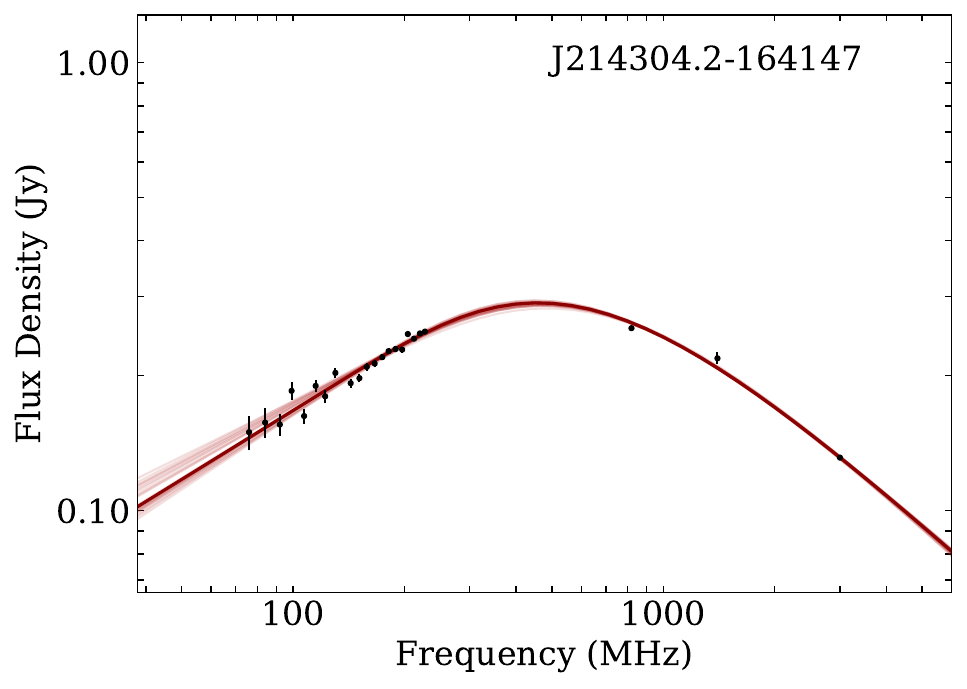}
    \includegraphics[width=0.45\linewidth]{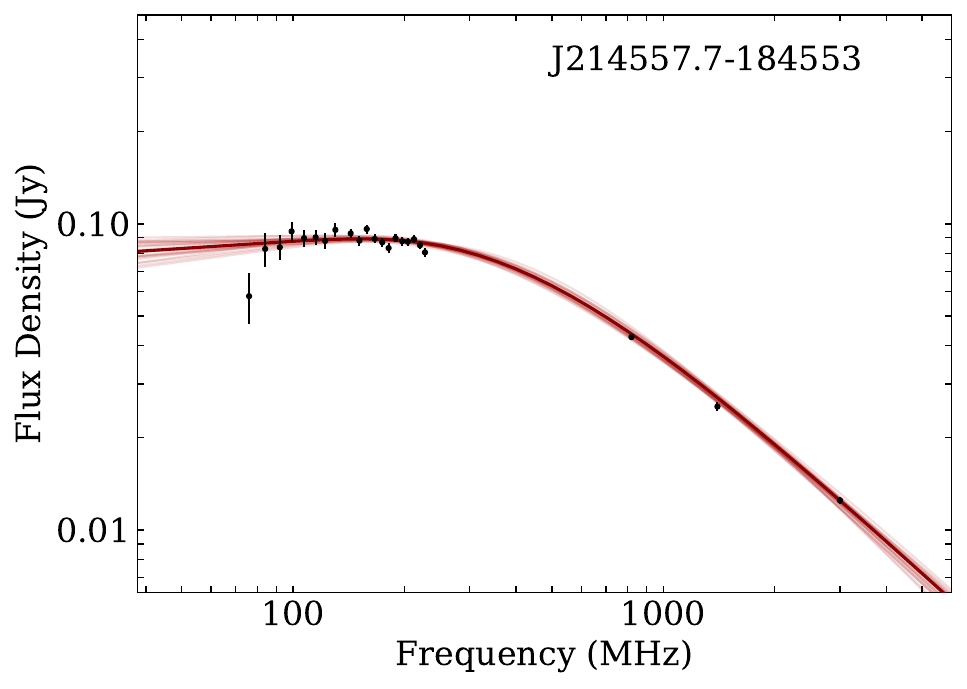}

    \caption{Radio SED fits for all of the slightly resolved sources in the final sample that had PS SEDs.}
    \label{fig:extended_peaked_seds}
\end{figure*}

Figure \ref{fig:extended_peaked_seds} shows the radio SED fitting for every source in the final sample that was classed as slightly resolved ($0.4 < \text{NSI} < 0.8$ from ASKAP), and had a peaked SED. These were included as sometimes there can be confusion in the beam if there are nearby bright sources.

These sources have an ASKAP NSI between 0.59 and 0.76, so while they are below the cutoff of 0.8, they are still relatively compact.

\end{document}